\documentclass{aa}  

\usepackage{xcolor}
\usepackage{graphicx}
\usepackage[varg]{txfonts}
\usepackage[unicode,colorlinks,allcolors=blue]{hyperref}
\usepackage[english]{babel}
\usepackage{float}
\usepackage{caption}
\usepackage{subcaption}
\bibpunct{(}{)}{;}{a}{}{,} 

\begin{document}
\newcommand{\I}{SEDCSJ1301144$-$1147490}
\newcommand{\II}{SEDCSJ1301349$-$1146512}
\newcommand{\III}{SEDCSJ1300003$-$1144203}
\newcommand{\IV}{SEDCSJ1302415$-$1143157}
\newcommand{\V}{EDCSNJ1301344$-$1142380}
\newcommand{\VI}{EDCSNJ1301323$-$1141558}
\newcommand{\VII}{EDCSNJ1301441$-$1140589}
\newcommand{\VIII}{EDCSNJ1301383$-$1140011}
\newcommand{\IX}{EDCSNJ1301363$-$1138495}
\newcommand{\X}{EDCSNJ1301336$-$1138071}
\newcommand{\Xb}{EDCSNJ1301336$-$1138090}
\newcommand{\XI}{EDCSNJ1301524$-$1138043}
\newcommand{\XII}{SEDCSJ1302036$-$1137519}
\newcommand{\XIII}{SEDCSJ1300324$-$1137445}
\newcommand{\XIV}{SEDCSJ1301216$-$1136480}
\newcommand{\XV}{SEDCSJ1302292$-$1133316}
\newcommand{\XVI}{SEDCSJ1300234$-$1132052}
\newcommand{\XVII}{SEDCSJ1301380$-$1127055}
\newcommand{\XVIII}{SEDCSJ1300340$-$1127269}
\newcommand{\XIX}{SEDCSJ1301240$-$1132137}
\newcommand{\XXII}{SEDCSJ1301044$-$1146232}
\newcommand{\XXIII}{SEDCSJ1301007$-$1147075}
\newcommand{\MS}{main-sequence}
\newcommand{\muhh}{$\mu_\mathrm{H_2}$}
\newcommand{\Mstar}{$M_{\rm star}$}
\newcommand{\cluster}{CL1301.7$-$1139}
\newcommand{\clusterb}{CL1411.1$-$1148}
\defcitealias{Sperone-Longin2021}{Paper~I}

 \title{SEEDisCS}
 \subtitle{II. Molecular gas in galaxy clusters and their large-scale structure: low gas fraction galaxies, the case of CL1301.7$-$1139}
  \titlerunning{SEEDisCS II. CL1301.7$-$1139}
  \author{D. Sp\'{e}rone-Longin\inst{\ref{inst1},}
  					\and P. Jablonka\inst{\ref{inst1}, \ref{inst2}}
  					\and F. Combes\inst{\ref{inst3}, \ref{inst4}}
  					\and G. Castignani\inst{\ref{inst1},\ref{inst5}}
  					\and M. Krips\inst{\ref{inst6}}
  					\and \\ G. Rudnick\inst{\ref{inst7}}
  					\and T. Desjardins\inst{\ref{inst7}} 
  					\and D. Zaritsky\inst{\ref{inst8}}
  					\and R. A. Finn\inst{\ref{inst9}}
  					\and G. De Lucia\inst{\ref{inst10}}
  					\and V. Desai\inst{\ref{inst11}}
  					}
  \authorrunning{Sp\'{e}rone-Longin et al.}
  \institute{Institute of Physics, Laboratory of Astrophysics, \'{E}cole Polytechnique F\'{e}d\'{e}rale de Lausanne (EPFL), 
  					  Observatoire de Sauverny, CH-1290 Versoix, Switzerland \label{inst1}
  					  \and GEPI, Observatoire de Paris, PSL University, CNRS, 5 Place Jules Janssen, 92190 Meudon, France \label{inst2}
 					  \and Observatoire de Paris, LERMA, CNRS, Sorbonne University, PSL Research Universty, 75014 Paris, France \label{inst3}
  					  \and Coll\`{e}ge de France, 11 Place Marcelin Berthelot, 75231 Paris, France \label{inst4}
  					  \and Dipartimento di Fisica e Astronomia, Alma Mater Studiorum Università di Bologna, Via Gobetti 93/2, I-40129 Bologna, Italy \label{inst5}
  					  \and IRAM, Domaine Universitaire, 300 rue de la Piscine, 38406 Saint-Martin-d’H\`{e}res, France \label{inst6}
  					  \and Department of Physics and Astronomy, The University of Kansas, Lawrence, KS, USA \label{inst7}
  					  \and Steward Observatory and Department of Astronomy, University of Arizona, Tucson, AZ, USA \label{inst8}
   					  \and Department of Physics and Astronomy, Siena College, Loudonville, NY, USA \label{inst9}
  					  \and INAF – Osservatorio Astronomico di Trieste, Via G. B. Tiepolo 11, 34143 Trieste, Italy \label{inst10}
 					  \and IPAC, Mail Code 100-22, Caltech, 1200 E. California Boulevard, Pasadena, CA, USA \label{inst11}
 					  }
\abstract
	{This paper is the second of a series that tackles the properties of molecular
gas in galaxies residing in clusters and their related large-scale
structures. Out of 21 targeted fields, 19 galaxies were detected in CO(3-2) with the Atacama Large Millimeter Array (ALMA),
including two detections within a single field. These galaxies are
either bona fide members of the \cluster\ cluster ($z=0.4828$, $\sigma_{cl}=681$\,km\,s$^{-1}$), 
or located within $\sim 7 \times R_{200}$, its virial radius. They have been selected to sample the
range of photometric local densities around \cluster, with stellar masses above
log(\Mstar) = 10, and to be located in the blue clump of star-forming galaxies
derived from the $u$, $g$, and $i$ photometric bands. Unlike previous works, our
sample selection does not impose a minimum star formation rate or
detection in the far-infrared. As such and as much as possible, it delivers an
unbiased view of the gas content of normal star-forming galaxies at $z \sim 0.5$.
Our study highlights the variety of paths to star formation quenching, and most
likely the variety of physical properties (i.e. temperature, density) of the
corresponding galaxy's cold molecular gas. Just as in the case of \clusterb,
although to a smaller extent, we identify a number of galaxies with lower
gas fraction than classically found in other surveys. These galaxies can still be on the star-forming main sequence. 
When these galaxies are not inside the cluster virialised region, 
we provide hints that they are linked to their infall regions within $\sim 4 \times R_{200}$.}
   \maketitle


\section{Introduction\label{section1}}

It has long been established that the fraction of ellipsoidal and passive galaxies is dramatically increased in galaxy clusters 
in comparison with the field \citep{Dressler1980, Smith2005}. 
There is now a growing body of evidence, from both observations and numerical simulations, 
that the related decrease in galaxy star formation activity already begins far beyond the cluster centres 
\citep[r $\ge$ 2--4 virial radii;][]{Gomez2003, Haines2015, Gouin2020},
possibly in groups, which are later accreted \citep{Balogh2004, McGee2009, Bianconi2018}, and in filaments \citep{Bahe2013}.

We have entered a precision era in which the interplay between the growth of the cosmic structures and galaxy evolution can be investigated in detail.
The most massive or quiescent galaxies are found the closest to the filament axes \citep{Malavasi2016, Laigle2018}. 
This large-scale colour-density relation is seen even at $z\sim$1 \citep[e.g.,][]{Guzzo2018}. 
However, quenching cannot be related to density in a simple way. 
As a matter of fact, star-forming galaxies exist in environments that span 4 orders of magnitude in local density \citep{Peng2010}.
\citet{Song2021} have looked into the relative influence of local density and the proximity to filaments. 
They conclude that the high vorticity of the filament plays an important role in star formation quenching, by impeding gas transfer to the galaxies.

Identifying how and where cold gas is abundant or missing and determining where its properties can change, 
is at the heart of our understanding of the nature and 
operation modes of star formation at galactic scales \citep[][]{Bigiel2008, Leroy2008, Schruba2011}. 
Molecular gas is unlikely to be as easily stripped as \ion{H}{I} because it is denser and more centrally concentrated. 
Nevertheless, there are a number of spectacular examples of ram pressure stripping of molecular gas in the local Universe, 
such as in the Coma and Virgo clusters for example \citep[][]{Scott2015, Jachym2014, Jachym2017}. 
Yet, it is impossible to ascertain whether these examples reveal the general mechanism responsible for the suppression of star formation in dense environments. 
There is no established consensus on the possible dependence of the galaxy cold gas content on its environment \citep[e.g.,][]{Boselli2014, Koyama2017}.
\citet{Lee2017} suggest that one should instead investigate whether the gas properties (density, temperature) are modified.

So far studies of the cold molecular gas content of galaxies at intermediate redshifts have mostly focussed on distinct environments, 
isolating the field \citep{Gao2004, Abdo2010, Daddi2010, Tacconi2010, 
Tacconi2013, Garcia-Burillo2012, Baumgartner2013, Bauermeister2013, Bauermeister2013a, Morokuma-Matsui2015, 
Saintonge2017, Hayashi2018, Spilker2018, Freundlich2019, Lamperti2020}, 
from  groups \citep{Boselli1996, Martinez2012, Lisenfeld2017} and clusters \citep{Geach2009a, Geach2011, 
Jablonka2013, Rudnick2017a, Noble2017, Noble2019, Castignani2020b}.
Very few of these cluster studies purposely target galaxies beyond the cluster virial radius \citep[e.g.,][]{Morokuma-Matsui2021, Castignani2021}.

This paper is the second of a series that tackles the properties of molecular gas in galaxies residing in clusters and their related large-scale structures. 
This survey concentrates on two spectroscopically well-characterised, intermediate-redshift, 
medium-mass clusters selected from the ESO Distant Cluster Survey \citep[EDisCS,][]{White2005}. 
The first results of the Spatially Extended EDisCS survey (SEEDisCS) were presented in \citet[][hereafter Paper I]{Sperone-Longin2021}.

This paper reports on the observations in CO(3-2) with the Atacama Large Millimeter Array (ALMA) 
of 22 galaxies in and around the second SEEDisCS galaxy cluster, \cluster, at $z=0.4828$, 
and with a velocity dispersion of $\sigma_{cl}=681$\,km\,s$^{-1}$. It is organised as follows: 
Section \ref{sample} details our selection of the sample and presents the ALMA observations.
In Sect. \ref{results} we present our results and our derivation of the galaxy parameters. 
We discuss our results in Sect. \ref{discussion}, 
and we give our conclusions in Sect. \ref{conclusion}. 

In the following, we assume a flat $\Lambda$ cold dark matter ($\Lambda$CDM) cosmology with $\Omega_{\rm m} = 0.3$,
$\Omega_\Lambda = 0.7$, and $H_{0} = 70$\,km\,s$^{-1}$\,Mpc$^{-1}$
\citep[see][]{Riess2019, Planck2018}, and we use a Chabrier initial mass
function (IMF) \citep{Chabrier2003}. All magnitudes are in the AB system.


\section{Sample and observations\label{sample}}

\subsection{Sample selection}

\begin{figure}[ht]
	\resizebox{\hsize}{!}{\includegraphics{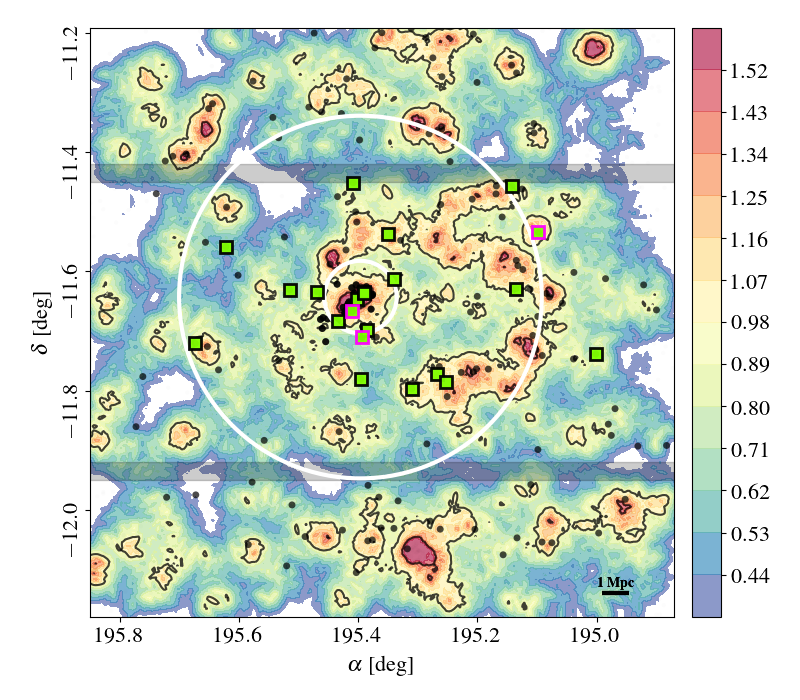}}
    \caption{Density map, based on $u,\,g,\,r,\,i$, and $z$ images, of the CFHT/MEGACAM 1$\degr\times$1$\degr$ field around
      \cluster. The colour-coding indicates the $\log_{10}$ of the density of galaxies
      averaged over the ten nearest neighbours. Black contours are at 1 and
      $3\sigma$ above the field mean density. The grey points identify all
      galaxies with a $z_\mathrm{spec}$ within $\mathrm{5\sigma_{cl}}$ of the
      cluster redshift. The inner and outer white circles are positioned at a $R_{200}$ and
      $5R_{200}$ radius, respectively. The grey bands indicate the gaps between the
      MEGACAM CCDs. The green squares show our ALMA targets. 
      The pink outline shows the position of the star-forming galaxies with low gas fractions 
      (see Fig. \ref{Ms_fg} and Sect. \ref{gas_fractions}).} 
    \label{targets}
\end{figure}

The 22 targets presented in this work are located inside \cluster\ and in its large-scale structures \citep{Milvang-Jensen2008}. 
\cluster\ was chosen from EDisCS \citep{White2005} because, at its intermediate redshift, 
environmentally induced galaxy transformation was in its heyday, 
and because its medium mass makes it a valid representative of the progenitors of typical nearby galaxy clusters \citep{Milvang-Jensen2008}. 
We first assembled CFHT/MEGACAM $u, g, r, i$, and $z$ deep imaging over a 1 deg$^2$ field of view, 
that is to say covering a region of radius r $\sim 10$\,Mpc around the cluster centre.
The $K_s$ deep imaging from GEMINI/NEWFIRM was obtained over an area of 0.24 deg$^2$, covering 5 cluster virial radii ($R_{200}$).
Photometric redshifts, $z_{\rm phot}$ were obtained with EAZY \citep{Brammer2008} and are very accurate, 
with a normalised median absolute deviation of $\sigma_{\rm NMAD} \sim 0.036$. 
They have enabled the identification of the filamentary structures and infalling groups around \cluster\ (see Fig. \ref{targets}).

The targets for the ALMA follow-up were selected based on the following four criteria: 
\textit{\textbf{i)}} they have spectroscopic redshifts from ESO/FORS2 or MMT/Hectospec, 
or at least a robust estimate from the IMACS Low Dispersion Prism \citep[$\sigma_z = 0.007$,][]{Just2019},
all falling within $\mathrm{3\sigma_{cl}}$; 
\textit{\textbf{ii)}} they are located within $7\times R_{200}$\footnote{For consistency with previous EDisCS works, 
R$_{200}$ was calculated from the cluster line-of-sight velocity dispersion, as in \cite{Finn2005}}, 
sampling the range of photometric local densities encountered around \cluster\ (see Fig.\ref{targets}); 
\textit{\textbf{iii)}} they have stellar masses above log(\Mstar)\,=\,10;
\textit{\textbf{iv)}} they cover the same sequence in $u-g$ versus $g-i$ which is typical of star-forming galaxies as in \clusterb\ 
at $z=0.5195$ \citepalias{Sperone-Longin2021}.
Table \ref{obs_table} provides the coordinates and optical redshifts of our targets, 
together with their stellar masses (\Mstar) and star formation rates (SFRs).

Figure \ref{targets} presents the position of our targets in the global photometric density map, 
as calculated from the $u,\,g,\,r,\,i$, and $z$ images, in the 1$\degr\times$1$\degr$ region centred on \cluster. Densities were
calculated within a photometric slice of $\pm 1 \delta_{\rm cl}=(1+z_{\rm cl})\times \sigma_{\rm NMAD} = 0.0533$ around the cluster redshift. 
Within this photometric redshift slice, we used a `nearest neighbour' approach, 
in which for any point $(x,y)$ one estimates the distance $r_N(x,y)$ to the $Nth$ nearest neighbour. 
The density is thus the ratio between the (fixed) $N$ and the surface defined by the adaptive distance: $\rho_N(x,y)=\frac{N}{\pi r^2_N(x,y)}$. 
We chose $N=10$, which corresponds to an average scale of about 0.8 Mpc and with 90\% of the values being smaller than $\sim1.5$\,Mpc. 

\begin{figure}[ht]
	\resizebox{\hsize}{!}{\includegraphics{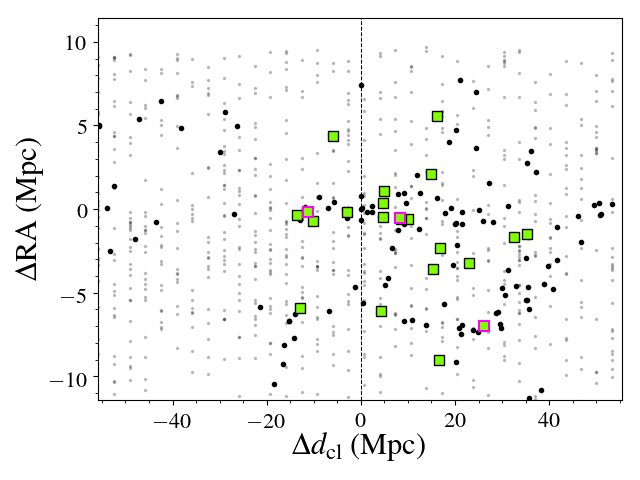}}
    \caption{Light cone centred on $z_{\rm clus}=0.4828$ and extending to 
    $\pm 1 \delta_{\rm cl}$ in redshift. In right ascension, $1\degr$ is covered.
    The vertical line is located at the cluster redshift, $z_{\rm cl}=0.4828$.
    The grey points are for the galaxies with a photometric redshift. Galaxies 
    with spectroscopic redshifts are in black. Our sample is in green, and
    lower \muhh\ (see Sect. \ref{gas_fractions}) galaxies are outlined in pink.
    Distances are expressed relative to the brightest cluster galaxy (BCG).} 
    \label{light_cone}
\end{figure}

Figure \ref{light_cone} provides another view of the spatial distribution of our ALMA targets, 
keeping the same 1$\degr\times$1$\degr$ MEGACAM field of view. 
The galaxy positions were calculated relative to the position of the brightest cluster galaxy (BCG) in redshift and right ascension (RA) \citep{White2005}. 
The galaxy relative position in redshift, $\Delta d_{\rm cl}$, was computed by taking the difference between the co-moving distances of the galaxy and of the BCG. 
The relative position in RA, $\Delta$RA, was obtained by transforming the angular separation between the BCG and the galaxy into a distance, using the angular distance at the redshift of the galaxy. 
Our full spectroscopic sample within $\pm 3\times\sigma_{\rm cl}$ of $z_{\rm cl}$ is presented as well as the photometric membership candidates. 
The finger-of-God structure due to the relative velocities of the galaxies in \cluster\ is clearly seen along the $\Delta d_{\rm cl}$-axis. 
Eight of our targets are genuine cluster members, within or at the cluster virial radius, and with v/$\sigma_{cl} \lesssim 1$. 
They are identified in Table \ref{obs_table} by their original EDisCS names.
Three galaxies, namely, \I, \XXII, and \XXIII, fall on the south-west photometric overdensity at RA $\sim195.3$ ($\Delta d_{\rm cl}\sim20$\,Mpc, 
$\Delta {\rm RA}\sim -3$). 
They have a mean redshift of 0.488 with a standard deviation of $\sigma_z = 9.8\times 10^{-4}$. 
Two other galaxies, with spectroscopic redshifts, however without CO observations, fall in this redshift interval as well, 
suggesting that these three ALMA targets indeed do belong to a group with $\sigma_{\rm g} \sim$\,150--200\,km\,s$^{-1}$. 
The rest of our ALMA sample is either located in low density regions or photometric overdensities, which has yet to be spectroscopically confirmed.

\begin{figure}[ht]
  \resizebox{\hsize}{!}{\includegraphics{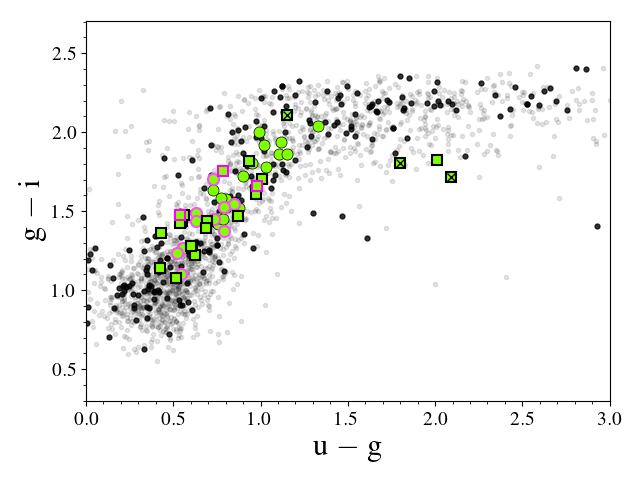}}
    \caption{Observed colour--colour diagram, $g-i$ as a function of
      $u-g$ for the galaxies in the \cluster\ region. Our ALMA sample is shown in green, 
      with squares for \cluster\ and with circles for \clusterb. In both cases, 
      pink borders indicate star-forming galaxies that have low gas fractions 
      for their stellar masses (see Sect. \ref{gas_fractions}).
      The small black dots show galaxies with spectroscopic redshifts 
      within $3\times\sigma_{\rm cl}$ of the \cluster\ redshift. 
      Galaxies that were not detected in CO are indicated with a cross.}
    \label{ugi}
\end{figure}

Figure \ref{ugi} displays the distribution of our targets in the $g-i$ vs $u-g$ colour--colour diagram, 
using total magnitudes. The ALMA targets in and around \cluster\ are shown with the green squares, 
and we show the ALMA targets for \clusterb\ with green circles for comparison. The \cluster\ galaxies span most of the blue clump, 
extending into the lower $g-i$ boundary of the red sequence. 
Three systems have $g-i$ colours slightly below the red sequence but their $u-g$ colours are consistent with them being on the red sequence.
 
\begin{figure}[ht]
	\resizebox{\hsize}{!}{\includegraphics{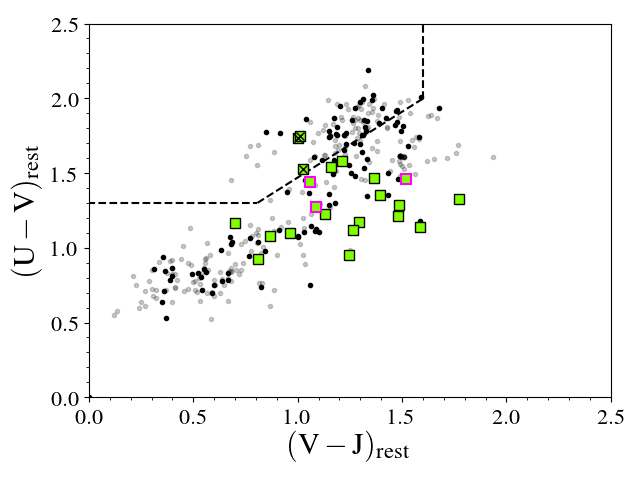}}
 	\caption{Rest-frame $UVJ$ colour--colour diagram. The dashed lines separate passive galaxies from star-forming ones \citep{Williams2009}. 
 	Our ALMA sample is in green. The black crosses mark the galaxies with CO upper limits. The squares with
 	the pink borders are for our low-\muhh\ galaxies (see Sect. \ref{gas_fractions}). The grey points are the photometric
 	redshift members with a $K$-band detection. The small black dots are the spectroscopic redshift galaxies,
 	within $3\times\sigma_{\rm cl}$ of the cluster redshift, and also with a $K$-band detection.}
    \label{rf_CCD}
\end{figure}

Figure \ref{rf_CCD} presents the rest-frame $U-V$ versus $V-J$ colour--colour diagram
which further helps discriminate between passive and star-forming galaxies
\citep{Williams2009}. The rest-frame colours were derived with EAZY
\citep{Brammer2008}. We used the Johnson-Cousins $U$ and $V$ bands, and the
2MASS $J$ band \citep{Skrutskie2006}, together with a set of six templates: five 
main component templates obtained following the \citet{Blanton2007a} algorithm
and one for dusty galaxies \citep{Brammer2008}.
From the four galaxies which were photometrically falling in the green valley as inferred from the $u$, $g$, and $i$ bands, 
three turned out to be located in the $UVJ$ region of passive systems. 
Two of them, namely \X\ and \Xb, were falling in the same ALMA field of view. \X\ was our primary target. Both are cluster members. 
The rest of our targets follow the $UVJ$ star-forming sequence, including \VII\ which was red in $u-g$ and at the blue $g-i$ border of the red sequence. 

\begin{table*}[ht]
\caption{Galaxy ID, coordinates, optical spectroscopic redshifts and SED-based
  estimates of the $M_{\rm star}$ and SFRs of our galaxy sample observed with
  ALMA.}
\label{obs_table}
\centering
\begin{tabular}{l c c l r r}
\hline \hline \\
IDs & R.A. (J2000) & Dec (J2000) & ${z_{\rm spec}}$
& \multicolumn{1}{p{1.5cm}}{\centering $M_{\mathrm{star}}$ \ ($\mathrm{10^{10}\,M_{\odot}}$)}
& \multicolumn{1}{p{1.5cm}}{\centering $\mathrm{SFR}_{SED}$ \ ($\mathrm{M_{\odot}\,yr^{-1}}$)} \\
\hline \\
\III\ & 13:00:00.3688 & $-$11:44:20.346 & 0.4878 & $4.68^{+1.83}_{-1.29}$ & $6.64^{+9.63}_{-4.59}$ \\
\XVI $^\dagger$ & 13:00:23.4345 & $-$11:32:05.222 & 0.4907 & $7.41^{+2.05}_{-2.22}$ & $8.85^{+9.78}_{-7.85}$ \\
\XIII\ & 13:00:32.4211 & $-$11:37:44.584 & 0.4841 & $3.63^{+1.34}_{-1.09}$ & $16.48^{+14.42}_{-6.45}$ \\
\XVIII\ & 13:00:34.0731 & $-$11:27:26.913 & 0.4789 & $3.98^{+1.19}_{-1.1}$ & $8.07^{+5.48}_{-4.65}$ \\
\XXIII\ & 13:01:00.7272 & $-$11:47:07.522 & 0.4770\tablefootmark{a} & $2.63^{+0.61}_{-0.67}$ & $4.86^{+2.91}_{-2.46}$ \\
\XXII\ & 13:01:04.4539 & $-$11:46:23.263 & 0.4880\tablefootmark{a} & $1.91^{+0.44}_{-0.39}$ & $10.76^{+7.06}_{-3.96}$ \\
\I & 13:01:14.4426 & $-$11:47:49.051 & 0.4879 & $2.82^{+0.84}_{-0.65}$ & $7.80^{+5.84}_{-4.31}$ \\
\XIV\ & 13:01:21.6371 & $-$11:36:48.027 & 0.4927 & $5.13^{+1.06}_{-1.06}$ & $12.22^{+7.6}_{-6.05}$ \\
\XIX\ & 13:01:24.0810 & $-$11:32:13.735 & 0.4790\tablefootmark{a} & $4.07^{+0.94}_{-0.75}$ & $14.86^{+9.58}_{-5.65}$ \\
\VI\ & 13:01:32.3246 & $-$11:41:55.846 & 0.4797 & $3.80^{+0.79}_{-0.70}$ & $10.52^{+5.57}_{-4.36}$ \\
\X $^*$ & 13:01:33.6127 & $-$11:38:07.126 & 0.4858 & $3.63^{+1.50}_{-0.50}$ & $0.36^{+0.24}_{-0.36}$ \\
\Xb $^*$ & 13:01:33.6420 & $-$11:38:09.025 & 0.4854 & $8.32^{+1.72}_{-2.49}$ & $0.52^{+0.46}_{-0.48}$ \\
\V $^\dagger$ & 13:01:34.4495 & $-$11:42:38.039 & 0.4853 & $1.82^{+0.38}_{-0.42}$ & $1.44^{+0.93}_{-0.81}$ \\
\II & 13:01:34.9086 & $-$11:46:51.222 & 0.4842 & $7.41^{+1.54}_{-2.05}$ & $4.09^{+3.3}_{-2.73}$ \\
\IX\ & 13:01:36.3447 & $-$11:38:49.502 & 0.4787 & $3.47^{+0.64}_{-0.80}$ & $4.00^{+3.45}_{-2.62}$ \\
\XVII\ & 13:01:38.0353 & $-$11:27:05.546 & 0.4819 & $7.59^{+2.10}_{-2.10}$ & $11.27^{+12.07}_{-10.12}$ \\
\VIII $^\dagger$ & 13:01:38.3445 & $-$11:40:01.197 & 0.4794 & $3.63^{+0.67}_{-0.75}$ & $5.09^{+2.75}_{-2.58}$ \\
\VII\ & 13:01:44.1493 & $-$11:40:58.995 & 0.4842 & $3.98^{+0.83}_{-1.28}$ & $1.89^{+1.44}_{-1.13}$ \\
\XI & 13:01:52.4694 & $-$11:38:04.391 & 0.4843 & $5.13^{+1.54}_{-1.42}$ & $6.41^{+5.09}_{-3.54}$ \\
\XII $^*$ & 13:02:03.6095 & $-$11:37:51.978 & 0.4873 & $3.09^{+0.64}_{-1.00}$ & $0.66^{+0.51}_{-0.53}$ \\
\XV\ & 13:02:29.2680 & $-$11:33:31.606 & 0.4810 & $5.50^{+1.64}_{-1.52}$ & $14.86^{+8.38}_{-7.36}$ \\
\IV & 13:02:41.5765 & $-$11:43:15.746 & 0.4877 & $4.37^{+1.91}_{-1.51}$ & $10.76^{+22.42}_{-6.32}$ \\
\hline
\end{tabular}
\tablefoot{
	Galaxies with low \muhh\ are identified with $^\dagger$. Galaxies with CO upper-limits are 
	identified with $^*$.
	\tablefoottext{a}{Galaxy with $z_{\rm LDP}$ as $z_{\rm spec}$.}	
	}
\end{table*}


\subsection{ALMA observations \label{alma_obs}}

The observations were conducted in the CO(3-2) line (233 GHz at $z\sim 0.48$), falling in the ALMA Band 6,
in compact configurations C43$-$2, for a beam size of 1.2\arcsec$\times$ 0.95\arcsec\, 
during Cycle 5 (program 2017.1.00257.S). The integration time on targets was 6.75 hours, 11h with overheads. 
The rms noise was computed at the fixed resolution of 50.7 km\,s$^{-1}$ and ranges from 29 to 84 $\mu$Jy/beam.
The large range of values is mainly due to different observing conditions for each of our targets.

The same data reduction procedure as used for \clusterb\ \citepalias{Sperone-Longin2021} was
applied, with the CASA ALMA Science Pipeline \citep{McMullin2007}. 
Problems with antennas and runs were flagged ($<3\%$), the continuum was subtracted, 
while excluding the spectral channels of the line emission during the fit of the continuum -- as we only focus on the CO line --
and the primary beam was corrected in order to form an astronomically correct image of the sky. 
The final continuum-subtracted and primary-beam-corrected maps were exported in order to be analysed using GILDAS\footnote{\url{http://www.iram.fr/IRAMFR/GILDAS}}. 
The $i$-band images of our targets, the CO maps and spectra are shown in Fig. \ref{maps}.


\section{Derived parameters \label{results}}


\subsection{CO flux and molecular gas mass}

Fluxes, $S_\mathrm{CO}\,\Delta V$, were extracted by integrating the CO(3-2)
emission over the full spatial extent of the source using circular apertures
with radii between 0.8\arcsec\ and 1.4\arcsec\ depending on the size of the
galaxy. Following \cite{Lamperti2020}, the error on the flux is defined as
\begin{equation}
\epsilon_{\rm CO} = \frac{\sigma_{\rm CO}\Delta V}{\sqrt{\Delta V \Delta w_{ch}^{-1}}},
\end{equation}
where $\sigma_{\rm CO}$ is the rms noise (in Jy) calculated in units of spectral resolution 
$\Delta w_{ch}$, and $\Delta V$ (in km\,s$^{-1}$) is the width of the 
spectral window in which the line flux was calculated, $\Delta w_{ch} = 50.7$\,km\,s$^{-1}$. 
All intensity maps and integrated spectra are shown in Fig \ref{maps} of the Appendix. 
The full widths at half maximum (FWHM) were derived from single Gaussian fits of the emission lines. 
We obtain a median FWHM of 220\,km\,s$^{-1}$ with a standard deviation of 96\,km\,s$^{-1}$, 
which is compatible with the type of massive galaxies we are studying \citep{Freundlich2019}.

Even though the targets are just slightly more extended than the beam size, 
a few of them show a double peaked emission line, which is an indication of rotation. 
This will be analysed in a forthcoming paper.

The intrinsic CO luminosity associated with a transition between the levels $J$ and $J-1$ is expressed as
\begin{equation} \label{eq1}
    L'_\mathrm{CO(J\rightarrow J-1)}=3.25 \times10^{7} S_\mathrm{CO(J\rightarrow
      J-1)} \Delta V \, \nu_\mathrm{obs}^{-2} \, D_\mathrm{L}^{2} \, (1+z)^{-3},
\end{equation}
where $L'_\mathrm{CO(J\rightarrow J-1)}$ is the line luminosity expressed in
units of ${\rm K\,km\,s\textsuperscript{-1} pc^2}$; $S_\mathrm{CO(J\rightarrow
J-1)} \Delta V$ is the velocity-integrated flux in ${\rm Jy\,km\,s^{-1}}$;
$\nu_\mathrm{obs}$ is the observed frequency in GHz; $D_\mathrm{L}$ is the
luminosity distance in Mpc; and $z$ is the redshift of the observed galaxy
\citep{Solomon1997, Solomon2005}.
The flux for our upper-limit detections are defined as $3\times\epsilon_{\rm CO}$,
with $\Delta V=220$\,km\,s$^{-1}$, the median value of our sample.

The total cold molecular gas mass ($M_\mathrm{H_2}$) was then estimated as
\begin{equation} \label{eq2}
    M_\mathrm{H_2}=\alpha_\mathrm{CO} \frac{L'_\mathrm{CO(J\rightarrow J-1)}}{r_{J1}}, 
\end{equation}
where $\alpha_\mathrm{CO}$ is the CO(1-0) luminosity-to-molecular-gas-mass
conversion factor, considering a 36\% correction to account for interstellar
helium, and $r_{J1}=L^{\prime}_\mathrm{CO(J\rightarrow J-1)} / L^{\prime}_\mathrm{CO(1-0)}$ is the
corresponding line luminosity ratio.

Just as in \citetalias{Sperone-Longin2021} and for the sake of reliable comparison, 
we have considered $\alpha_\mathrm{CO}=4.36 \pm 0.9\,M_\odot\,{\rm (K\,km\,s^{-1}\,pc^2)^{-1}}$, 
which includes the correction for helium, as a good estimate for normal star-forming galaxies \citep{Dame2001, Grenier2005, Abdo2010, Leroy2011, Bolatto2013, Carleton2017}. 
For the same reason, we assume $r_{31}=0.5\pm0.05$ similarly to other intermediate to high-$z$ studies \citep{Bauermeister2013, Genzel2015, Chapman2015, Carleton2017, Tacconi2018}
The intrinsic CO(3-2) luminosity $L^{\prime}_\mathrm{CO(3-2)}$, the FWHM, 
the cold molecular gas mass $M_\mathrm{H_2}$, the corresponding gas-to-stellar-mass ratio
$\mu_\mathrm{H_2}=M_\mathrm{H_2}/M_\mathrm{star}$, and the redshift of the CO
emission of our sample galaxies are listed in Table \ref{properties_table}.

\begin{table*}[ht]
\caption{CO redshift, line-integrated flux, line width, luminosity of the CO(3-2) emission, 
cold molecular gas masses, cold molecular gas-to-stellar mass ratios, and gas depletion times of the ALMA targets.}
\label{properties_table}
\centering
\begin{tabular}{l l c r r r c c}
\hline \hline \\
IDs 
& $z_{\mathrm{CO}}$ 
& \multicolumn{1}{p{1.5cm}}{\centering $S_{\mathrm{CO(3-2)}}\Delta V$ \\ ($\mathrm{Jy\,km\,s^{-1}}$)} 
& \multicolumn{1}{p{1.5cm}}{\centering FWHM \\ ($\mathrm{km\,s^{-1}}$)} 
& \multicolumn{1}{p{1.5cm}}{\centering $L^{'}_{\mathrm{CO(3-2)}}$ \\ ($\mathrm{10^{8}\,L_{\odot}}$)} 
& \multicolumn{1}{p{1.5cm}}{\centering $M_{\mathrm{H_{2}}}$ \\ ($\mathrm{10^{9}\,M_{\odot}}$)} 
& $\mu_{\mathrm{H_{2}}}$ 
& \multicolumn{1}{p{1.5cm}}{\centering $t_{\rm depl}$ \\ (Gyr)} \\
\hline \\
\III\ & 0.4879 & 0.600 $\pm$ 0.020 & 440 $\pm$ 40 & 8.295 $\pm$ 0.280 & 7.23 $\pm$ 0.24 & $0.155^{+0.066}_{-0.048}$ & $1.089^{+1.581}_{-0.753}$ \\
\XVI $^\dagger$ & 0.4907 & 0.105 $\pm$ 0.014 & 230 $\pm$ 20 & 1.469 $\pm$ 0.198 & 1.28 $\pm$ 0.17 & $0.017^{+0.007}_{-0.007}$ & $0.145^{+0.161}_{-0.130}$ \\
\XIII\ & 0.4841 & 0.523 $\pm$ 0.013 & 190 $\pm$ 10 & 7.115 $\pm$ 0.176 & 6.20 $\pm$ 0.15 & $0.171^{+0.067}_{-0.055}$ & $0.376^{+0.330}_{-0.148}$ \\
\XVIII\ & 0.4789 & 0.722 $\pm$ 0.013 & 180 $\pm$ 10 & 9.606 $\pm$ 0.176 & 8.38 $\pm$ 0.15 & $0.210^{+0.067}_{-0.062}$ & $1.038^{+0.705}_{-0.598}$ \\
\XXIII\ & 0.4874\tablefootmark{a} & 0.222 $\pm$ 0.010 & 180 $\pm$ 10 & 3.063 $\pm$ 0.132 & 2.67 $\pm$ 0.11 & $0.102^{+0.028}_{-0.030}$ & $0.550^{+0.330}_{-0.279}$ \\
\XXII\ & 0.4898\tablefootmark{a} & 1.420 $\pm$ 0.016 & 310 $\pm$ 20 & 19.785 $\pm$ 0.235 & 17.25 $\pm$ 0.21 & $0.905^{+0.219}_{-0.198}$ & $1.603^{+1.052}_{-0.591}$ \\
\I & 0.4880 & 0.295 $\pm$ 0.013 & 240 $\pm$ 20 & 4.079 $\pm$ 0.181 & 3.56 $\pm$ 0.16 & $0.126^{+0.043}_{-0.035}$ & $0.456^{+0.342}_{-0.253}$ \\
\XIV\ & 0.4927 & 0.694 $\pm$ 0.013 & 260 $\pm$ 20 & 9.792 $\pm$ 0.192 & 8.54 $\pm$ 0.17 & $0.166^{+0.038}_{-0.038}$ & $0.699^{+0.435}_{-0.346}$ \\
\XIX\ & 0.4935\tablefootmark{a} & 0.691 $\pm$ 0.008 & 140 $\pm$ 10 & 9.780 $\pm$ 0.120 & 8.53 $\pm$ 0.10 & $0.209^{+0.051}_{-0.041}$ & $0.574^{+0.370}_{-0.218}$ \\
\VI\ & 0.4796 & 0.790 $\pm$ 0.010 & 180 $\pm$ 10 & 10.541 $\pm$ 0.135 & 9.19 $\pm$ 0.12 & $0.242^{+0.053}_{-0.048}$ & $0.874^{+0.463}_{-0.362}$ \\
\X $^*$ & 0.4869 & $<0.026$ & $...$ & $...$ & $<0.32$ & $<0.009$ & $<0.882$ \\
\Xb $^*$ & 0.4854 & $<0.026$ & $...$ & $...$ & $<0.32$ & $<0.004$ & $<0.607$ \\
\V $^\dagger$ & 0.4854 & 0.140 $\pm$ 0.012 & 210 $\pm$ 10 & 1.915 $\pm$ 0.163 & 1.67 $\pm$ 0.14 & $0.092^{+0.027}_{-0.029}$ & $1.160^{+0.754}_{-0.662}$ \\
\II & 0.4843 & 0.344 $\pm$ 0.024 & 450 $\pm$ 20 & 4.684 $\pm$ 0.321 & 4.08 $\pm$ 0.28 & $0.055^{+0.015}_{-0.019}$ & $0.999^{+0.808}_{-0.670}$ \\
\IX\ & 0.4785 & 0.295 $\pm$ 0.014 & 320 $\pm$ 20 & 3.917 $\pm$ 0.191 & 3.42 $\pm$ 0.17 & $0.099^{+0.023}_{-0.027}$ & $0.854^{+0.739}_{-0.562}$ \\
\XVII\ & 0.4819 & 0.465 $\pm$ 0.022 & 380 $\pm$ 20 & 6.266 $\pm$ 0.295 & 5.46 $\pm$ 0.26 & $0.072^{+0.023}_{-0.023}$ & $0.485^{+0.520}_{-0.436}$ \\
\VIII $^\dagger$ & 0.4799 & 0.145 $\pm$ 0.015 & 160 $\pm$ 20 & 1.937 $\pm$ 0.196 & 1.69 $\pm$ 0.31 & $0.047^{+0.017}_{-0.018}$ & $0.332^{+0.190}_{-0.179}$ \\
\VII & 0.4847 & 0.285 $\pm$ 0.008 & 160 $\pm$ 10 & 3.886 $\pm$ 0.114 & 3.39 $\pm$ 0.10 & $0.085^{+0.020}_{-0.030}$ & $1.793^{+1.363}_{-1.075}$ \\
\XI & 0.4845 & 0.365 $\pm$ 0.011 & 180 $\pm$ 10 & 4.973 $\pm$ 0.155 & 4.34 $\pm$ 0.13 & $0.085^{+0.028}_{-0.026}$ & $0.676^{+0.538}_{-0.374}$ \\
\XII $^*$ & 0.4873 & $<0.097$ & $...$ & $...$ & $<1.17$ & $<0.038$ & $<1.769$ \\
\XV\ & 0.4813 & 0.580 $\pm$ 0.011 & 130 $\pm$ 20 & 7.797 $\pm$ 0.147 & 6.8 $\pm$ 0.13 & $0.124^{+0.039}_{-0.037}$ & $0.458^{+0.258}_{-0.227}$ \\
\IV & 0.4877 & 0.295 $\pm$ 0.009 & 120 $\pm$ 10 & 4.074 $\pm$ 0.129 & 3.55 $\pm$ 0.11 & $0.081^{+0.038}_{-0.031}$ & $0.330^{+0.688}_{-0.194}$ \\
\hline
\end{tabular}
\tablefoot{
	Galaxies with low \muhh\ are identified with $^\dagger$. Galaxies with CO upper-limits are identified with $^*$.
	\tablefoottext{a}{Galaxy with $z_{\rm LDP}$ as $z_{\rm spec}$.}	
	}
\end{table*}


\subsection{Stellar masses and SFRs \label{MS_SFR}}

Stellar masses and SFRs were derived with MAGPHYS\footnote{\url{http://www.iap.fr/magphys/index.html}}
\citep{daCunha2008}, based on the $u$, $g$, $r$, $i$, $z$, and $K_s$ total magnitudes.
The stellar populations and dust extinction models are those of \citet{Bruzual2003} and
\citet{Charlot2000}. MAGPHYS provides probability density functions
(PDFs) for each parameter, such as the SFR, \Mstar, dust mass, and dust
temperature. Their values in this work correspond to the peak values of
the PDFs. The uncertainties are the 68\% confidence interval of
the PDFs. The photometric wavelength coverage does not allow for
the identification of active galactic nuclei (AGNs), which could affect the SFRs in particular. 
However, the analysis of the galaxy spectra, in particular, 
the [\ion{O}{II}] to H$\beta$ line ratio suggests that they are not typical AGNs \citep{Sanchez-Blazquez2009, Rudnick2017}. 
These are anyway rare in clusters, comprising approximately 3\% of the overall population \citep{Miller2003, Kauffmann2004, Mishra2020}.

\begin{figure}[ht]
\centering
\resizebox{\hsize}{!}{\includegraphics{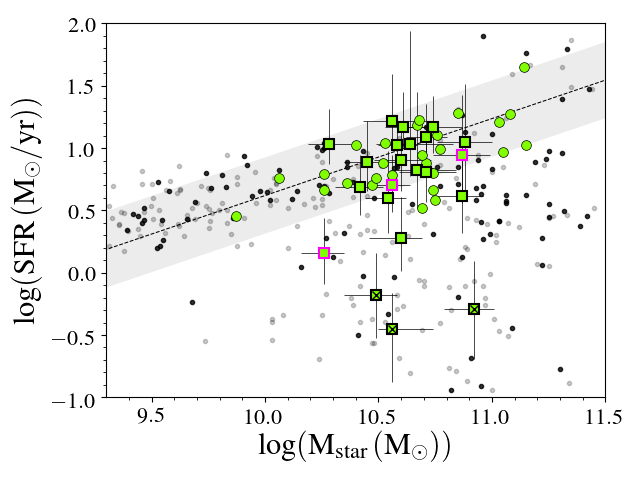}}
    \caption{Location of the CL1301 (grey) and 
    ALMA (green squares) galaxies in the \Mstar--SFR plane.
    The galaxies with only upper limits in CO are identified with a cross.
    Squares with pink borders show the low-\muhh\ galaxies, see Sect. \ref{gas_fractions}. 
    Galaxies in black and grey are the spectroscopic and photometric samples, respectively, 
    at the redshift of \cluster. For comparison, the light green circles show the \clusterb\ ALMA galaxies.
    The dashed black line is the \citet{Speagle2014} MS, corrected for a Chabrier IMF at $z=0.4828$, 
    with the corresponding $\pm$0.3\,dex scatter being the grey shaded area.}
    \label{MScl1301}
\end{figure}

Figure \ref{MScl1301} presents the position of our ALMA targets relative to
the main sequence (MS) of normal star-forming galaxies at the cluster redshift as inferred by
\citet{Speagle2014}. Only galaxies with spectroscopic or photometric redshift and with $K_s$ photometry are shown. 
Indeed the near-infrared (NIR) flux allows for the most robust mass and SFR estimates.
Almost two-thirds (65\%) of our ALMA targets fall within the $\pm0.3$\,dex dispersion of the MS.
The following three ALMA targets fall far below the $-0.3$\,dex boundary, in between the MS and the red sequence: \II, \V, and \VII.
These are systems in the transition region between star-forming and passive systems.
Two of those are bona fide cluster members (\V\ and \VII), while \II\ is located at a distance of $\sim2R_{200}$, 
however it is still within v/$\sigma_{cl}<1$ and is therefore most probably an infalling system \citep[][]{Mahajan2011}. 
All but three were detected in CO. 
Finally, three galaxies turned out to be on the passive sequence (log(SFR) $\lesssim -0.2$). 
Two of them (\X\ and \Xb\ are cluster members), 
while \XII\ is an infalling system that is $\sim1.5 R_{200}$ and still within v/$\sigma_{cl}\sim1.5$. 
We could only place upper limits on their CO fluxes. 
It would be tempting to attribute their lack of star formation activity and their CO deficiency to their early-type morphologies. 
However, similar systems without evidence of disks in Fig. \ref{maps} (\I, \V\ and \XIV) have nevertheless been detected in CO.

\begin{figure}[ht]
  \resizebox{\hsize}{!}{\includegraphics{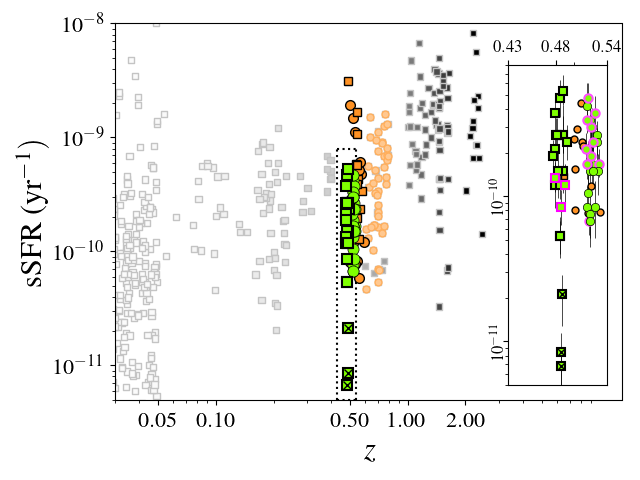}}
    \caption{Specific SFRs as a function of redshift.
    	The green markers identify the ALMA targets from both of our clusters, 
    	with the circles for \clusterb\ and the squares for \cluster. 
    	The pink outlines show the galaxies with a low gas fraction, as defined in Sect. \ref{gas_fractions}. 
    	The orange circles stand for the PHIBSS2 galaxies, with a dark colour for
    	the systems at $0.49 < z \leq 0.6$ and a lighter one for the
    	galaxies at $0.6 < z \leq 0.8$. The orange squares are for the 
    	\citet{Castignani2020b} luminous infrared galaxies (LIRGs) with $0.45 < z < 0.55$. 
    	The symbols in shades of grey are for the samples we took from the literature at different redshifts.
    	We provide a zoom-in of the region delineated by the dotted lines (see inset), 
       around the redshift of \cluster. \label{z_sSFR}}
\end{figure}

Figure \ref{z_sSFR} provides an overview of the galaxy specific star formation rates (sSFR) as a function of redshift 
from $\sim0.02$ to $\sim2$ for galaxies with existing CO line fluxes. 
The list of our comparison samples is identical to that of \citetalias{Sperone-Longin2021} where their exact composition, redshift range, 
selection and CO line transitions are detailed \citep{Gao2004, Garcia-Burillo2012,Saintonge2017,Baumgartner2013, 
Lamperti2020,Abdo2010, Bauermeister2013a,Bauermeister2013,Morokuma-Matsui2015,cybulski2016,
Castignani2020b,Jablonka2013,Geach2009a,Geach2011,Tacconi2010,Tacconi2013,
Tacconi2018,Freundlich2019,Spilker2018,Hayashi2018,Daddi2010,Noble2017,Noble2019,Rudnick2017a}.
The galaxy stellar masses of these comparison samples were originally derived by SED fitting, 
assuming either a `Chabrier' or a `Kroupa' IMF \citep{Kroupa2001,Chabrier2003}. 
We converted all masses to a Chabrier IMF using 
the relation of \citet{Zahid2012}: $M_{{\rm star},\,C}=0.94 \times M_{{\rm star},\,K}$.

Most existing datasets have focussed on field galaxies, with the exception of \citet{Geach2009a,Geach2011}, \citet{Jablonka2013},
\citet{cybulski2016}, \citet{Rudnick2017a}, \citet{Hayashi2018}, \citet{Noble2017,Noble2019} and \citet{Castignani2020b}.
The present analysis provides the second largest sample centred on galaxy clusters at a fixed given redshift, 
the other one being our analysis of galaxies related to the \clusterb\ environment. 
Both samples cover the same range of specific SFRs and uniquely trace galaxies in 
interconnected cosmic structures around galaxy clusters.

\begin{figure*}[h!]
\centering
\begin{subfigure}[t]{.49\textwidth}
	\resizebox{\hsize}{!}{\includegraphics{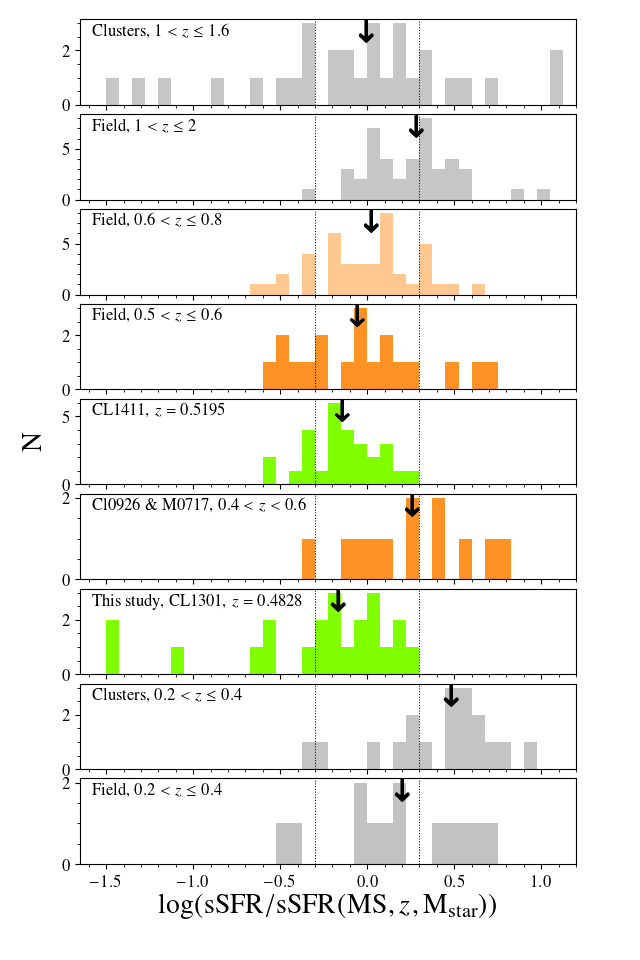}}
\end{subfigure}
\begin{subfigure}[t]{.49\textwidth}
	\resizebox{\hsize}{!}{\includegraphics{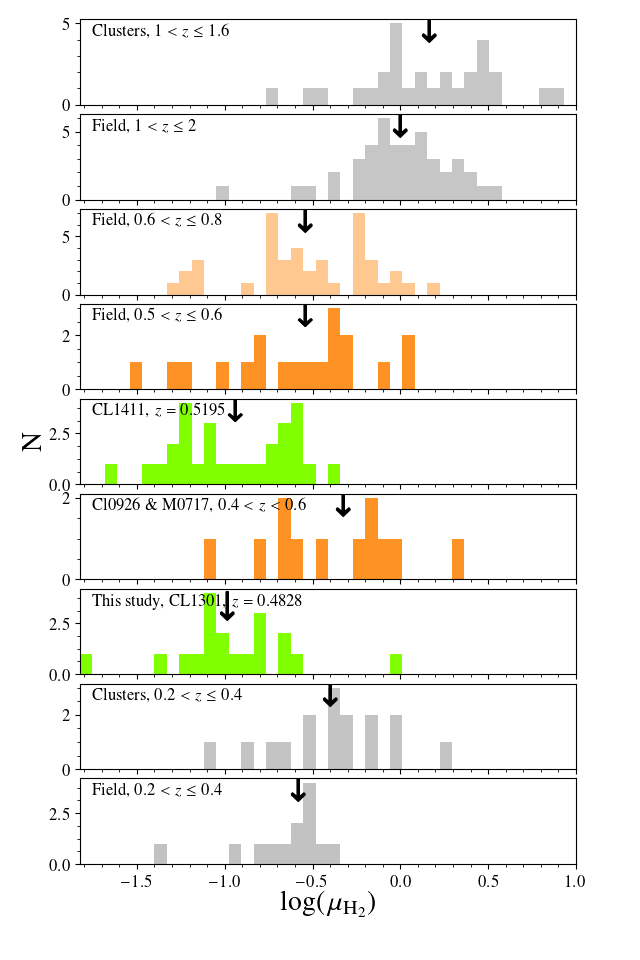}}
\end{subfigure}
\caption{Distribution of the sSFR normalised to the sSFR of the MS, in the left panel, 
    and of the molecular gas-to-stellar mass ratio, \muhh, in the right panel, for different redshift ranges and galaxy samples. 
    Both of our ALMA samples \clusterb\ and \cluster\ are in green. The PHIBSS2 sample, constituted of field galaxies, 
    is divided into two redshift sub-samples, one at $0.5 \leq z \leq 0.6$ and the other one at $0.6 < z \leq 0.8$. 
    A sub-sample of the \citet{Castignani2020b} LIRGs within clusters Cl 0926+1242 and MACS J0717.5+3745 is provided. 
    The cluster sample at lower redshifts comprises the rest of the \citet{Castignani2020b} LIRGs, as well as \citet{Geach2011}
    and \citet{cybulski2016} cluster galaxies. The field samples at the bottom and top part of both panels, 
    as well as the cluster sample on top, contain galaxies from the literature at the corresponding redshifts.
    The black arrows correspond to the medians of each sample. The same value of $\alpha_{\rm CO}$ has been used for all samples.}
\label{histograms}
\end{figure*}

One hypothesis raised in \citetalias{Sperone-Longin2021} is that the selection criteria of the different CO galaxy surveys could impact 
our understanding of the relationship between the galaxy stellar mass, star formation activity, and cold molecular gas content.
In order to help shed light on this issue, the left panel of Fig. \ref{histograms} presents the galaxy sSFRs normalised to the 
position of the galaxies on the MS (sSFR/sSFR(MS, $z$, \Mstar) as originally put forward by \citet{Genzel2015}. 
Galaxies are grouped in redshift slices and we distinguish between field and clusters samples.

While, as mentioned above, the galaxy stellar masses are derived in a relatively homogeneous way, 
the estimates of the SFRs rely on a much larger set of methods. 
The most common one is based on the galaxy far-infrared (FIR) luminosity, 
which sometimes takes the UV emission into account \citep{Daddi2007, Daddi2010, Tacconi2013, cybulski2016, 
Hayashi2018, Noble2017, Noble2019, Freundlich2019}.
Another group of studies have used galaxy SED fits \citep{Castignani2020b, Rudnick2017a, Sperone-Longin2021}. 
\citet{Wuyts2011a} showed that in the low- to intermediate-SFR regime, 
SFR $\lesssim 50\,M_\sun {\rm yr}^{-1}$, the values obtained by the two above methods, 
SFR$_{\rm UV + FIR}$ and SFR$_{\rm SED}$, agree with each other within uncertainties. 
Finally, SFRs can be estimated from a set of emission lines, such as [\ion{O}{II}], [\ion{O}{III}], H$\alpha$, 
[\ion{N}{II}], and [\ion{S}{II}] \citep{Bauermeister2013a, Bauermeister2013}. 
The slope and width of the MS provided by \citet{Speagle2014} take this variety of methods into account. 
Therefore the comparison of the different surveys and specifically the position of the galaxies in sSFR/sSFR(MS, $z$, \Mstar) is meaningful.

The difference in galaxy population between surveys is conspicuous in Fig.\ref{histograms}. 
It does not depend on field or cluster environments, but rather on the target selection criteria, 
which results in a general over-representation of galaxies above the MS As a matter of fact,
\citet{Bauermeister2013a, Bauermeister2013} explicitly selected their targets with sSFR between 1 and 4 times that of MS galaxies. 
Similarly, prior detection at 24$\mu$m \citep[e.g.,][]{Daddi2004,cybulski2016}, 
or some of the \texttt{Herschel} bands \citep[e.g.,][]{Daddi2010,Castignani2020b} does play a role in this bias towards actively star-forming systems, 
unless it is purposely controlled to include systems below the MS,
as it was done by the PHIBSS1/2 field surveys at $0.5 < z < 0.8$ \citep[e.g.,][]{Tacconi2010,Tacconi2013,Freundlich2019}.
Very noticeably, our \cluster\ and \clusterb\ samples, without any prior on previous FIR detection and in the vicinity of high density environments, are devoid of galaxies above the MS.


\section{Results\label{discussion}}


\subsection{Gas fractions \label{gas_fractions}}

The right panel of Fig. \ref{histograms} presents the distribution of the galaxy gas fraction, \muhh\,$=M_{\mathrm{H_{2}}} / $\Mstar, 
corresponding to the distribution in normalised sSFRs of the left panel.
The cold molecular gas masses were derived using the same value of $\alpha_{\rm CO}$, 
$\alpha_{\rm MW}=4.36\,M_\odot{\rm (K\,km\,s^{-1}\,pc^2)^{-1}}$, for most samples \citep{Daddi2010, Geach2011, 
Tacconi2010, Tacconi2013, Bauermeister2013a, Bauermeister2013, cybulski2016, Rudnick2017a, Hayashi2018, Noble2017, Noble2019}.
The other samples have linked the value of their CO-to-H$_2$ conversion factor either to metallicity \citep{Freundlich2019}, or
to the position of the galaxy with respect to the MS \citep{Castignani2020b}. 
Considering $\alpha_{\rm MW}$ for these two samples slightly redistributes the position of the galaxies within the distributions, 
but does not modify their global shapes and mean positions.
It stands out from Fig. \ref{histograms} that samples that are skewed towards high sSFRs also have higher galaxy gas mass fractions. 
This is particularly the case at $z \le 0.8$. 
At $z > 1$, the galaxy star formation activity is not sufficient to explain the very significant enhancement of the galaxy cold gas content
and evolution with the lookback time which is an important factor.

\begin{figure}[ht]
	\resizebox{\hsize}{!}{\includegraphics{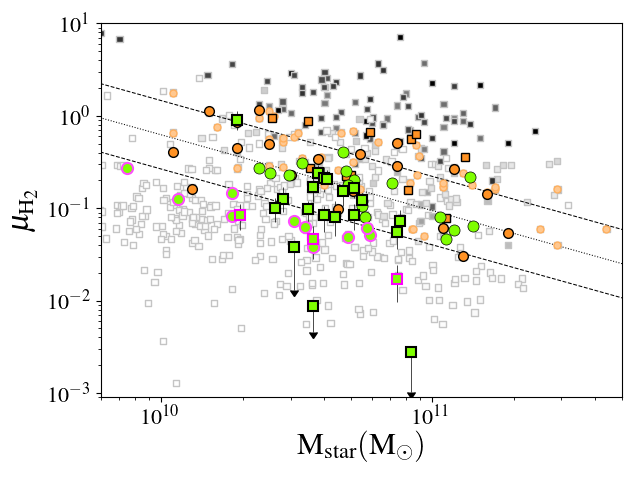}}
	\caption{Fraction of cold molecular gas as a function of the galaxy stellar mass. The colours and shapes of the 
	markers are the same as in Fig. \ref{z_sSFR}. The dotted line is the fit of the \Mstar--\muhh\ 
	relation at $z\sim0.45$ and derived from the relation for the PHIBSS2 galaxies at $z\sim 0.5$, 
	with its dispersion being represented by the two dashed lines. 
	The pink outlined green markers are for the CL1301 low-\muhh\ galaxies located below 
	the $1\times$ $\sigma_{\rm H_2}$ line of the \Mstar--\muhh\ relation for the PHIBSS2 field galaxies.}
    \label{Ms_fg}
\end{figure}

Our \cluster\ and \clusterb\ samples do not cover the high-\muhh\ tail of the other distributions, 
but they do not reach particularly low values either, in particular with respect to PHIBSS2, 
which is the best field counterpart to our study. 
Just as in the case of \clusterb, differences in \muhh\ appear when it is matched with the galaxy \Mstar. This is done in Fig. \ref{Ms_fg}.

In \citetalias{Sperone-Longin2021}, we had identified a subset of star-forming galaxies with low gas mass fractions for their stellar mass.
More specifically, these galaxies were falling below the $1\times\sigma_{\rm H_2}$ dispersion 
of the \Mstar--\muhh\ relation derived at $z=0.55$ from the PHIBSS2 field galaxies at $0.5<z\leq0.6$. 
In order to look for similar types of galaxies around \cluster, we first needed to set the relation between \Mstar\ and \muhh\ at $z=0.45$.
\citet{Tacconi2018} have shown that the slope, in logarithmic scale, of the \Mstar--\muhh\ relation for MS galaxies
does not vary with redshift and that only a zero point shift should be taken into account. 
Therefore, the \Mstar--\muhh\ relation for field MS systems is shown in Fig. \ref{Ms_fg} with a constant slope of $-0.82$ and a shift of $-0.08$\,dex, 
induced by the transition from $z=0.55$ to $z=0.45$. 
This new relation is shown together with the $\pm 1 \times \sigma_{\rm H_2}$ interval, 
with $\sigma_{\rm H_2}=0.37$ the variance of the \Mstar--\muhh\ relation calculated in \citetalias{Sperone-Longin2021}.

We stress that the identification of the low-\muhh\ galaxies is drawn relative to field surveys. 
It does not necessarily imply that these star-forming galaxies have a low cold gas content in absolute terms, 
but rather that they were at least missing from an earlier investigation. 
We explore to what extent their properties could be linked to the high density environments, 
themselves, or their close vicinity, on which we are focussing.

The three passive galaxies of Fig. \ref{MScl1301} are not considered in this analysis because we focus on systems that are still active. 
Therefore, following the same criterion as in \citetalias{Sperone-Longin2021}, 
three star-forming galaxies fall below the $-1\times \sigma_{\rm H_2}$ line of the \Mstar--\muhh\ relation. 
They are highlighted with a pink outer edge in all figures. This is formally a much smaller fraction (14\%) than for \clusterb\ (37\%). 
Nevertheless, in a similar way as for \clusterb, with the exception of one galaxy, 
all our targets fall on the low $1\times \sigma_{\rm H_2}$ part of the \Mstar--\muhh\ relation, 
possibly reflecting the fact that our sample does not contain galaxies with SFRs above the MS (Figs. \ref{MScl1301} and \ref{histograms}).

\begin{figure}[ht]
	\resizebox{\hsize}{!}{\includegraphics{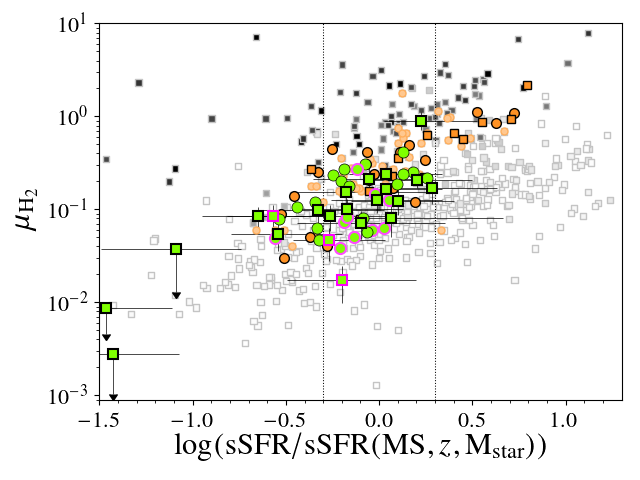}}
	\caption{Fraction of cold molecular gas as a function of the normalised sSFR. The colours and shapes of the 
	markers are the same as on Fig. \ref{z_sSFR}. The dotted vertical lines represent the extent of the MS. 
	The pink outlined green markers are for the low-\muhh\ galaxies from both ALMA samples.}
    \label{sSFR_fg}
\end{figure}

Figure \ref{sSFR_fg} presents the relation between \muhh\ and sSFR normalised to the galaxy position on the MS. 
Two of the low-\muhh\ galaxies are located on the MS, and one falls below it. This latter system, \V, is a cluster member. 
It is interesting to see that while its star formation activity is diminished and its gas content is correspondingly low for its stellar mass, 
it lies however at the same level of \muhh\ as other (more massive) MS galaxies in Fig. \ref{sSFR_fg}. 
However, cluster members are not necessarily depleted in cold gas. Indeed, as alluded in Section \ref{MS_SFR}, 
the galaxies below the MS are linked to the cluster environment, and this is before their reservoir of cold gas is impacted for two of them. 

The other low-\muhh\ cluster member is \VIII, and still is on the MS, but on the lower edge. 
The third low-\muhh\ galaxy is \XVI\, again on the MS and on the lower edge. 
It is located far from the cluster core (r $> 5.5 R_{200}$) and does not belong to any identified overdensity.

We performed a Anderson--Darling (A-D) test \citep{Scholz1987}
between the \muhh\ distributions of the MS PHIBSS2 and \clusterb\ galaxies. 
It results in $p = 0.021$, meaning that there is only a 2.1\% chance that the two samples arise from the same \muhh\ distribution. 
Similarly, combining both \cluster\ and \clusterb\ gives p-value of 0.011, 
suggesting that these samples are not drawn from a PHIBSS2-like population at a 98.9\% confidence level.

\begin{figure}[ht]
\resizebox{\hsize}{!}{\includegraphics{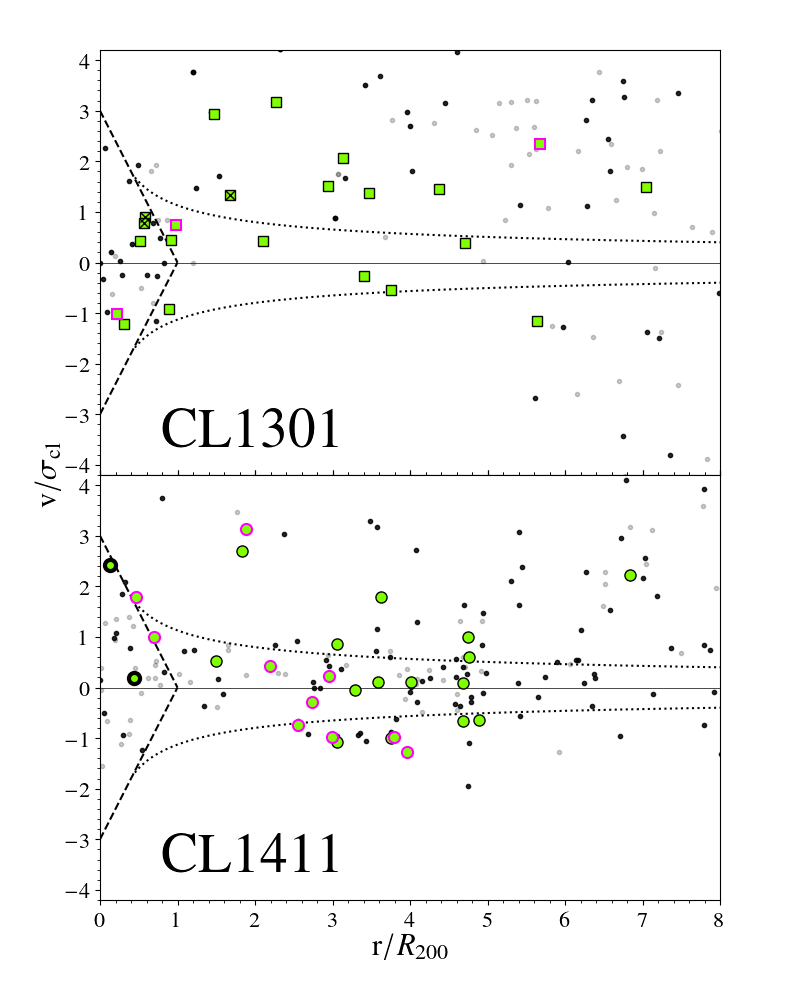}}
\caption{Projected phase-space diagrams \cluster\ (top panel) and \clusterb\ (bottom panel). 
The star-forming (black dots) and passive (grey points) galaxies with spectroscopic redshifts are identified in the $UVJ$ diagram for \clusterb. 
As to \cluster, when $Ks$ band photometry is missing, the distinction between the two types of galaxies is also based on the positions of the red sequence,
and the blue clump in the colour-magnitude diagrams using the $u,\,g$, and $i$ bands.
Assuming a Navarro-Frenck-White (NFW) halo \citep{Navarro1996}, the area under the influence of the cluster potential, 
either from small relative velocities or small distance to the cluster cores, is shown with the dotted black line.
The dashed black lines indicate the region in which the galaxies are considered to be cluster members \citep[i.e. virialised region,][]{Mahajan2011}.
}

\label{psd}
\end{figure}


 \subsection{Comparison with CL1411.1$-$1148 \label{comparison}}

The difference in the fraction of low-\muhh\ galaxies in \cluster\ and \clusterb\ is puzzling, 
given their similar properties in mass, redshift, and the identical target selection criteria that we applied. 
As seen in Fig. \ref{MScl1301}, the ALMA targets linked to \clusterb\ span a slightly more extended stellar mass range than those linked to \cluster. 
However their distributions have the same median, log(\Mstar/$M_\sun$)\,=\,10.69 and 10.6, respectively.
Some of the \cluster\ ALMA targets have diminished SFRs, placing them further below the MS than in the case of \clusterb, 
but these galaxies do not form the bulk of the low-\muhh\ systems. Therefore, the origin of the difference between the two samples must be found elsewhere.

Figure \ref{psd} shows the projected phase-space diagrams of the two clusters. 
Following \citet{Jaffe2015}, we assumed a NFW halo profile with a concentration of ${\rm c} = 6$ to compute the escape velocity for our clusters (dotted lines). 
Here, we only look at star-forming galaxies as identified from the $UVJ$ or $ugi$ diagrams (black dots), 
because they are the galaxies for which we can define the low-\muhh\ systems. 
The spectroscopic members, including the ALMA targets, within 4\,$\sigma_{cl}$ and $8 \times R_{200}$, 
are seen relatively to the virialised, infall, and escape region boundaries \citep[see also][]{Mahajan2011}. 
Galaxies are considered infalling onto the clusters when they are located between 1 and $\sim 5$ virial radii \citep{Oman2013, Albaek2017}. 
The difference between the two samples stands out: most of the ALMA targets are located inside the \clusterb\ infall region, 
while they are much more spatially scattered in the case of \cluster. 
Furthermore, all low-\muhh\ galaxies are located either inside or very close to the cluster core, or again within or very close to the \clusterb\ infalling region. 
Leaving aside the clusters themselves, these low-\muhh\ galaxies represent 44\% of the full infalling population, and 66\% of the galaxies located between 1 and $4 \times R_{200}$. 
Applying these fractions to \cluster\, we could in principle have expected 1 or 2 low \muhh\ infalling galaxies however, we found none. 
This absence most likely reflects the fluctuations induced by low number statistics, given that we only observed three \cluster\ star-forming galaxies between 1 and $4 \times R_{200}$. 
This is a strong hint that the population of low-\muhh\ galaxies are preferentially induced by the cluster environment. 
The interplay between the removal of the cold gas reservoir, or at least the change in its properties, and the decrease in SFR is subtle and difficult to catch in the act. 
The cluster infall regions seem to be the best place to look for them and identify the physical processes at play.
Future follow-up studies of the same type will assess the significance and origin of the transformation of galaxies along their path to the cluster cores.


\section{Conclusion\label{conclusion}}

We have presented the analysis of the molecular gas content, derived from ALMA CO(3-2) line observations, 
of a sample of 22 galaxies located within \cluster\ ($z=0.4828$, $\sigma_{\rm cl}=681$\,km\,s$^{-1}$) and in its surrounding large-scale structure. 
Unlike previous works, our sample selection does not impose a minimum galaxy SFR or detection in FIR.
As such and as much as possible, it delivers an unbiased view of the gas content of normal star-forming galaxies at $z\sim0.5$. 
At the same time, it offers insight into the status of the molecular gas content of galaxies in interconnected cosmic structures.
Our study highlights the variety of paths to star formation quenching, 
and most likely the variety of physical properties (i.e. temperature, density) of the corresponding galaxy cold molecular gas.

Similar to our first results on the analysis of the large-scale structure around \clusterb\ presented in \citetalias{Sperone-Longin2021}, 
although to a smaller extent, our observations in \cluster\ reveal a larger number of star-forming galaxies with lower gas fraction, 
at fixed stellar mass, than what had been found in previous surveys at comparable redshifts. 

The cluster environment does not necessarily affect the galaxy molecular gas content. 
Eight of our ALMA targets are bona fide \cluster\ members (r $\lesssim R_{200}$ and v/$\sigma_{cl} \lesssim 1$). 
From those, only two show signs of lower gas fractions for their stellar masses when compared to the  relation between these two quantities derived for field galaxies.
One of them is nevertheless still on the star-forming MS, hence with a normal activity. The second system shows evidence of transition towards the passive sequence.

Star formation and cold gas content are indeed not always linked in a straightforward manner. 
Three of our ALMA targets fall below the $-0.3$\,dex boundary of the star-forming MS. 
These are systems in the transition region between star-forming and passive systems. 
While two of those are bona fide cluster members, the third one is infalling. 
Only one of the cluster members mentioned above has a low gas fraction for its stellar mass, 
the other two galaxies are normal despite their lowered star-forming activity and irrespective of their location.

Three galaxies turn out to be on the passive sequence (log(SFR)\,$\lesssim-0.2$) and only have upper limits on their CO fluxes. 
The lack of star formation activity and their CO deficiency does not seem to be driven by morphology (internal quenching). 
Indeed, similar systems in our sample, without evidence of disks nevertheless have well measured CO fluxes.

The comparison with \clusterb\ highlights the enhanced fraction of galaxies with low gas fraction, 
compared to the field and at a fixed stellar mass in the cluster infall region. 
This provides a strong hint of environmental dependence and stresses the need for the extension of this type of investigation to a larger number of clusters and their related large-scale structures.


\begin{acknowledgements}
This paper makes use of the following ALMA data: ADS/JAO.ALMA\#2017.1.00257.S.
ALMA is a partnership of ESO (representing its member states), NSF (USA) and NINS (Japan), together with NRC (Canada), 
MOST and ASIAA (Taiwan), and KASI (Republic of Korea), in cooperation with the Republic of Chile. 
The Joint ALMA Observatory is operated by ESO, AUI/NRAO and NAOJ.
The authors are indebted to the \textit{International Space Science Institute} (ISSI), Bern, Switzerland, 
for supporting and funding the international team `The Effect of Dense Environments
on Gas in Galaxies over 10 Billion Years of Cosmic Time'.
We are grateful to the \texttt{Numpy} \citep{numpy:2006, numpy:2011}, 
\texttt{SciPy} \citep{scipy:2020}, \texttt{Matplotlib} \citep{matplotlib:2007},
\texttt{IPython} \citep{ipython:2007}, and 
\texttt{Astropy} \citep{astropy:2013,astropy:2018} teams for providing the scientific community with essential Python tools.
\end{acknowledgements}


\bibliographystyle{aa}
\bibliography{library}


\begin{appendix}


\section{ALMA maps and spectra of our galaxies \label{section-maps}}


In this appendix, we present the $i$-band images, the ALMA intensity maps and the spectra of all of our targets.
The low \muhh\ targets are indicated as such by a label on the bottom left of their $i$-band image.

\begin{figure*}[ht]
    \centering	
	\includegraphics[scale=0.3]{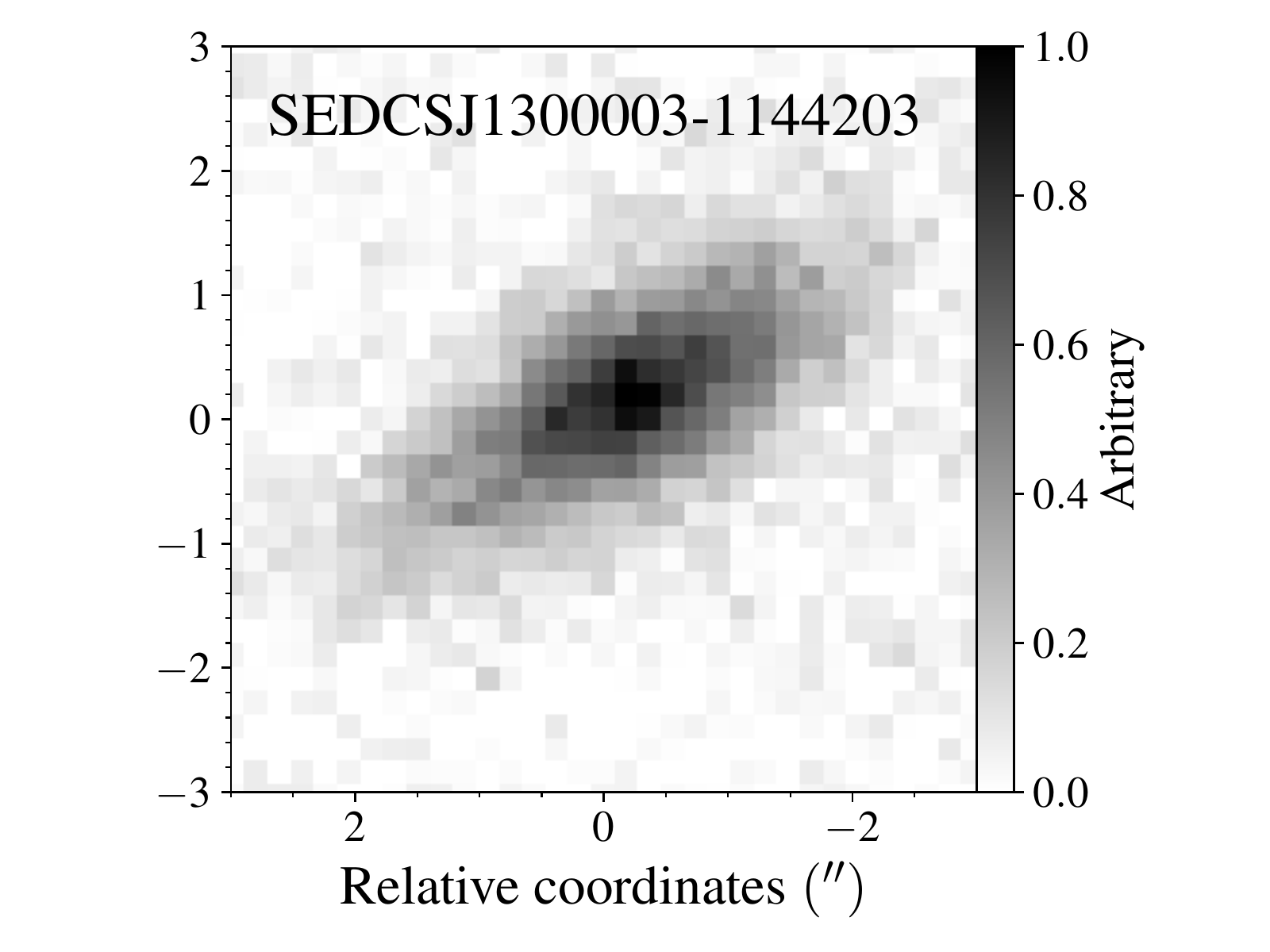}
	\includegraphics[scale=0.3]{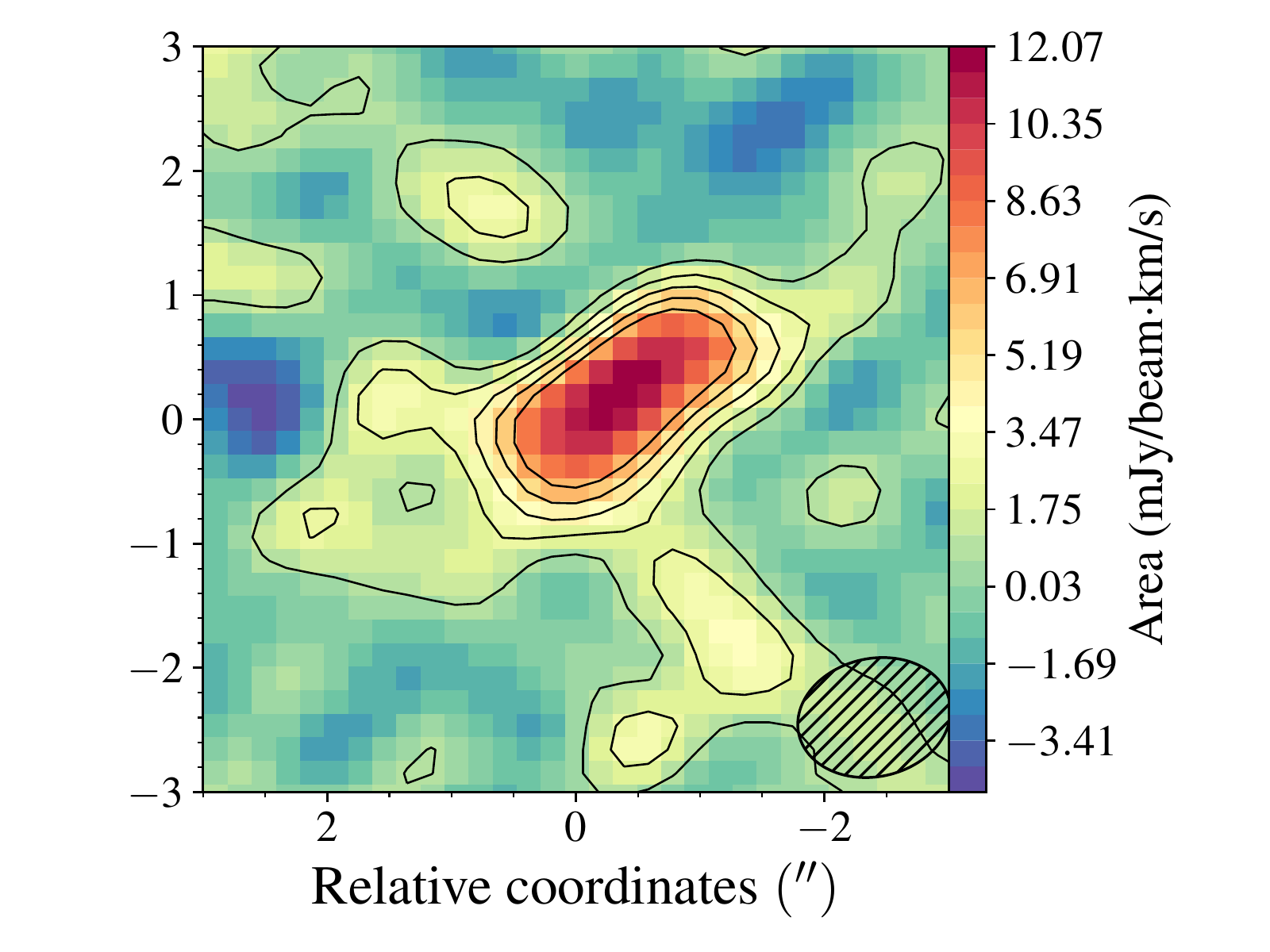}
	\includegraphics[scale=0.3]{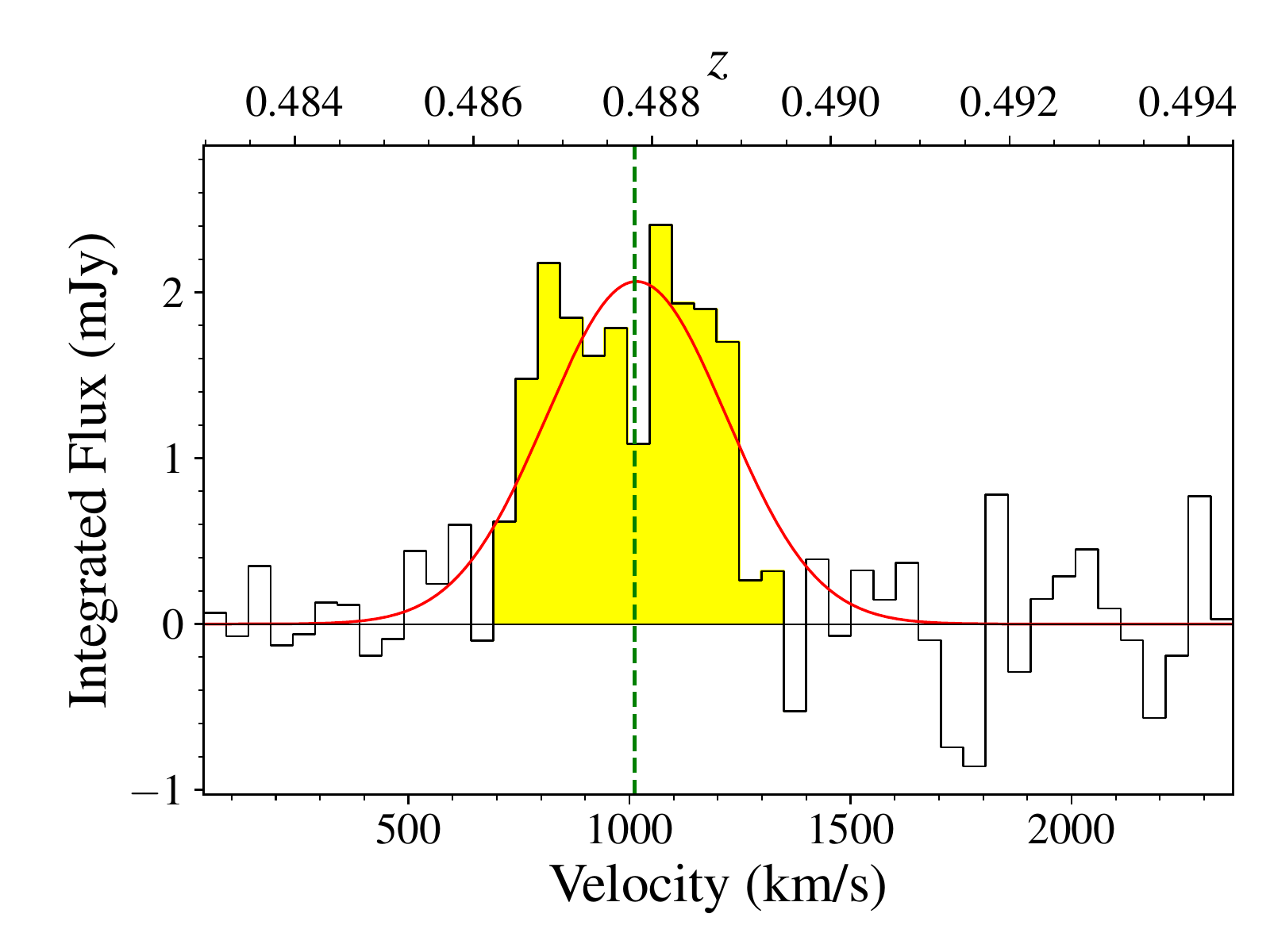}

	\includegraphics[scale=0.3]{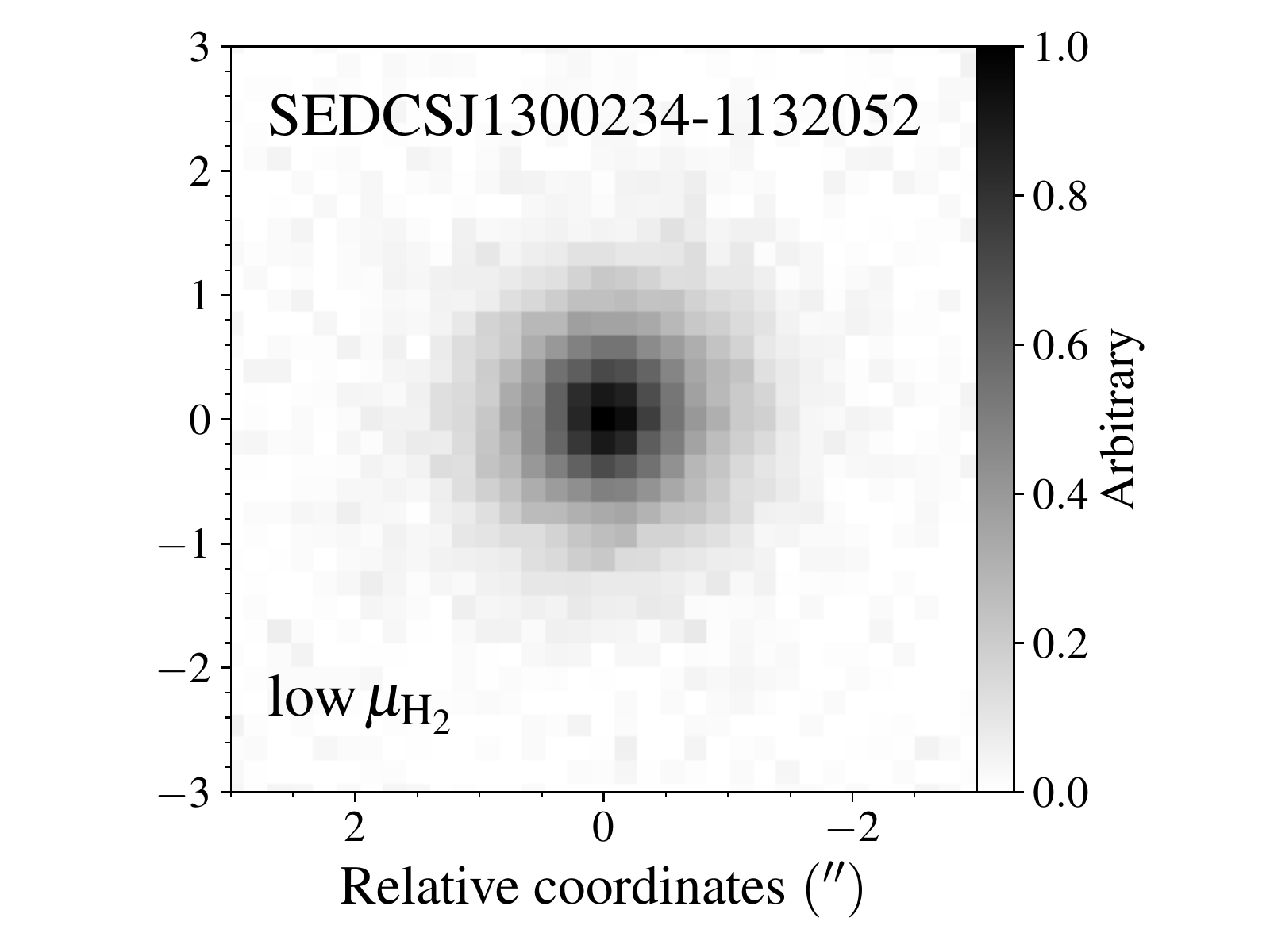}
	\includegraphics[scale=0.3]{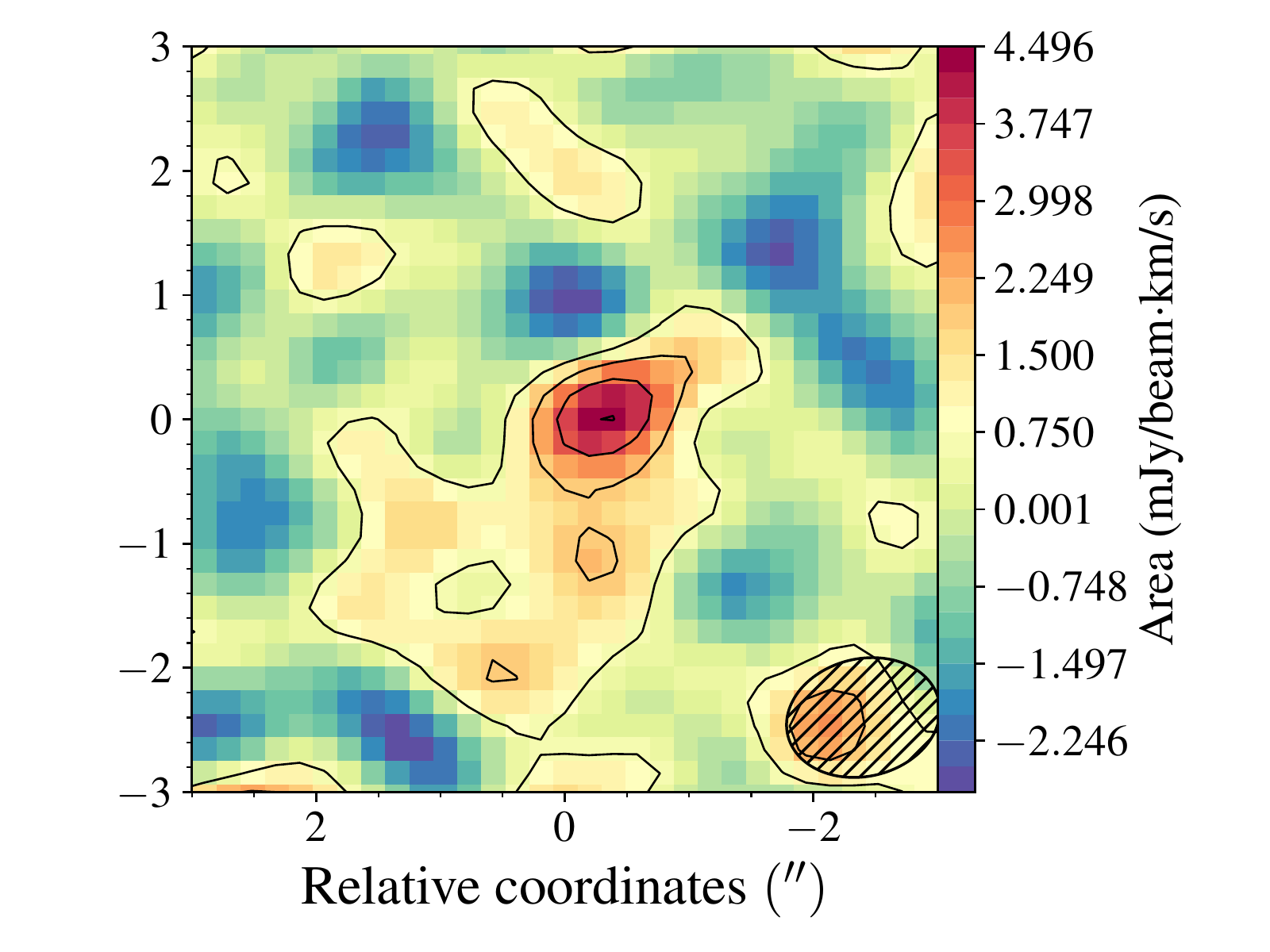}
	\includegraphics[scale=0.3]{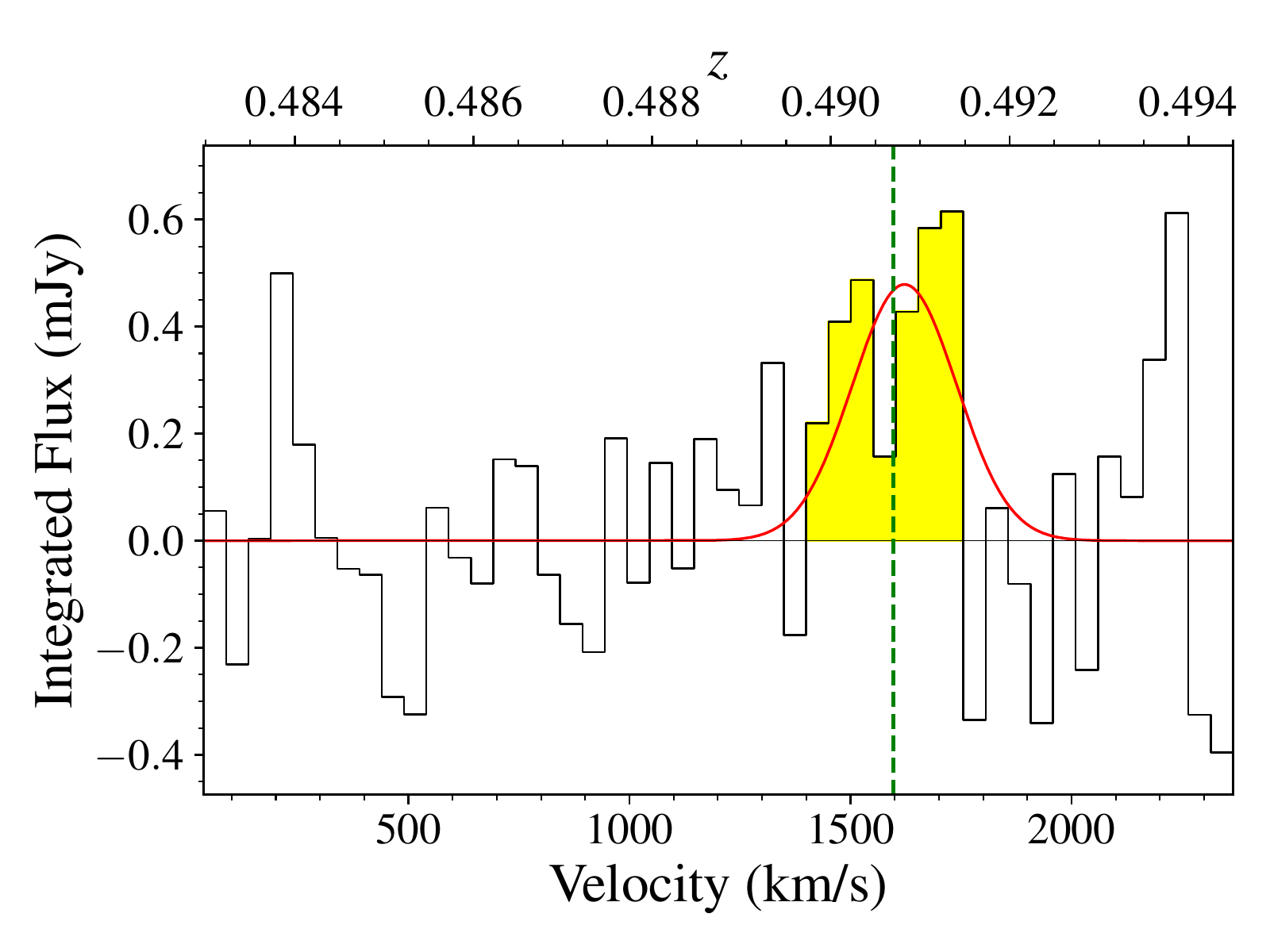}

	\includegraphics[scale=0.3]{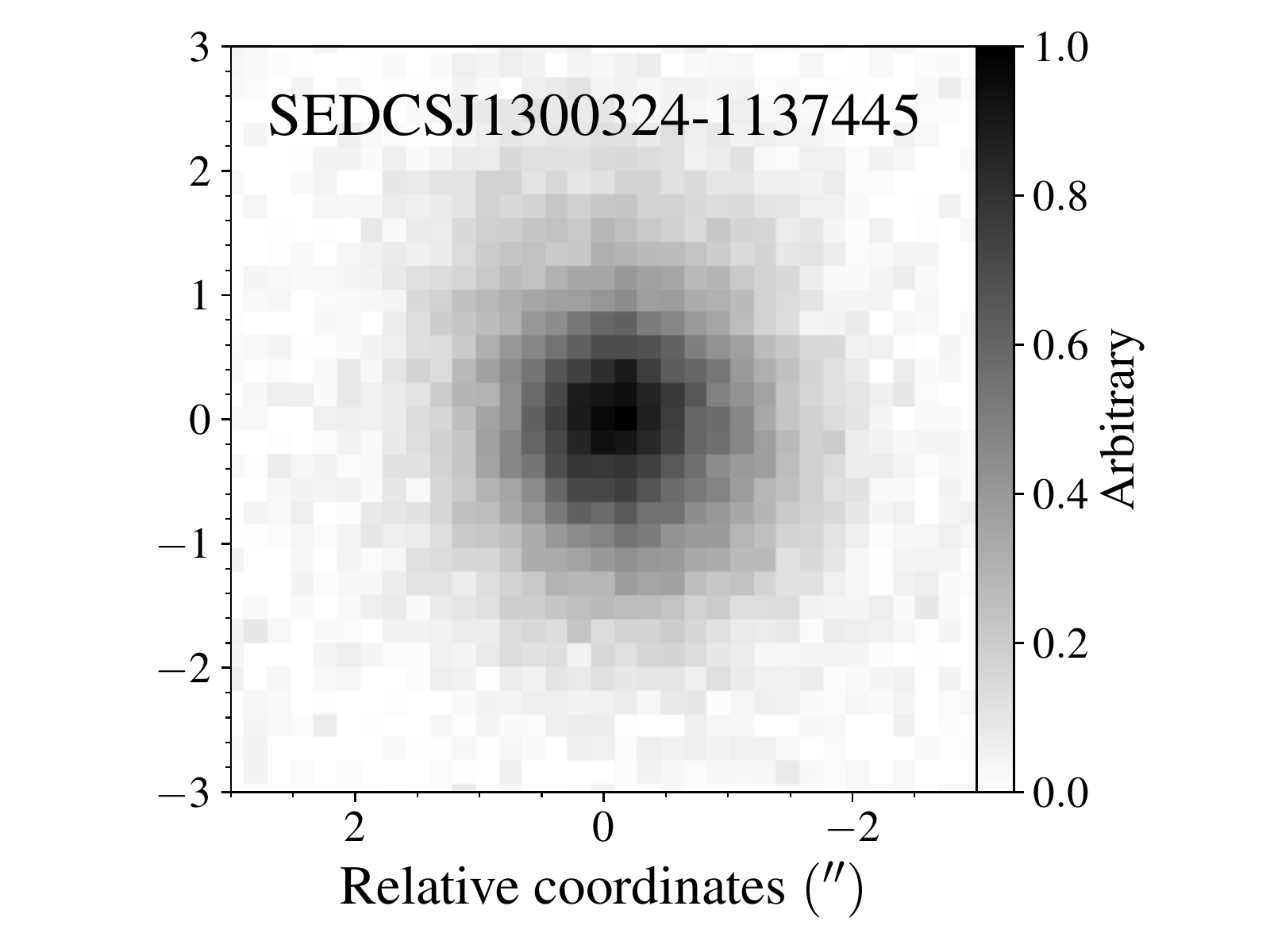}
	\includegraphics[scale=0.3]{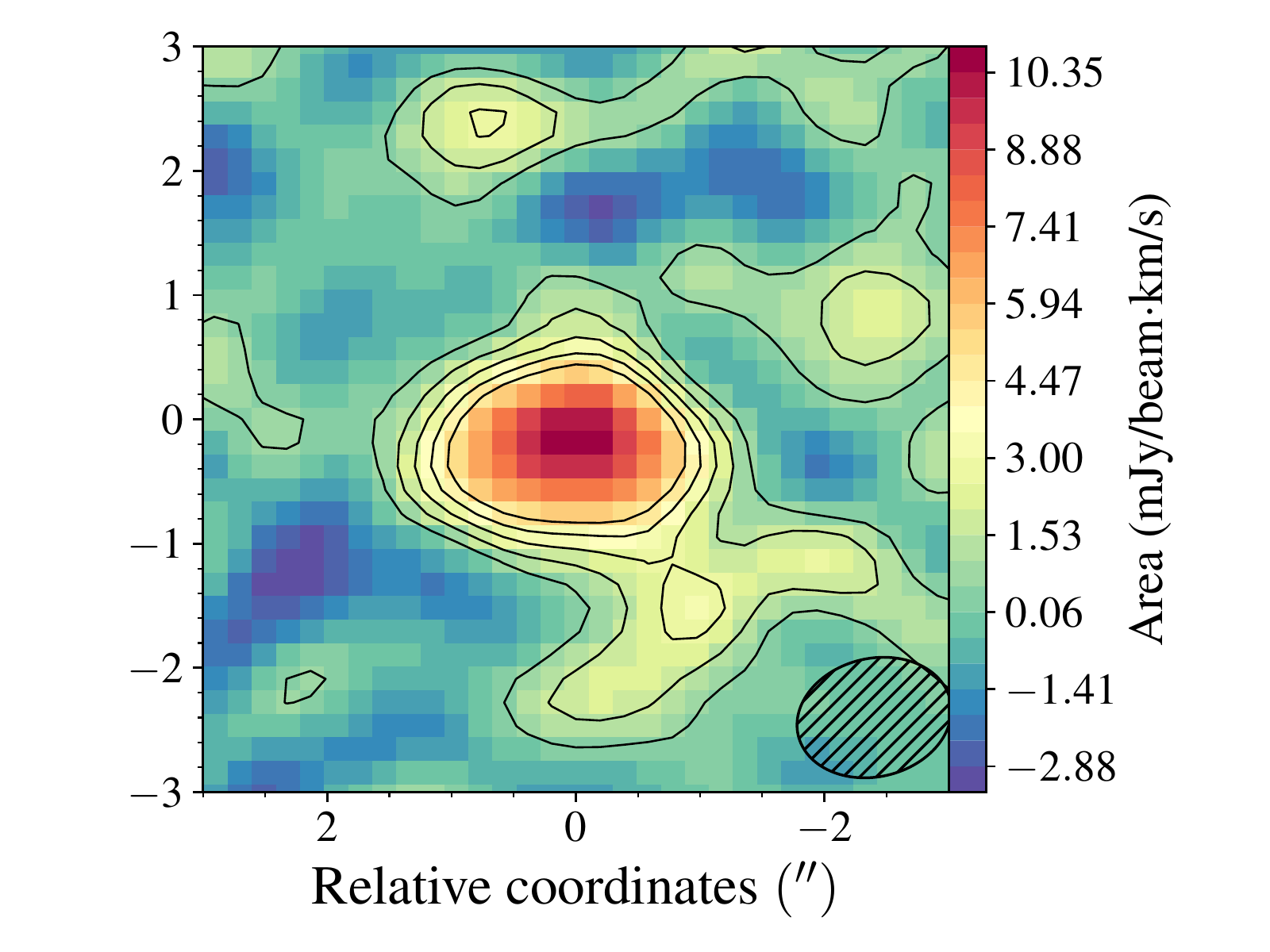}
	\includegraphics[scale=0.3]{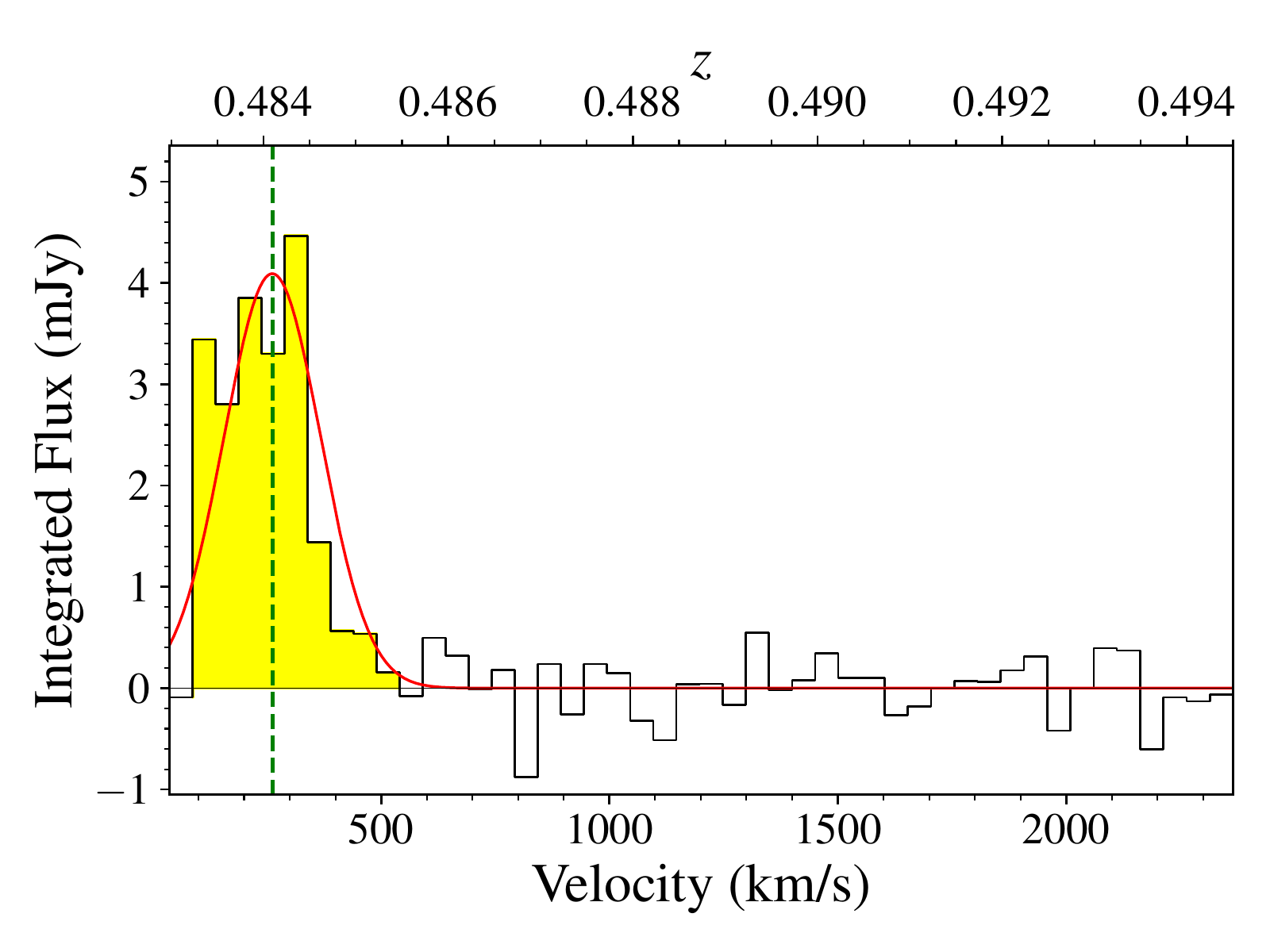}

	\includegraphics[scale=0.3]{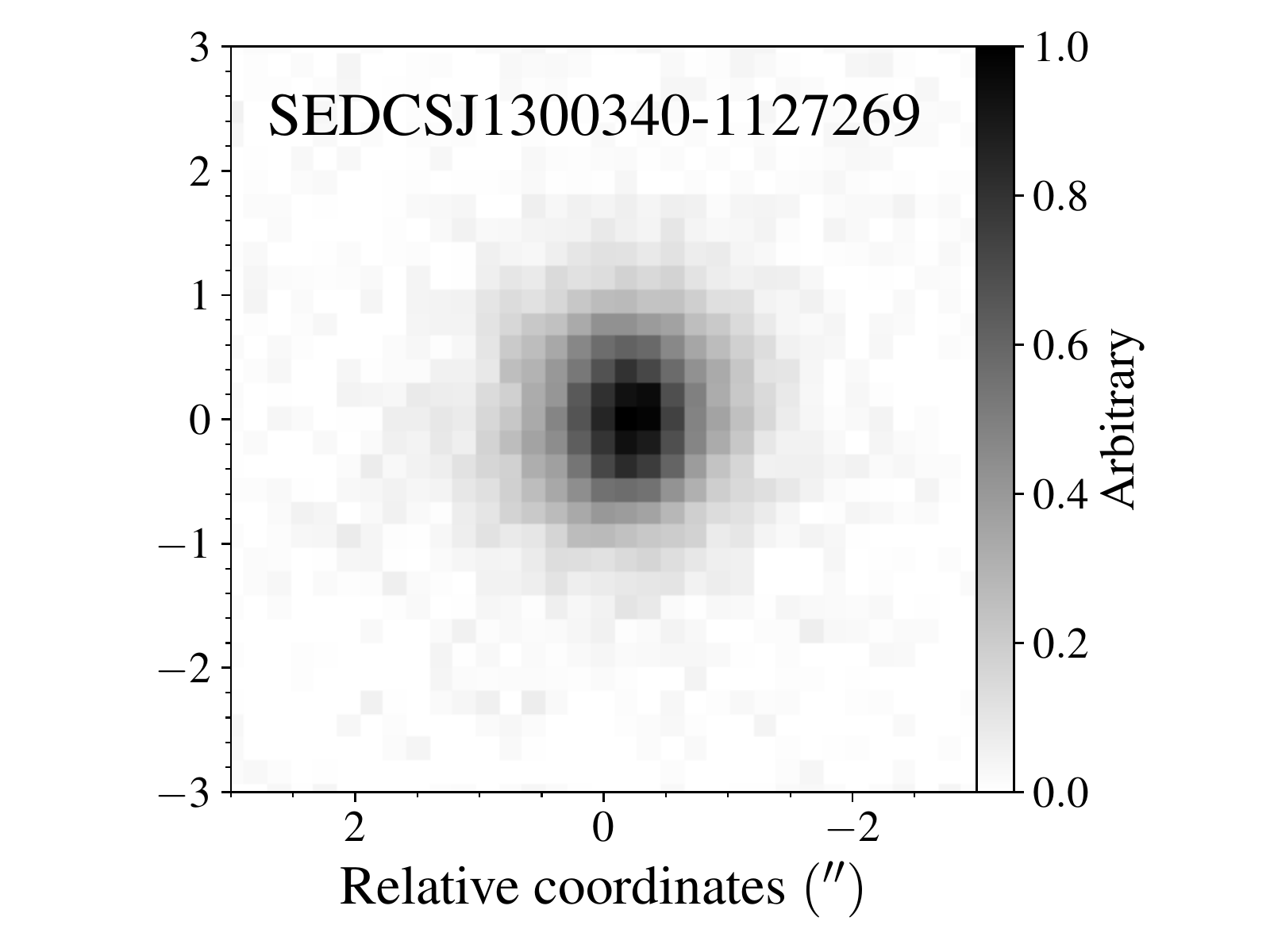}
	\includegraphics[scale=0.3]{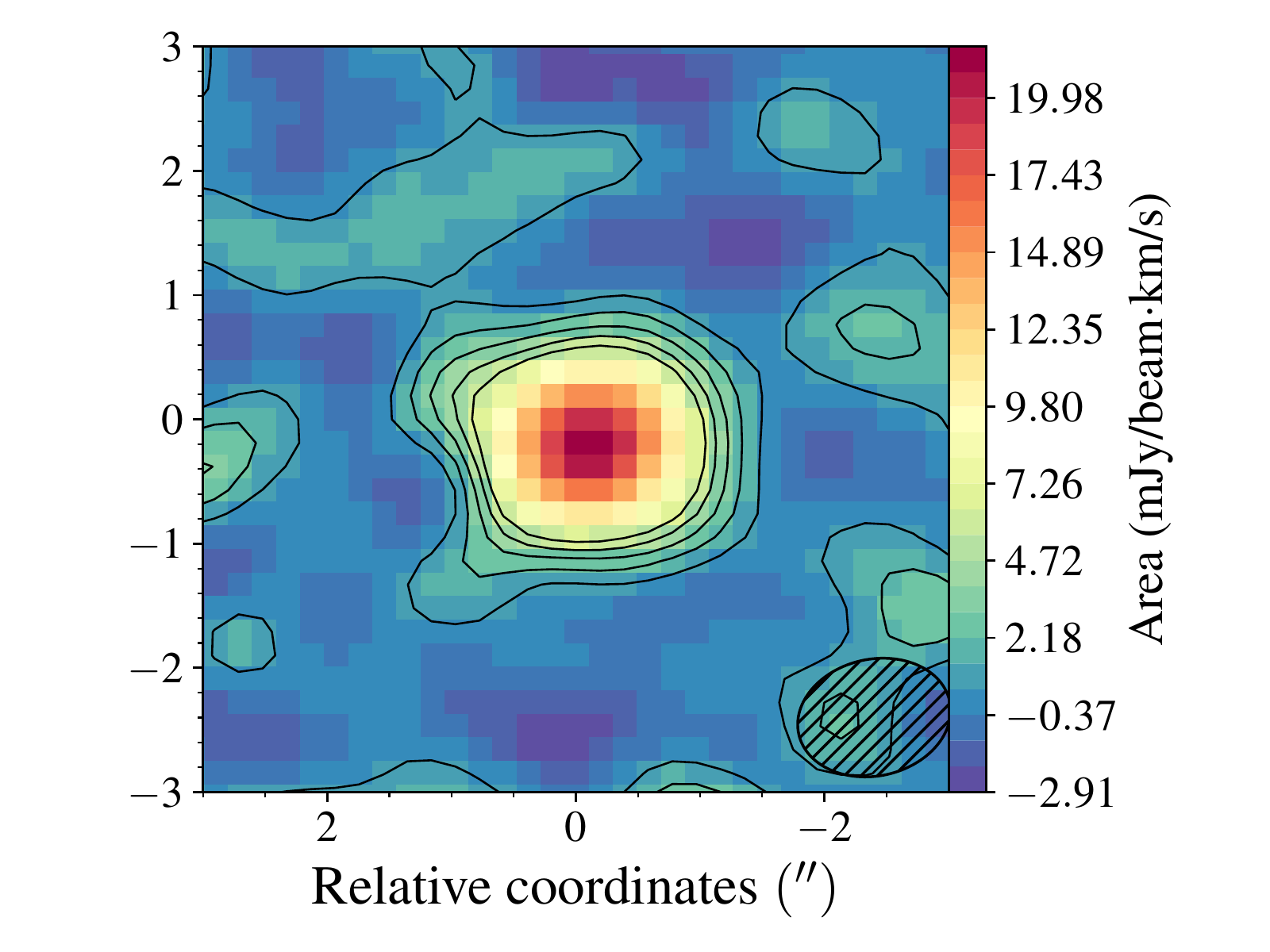}
	\includegraphics[scale=0.3]{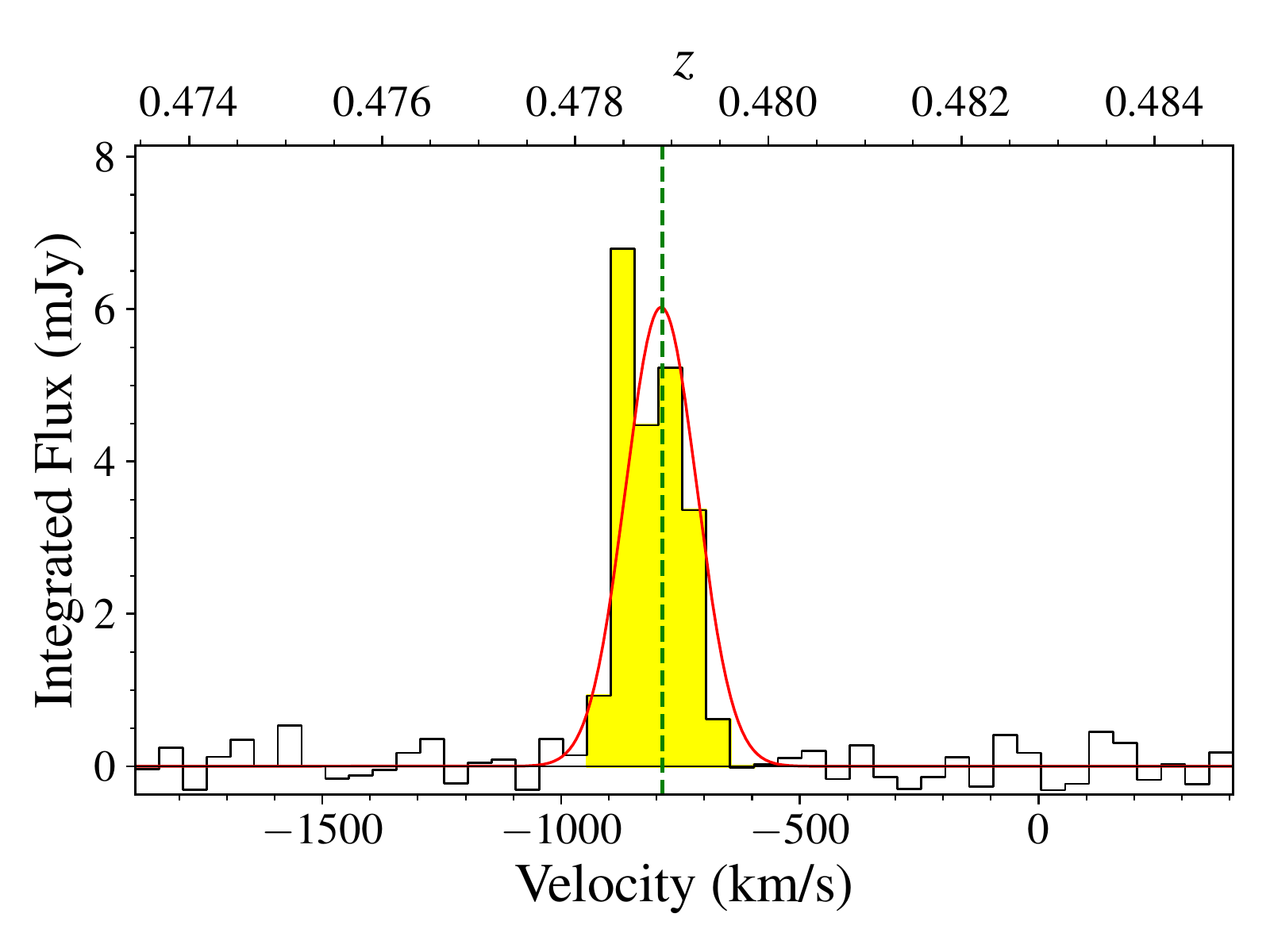}

	\includegraphics[scale=0.3]{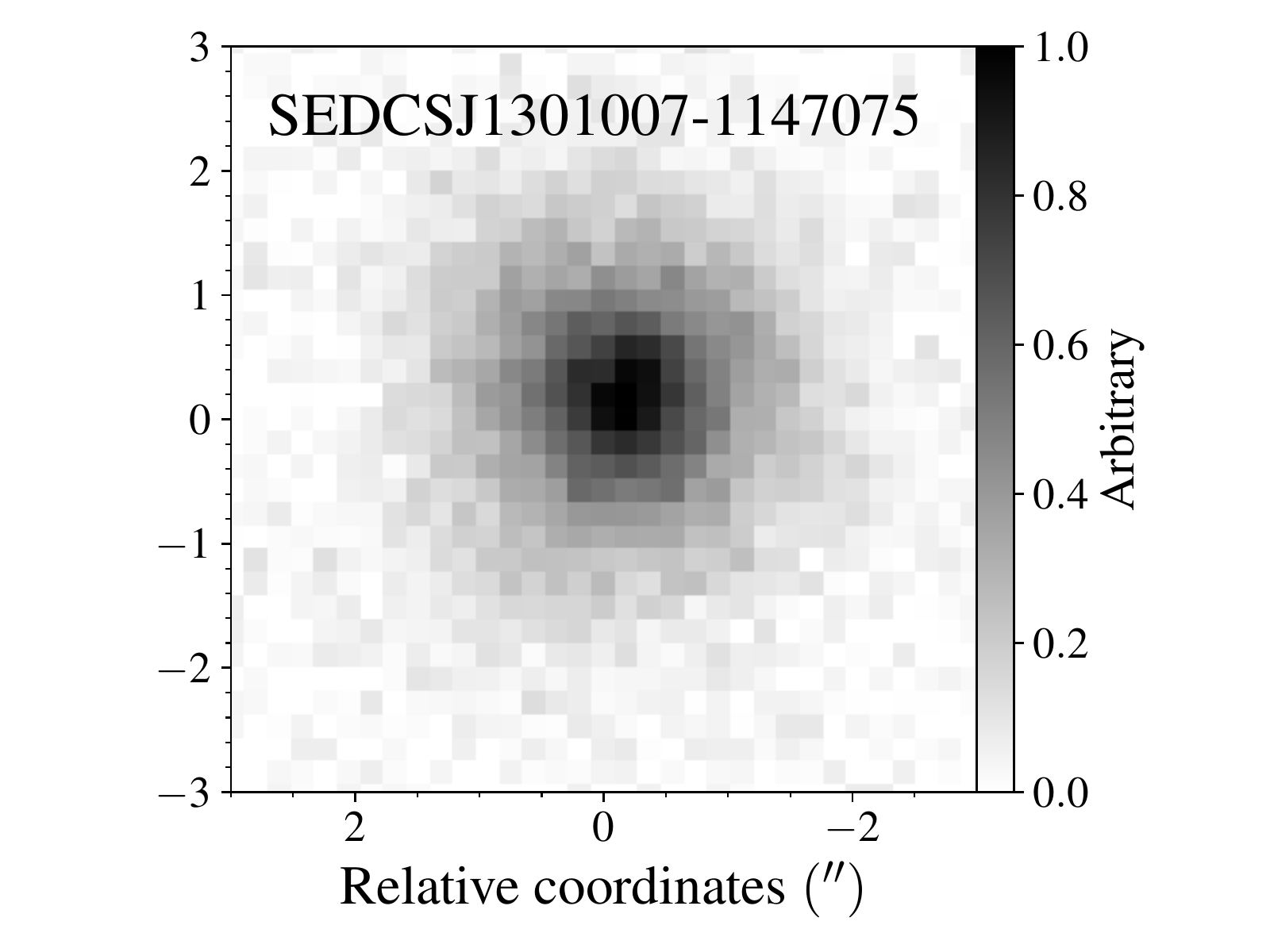}
	\includegraphics[scale=0.3]{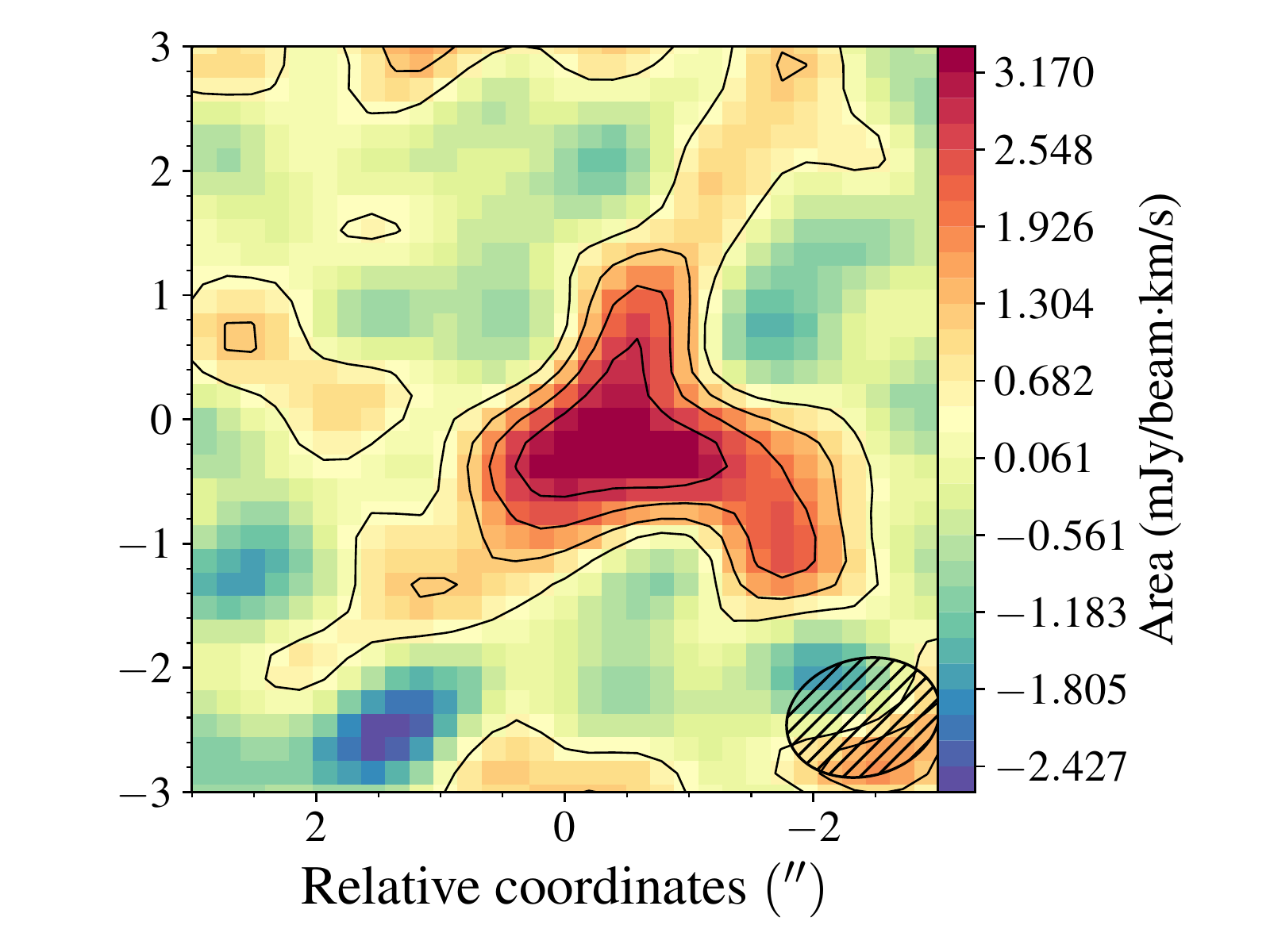}
	\includegraphics[scale=0.3]{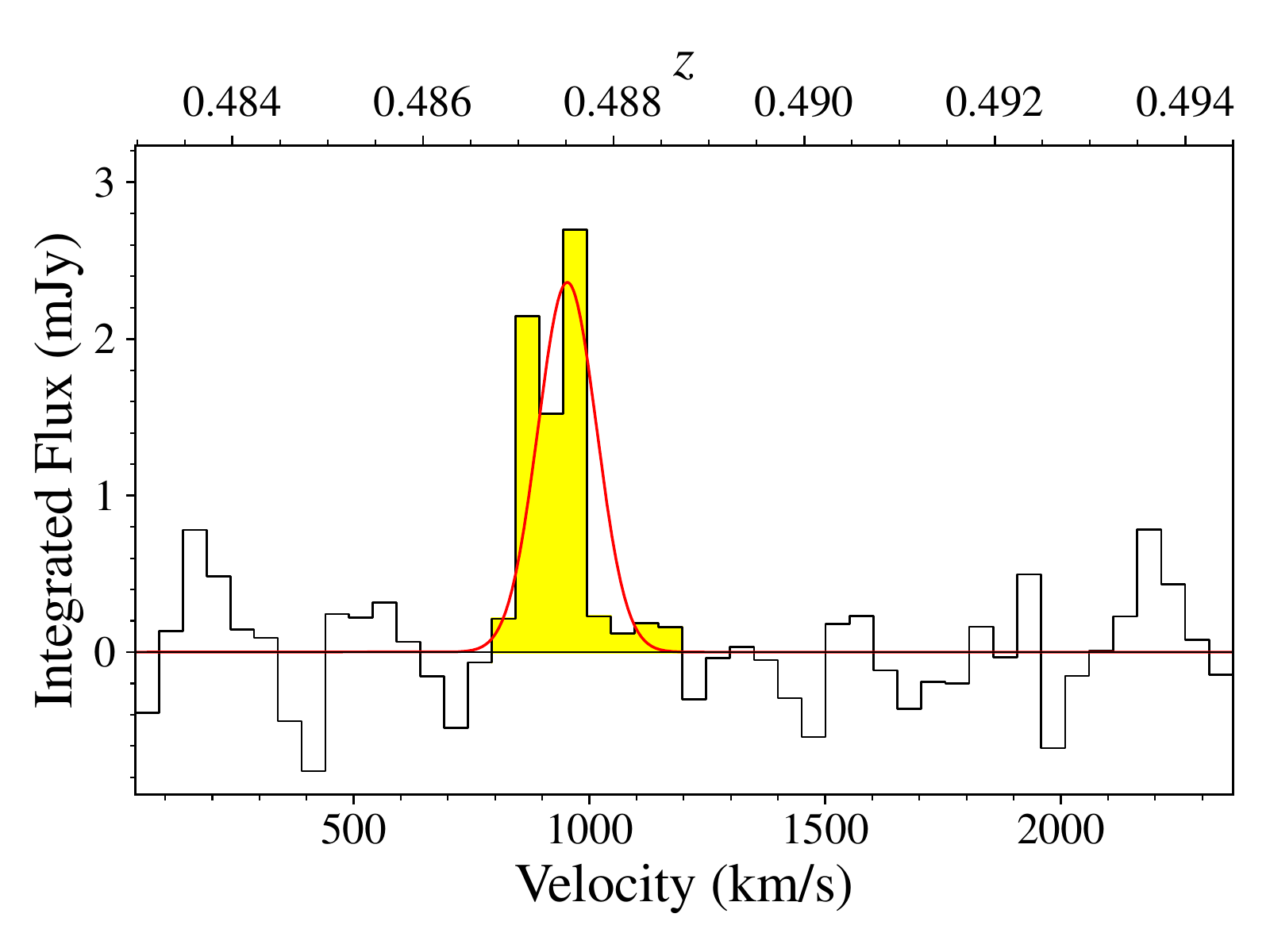}

    \caption{\textit{Left}: CFHT/MEGACAM $i$-band images of our galaxies
    in a $6\arcsec \times 6\arcsec$ snapshot, centred on the galaxies coordinates. 
    \textit{Middle}: ALMA map of the CO(3-2) emission around our galaxies. 
    The spatial scale is the same as in the left panel.
    The colour wedge of the intensity map is in mJy$/$beam\,km\,s$^{-1}$. The 
    contours are defined such that they are spaced by 2 times the rms and are between 1 and 9. 
    In the bottom right corner is the beam size. 
    \textit{Right}: Spectra show the flux, $S_\mathrm{CO}$, spatially 
    integrated as indicated in Sect. \ref{alma_obs}, of the source 
    in mJy in function of the velocity in km\,s$^{-1}$, with respect to the cluster redshift. 
    The Gaussian profiles are fits of the emission lines from which we derived our FWHMs. 
    The yellow filled zones correspond to the spectral extent of the emissions.
    The green vertical line corresponds to the spectroscopic redshift.}
 \label{maps}
\end{figure*}    

\begin{figure*}[htbp]\ContinuedFloat
	\centering			
	\includegraphics[scale=0.3]{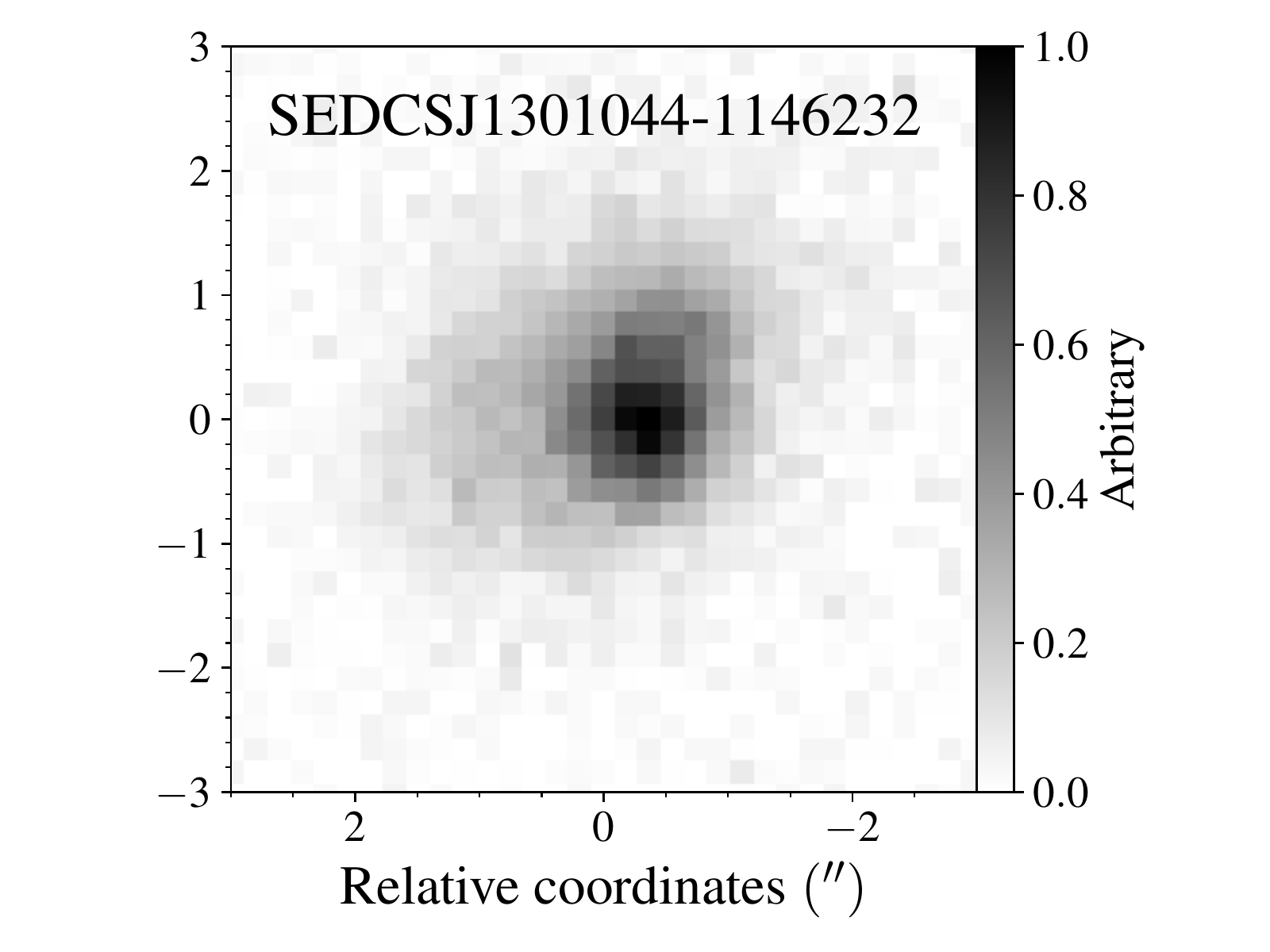}
	\includegraphics[scale=0.3]{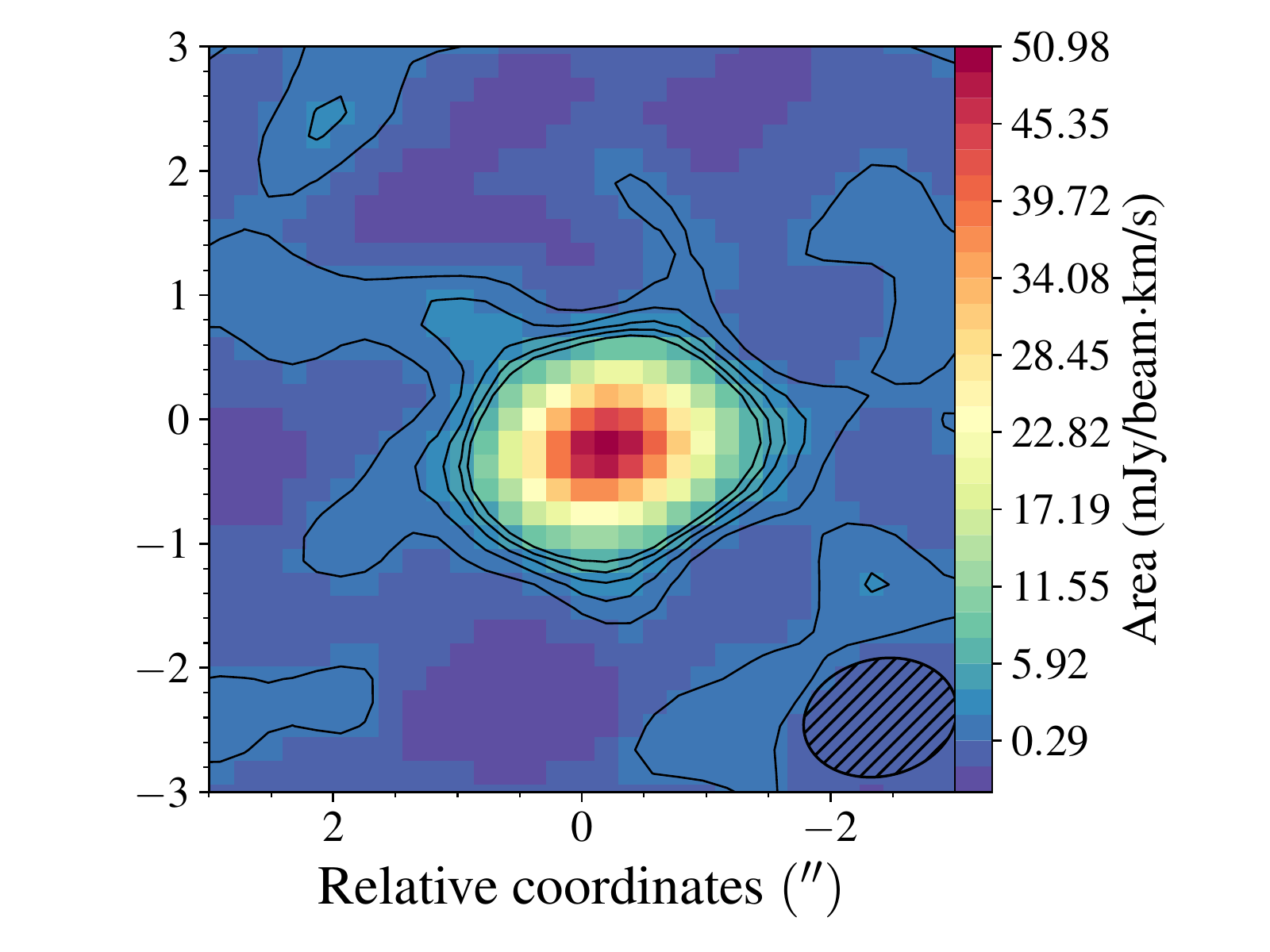}
	\includegraphics[scale=0.3]{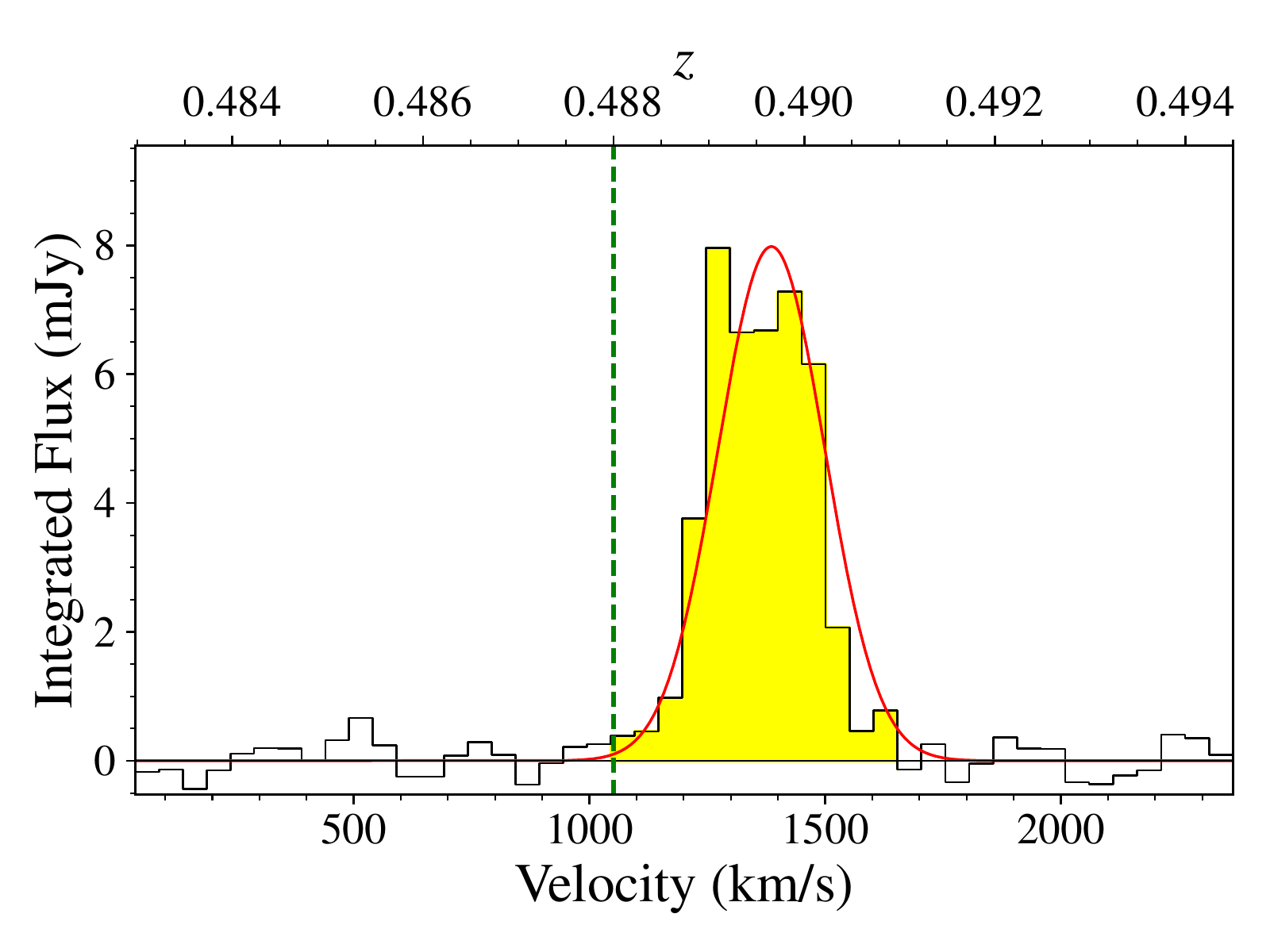}

	\includegraphics[scale=0.3]{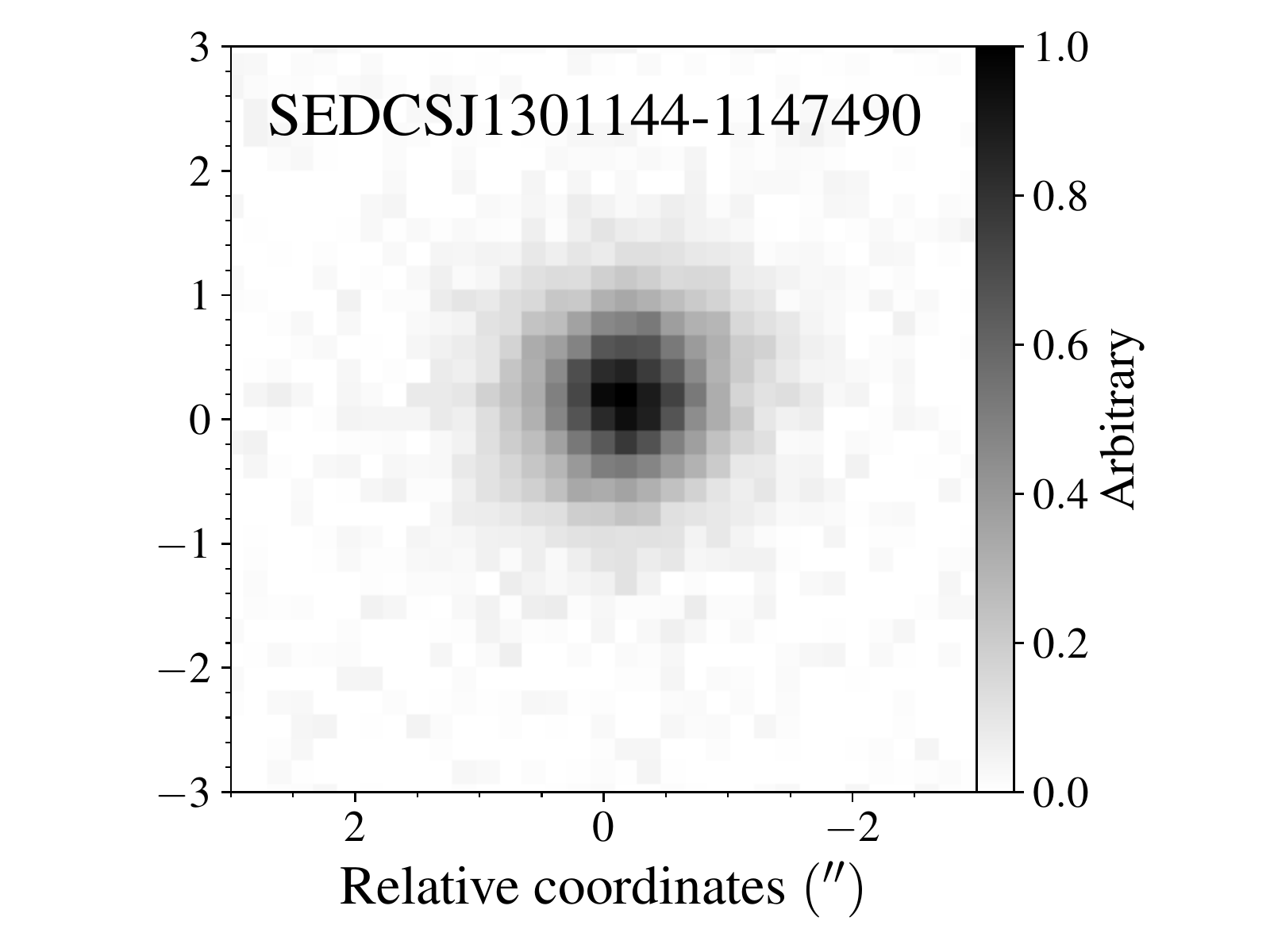}
	\includegraphics[scale=0.3]{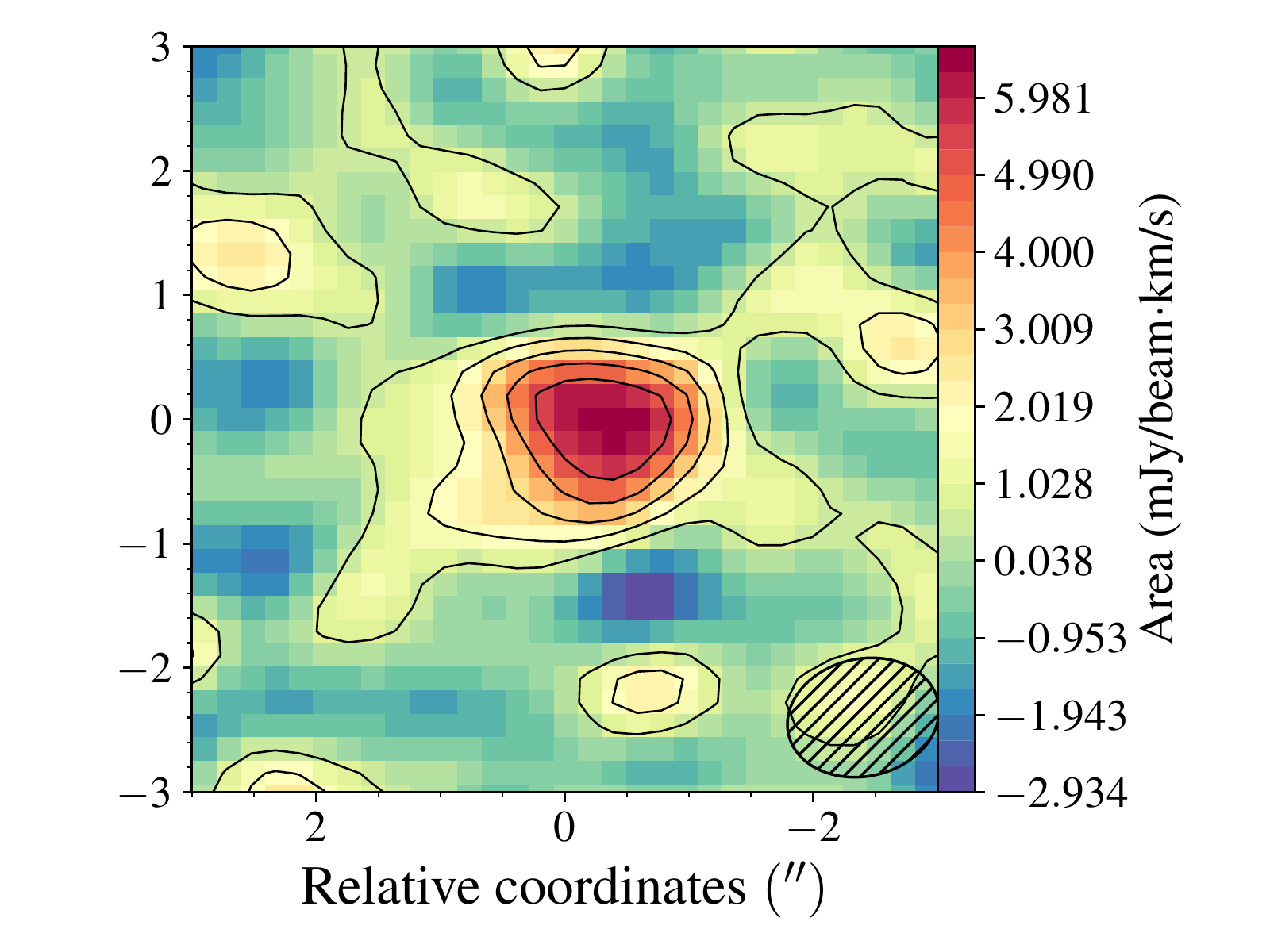}
	\includegraphics[scale=0.3]{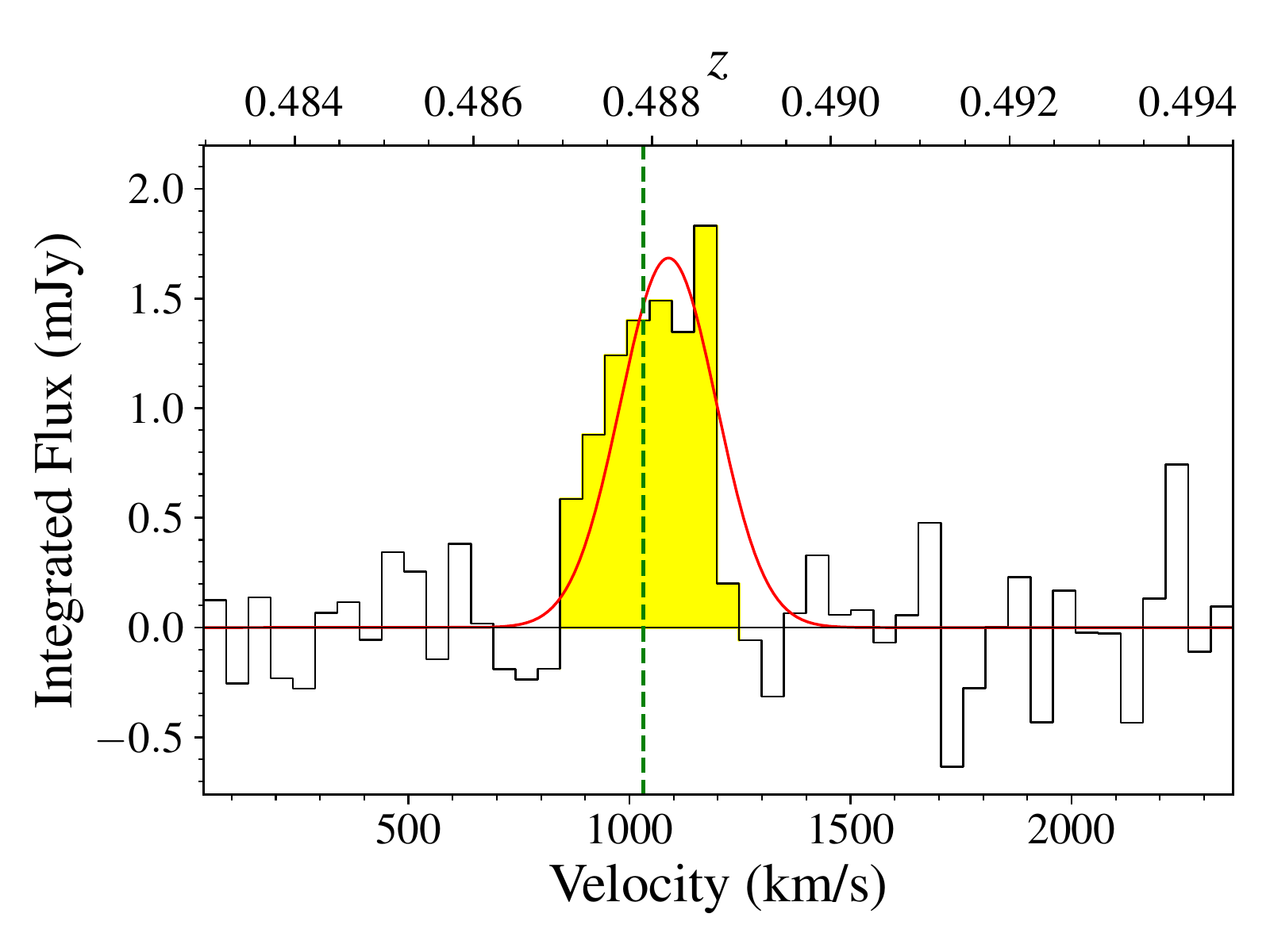}

	\includegraphics[scale=0.3]{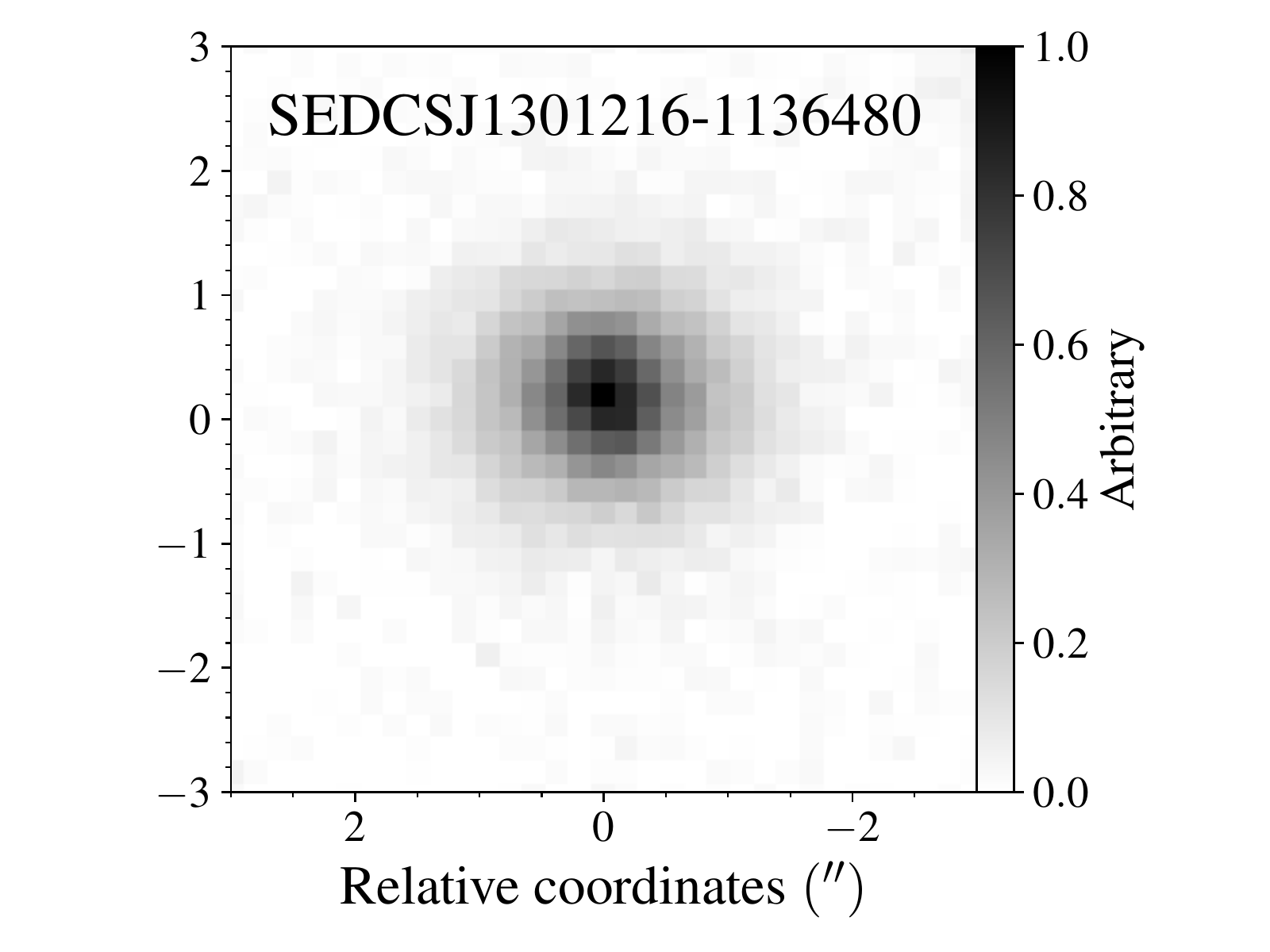}
	\includegraphics[scale=0.3]{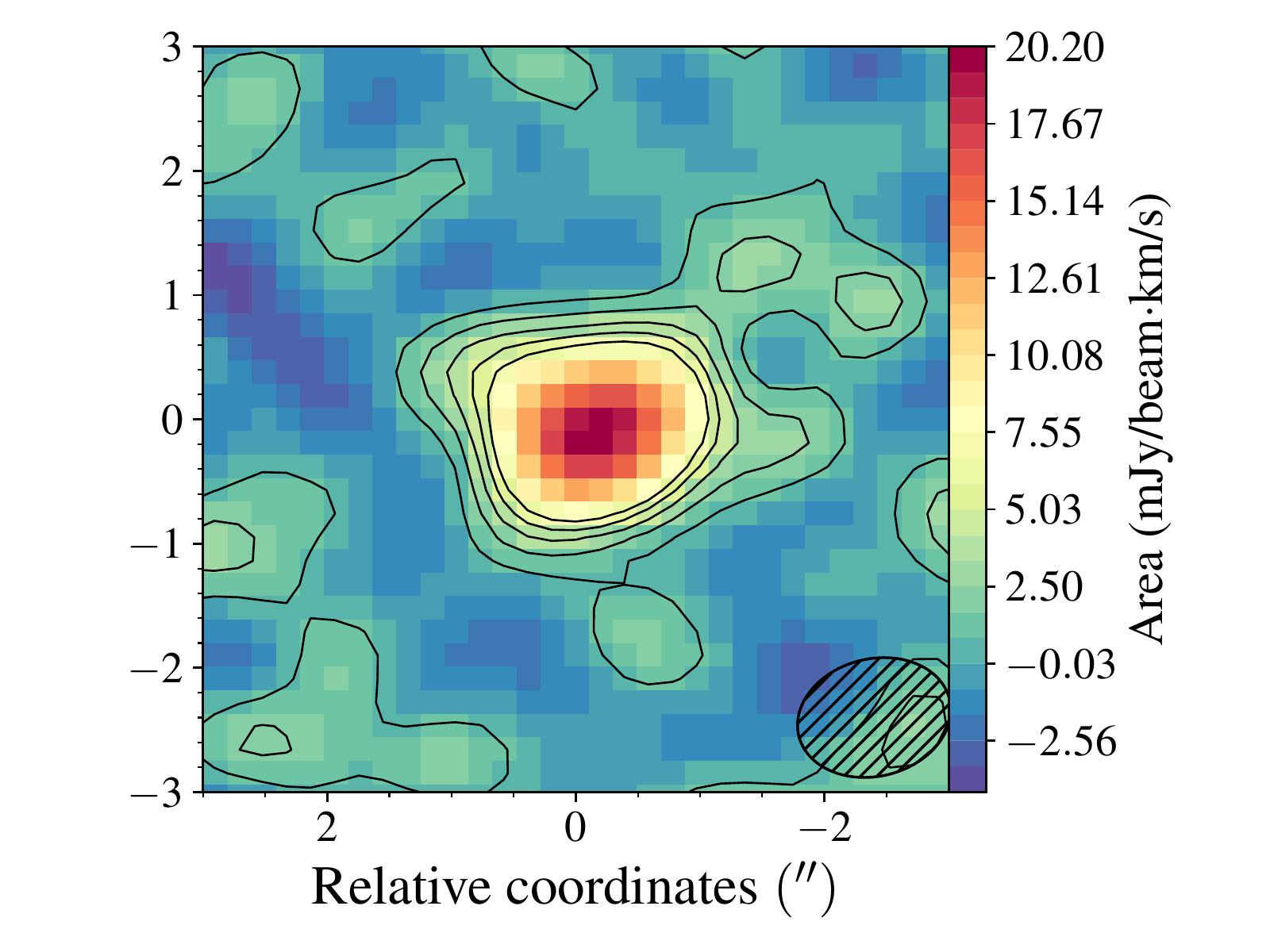}
	\includegraphics[scale=0.3]{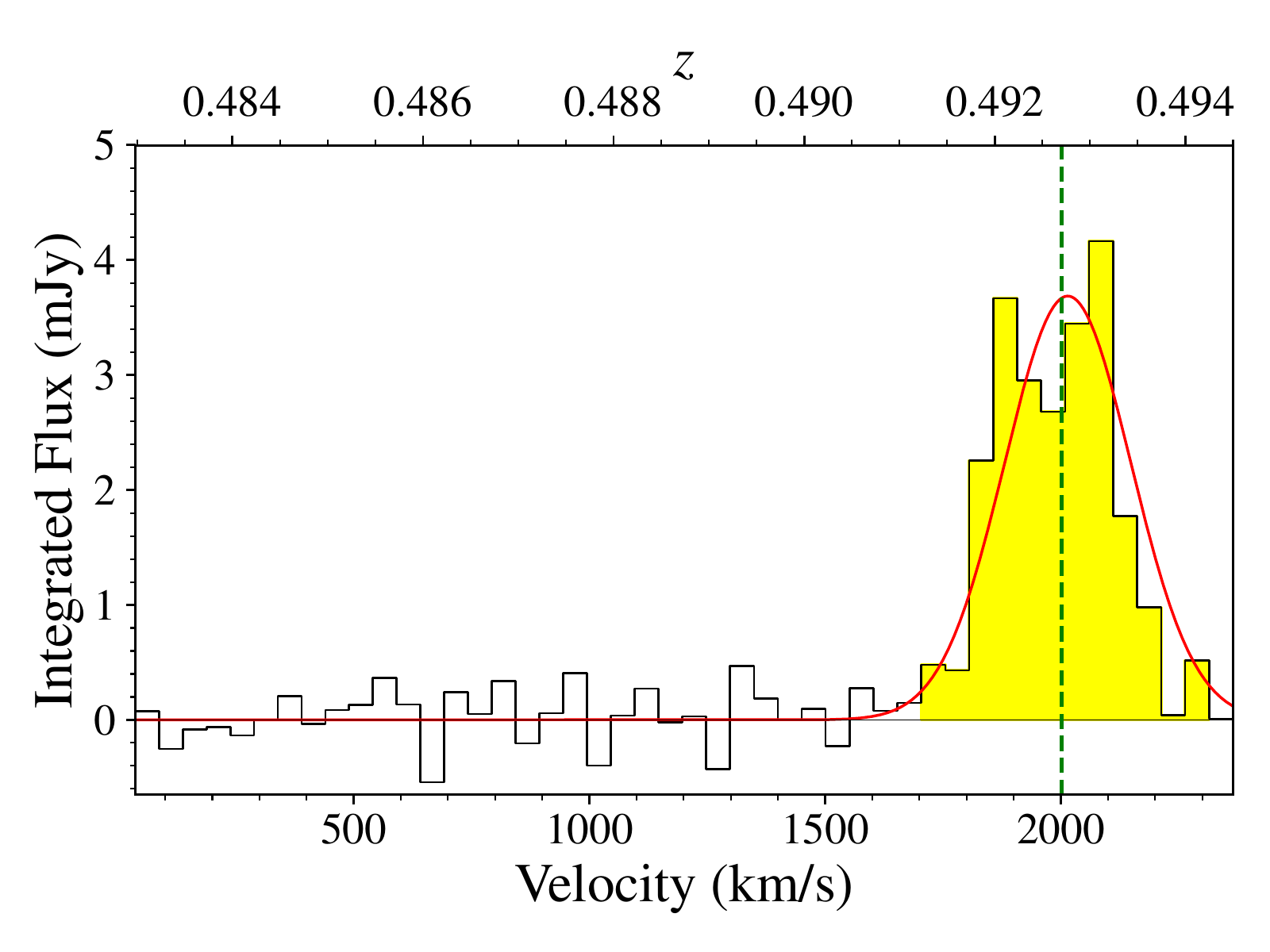}

	\includegraphics[scale=0.3]{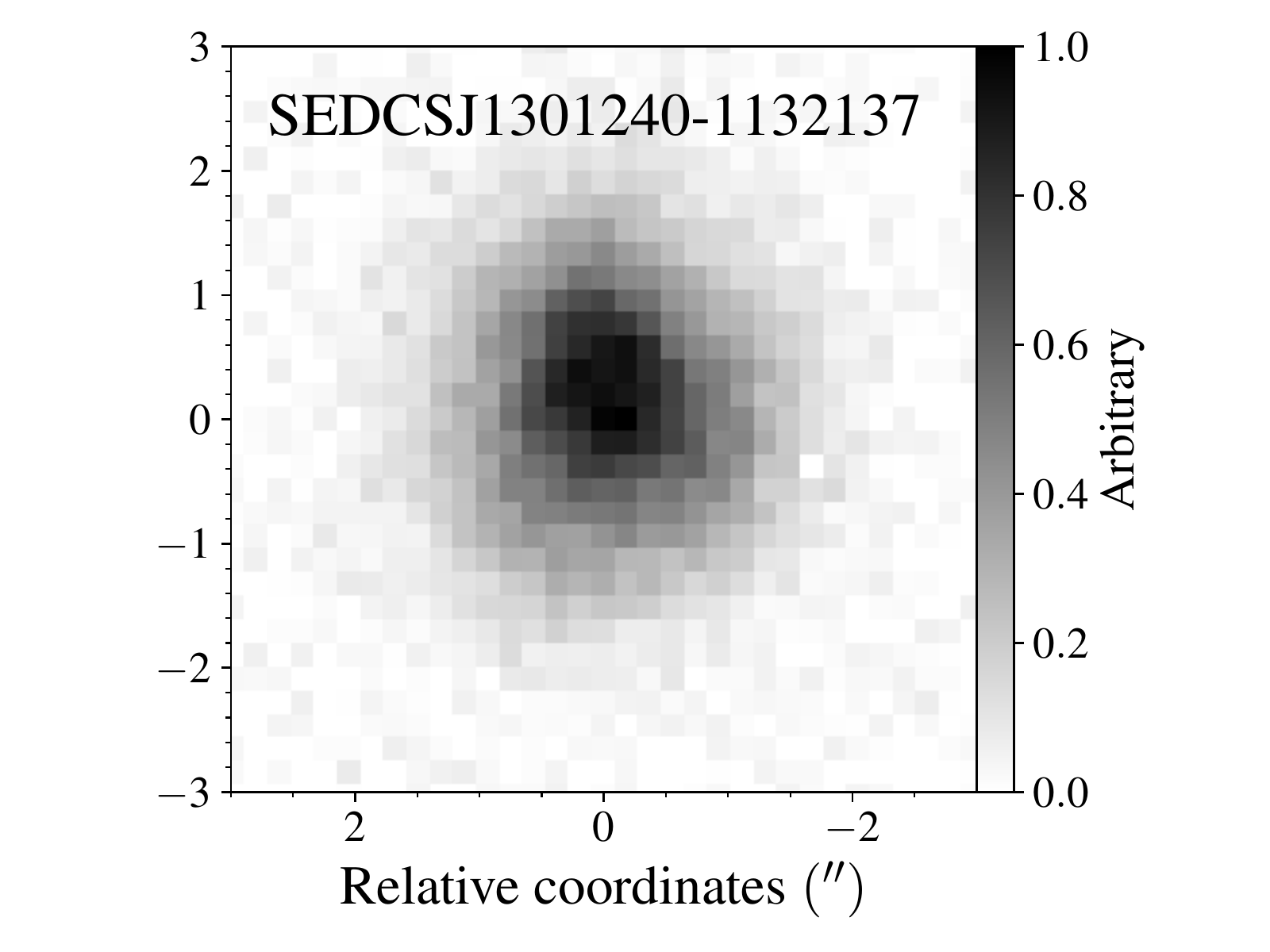}
	\includegraphics[scale=0.3]{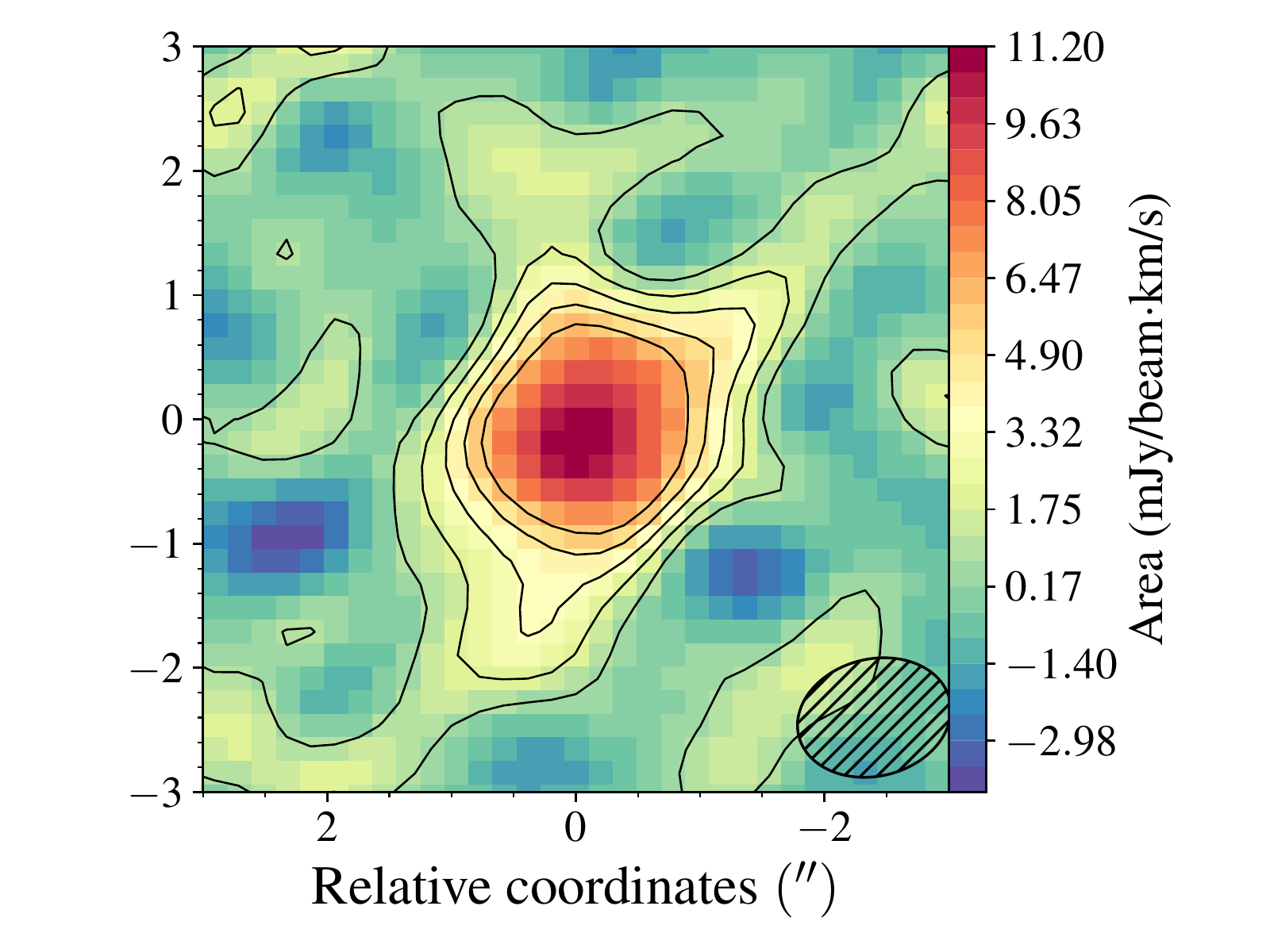}
	\includegraphics[scale=0.3]{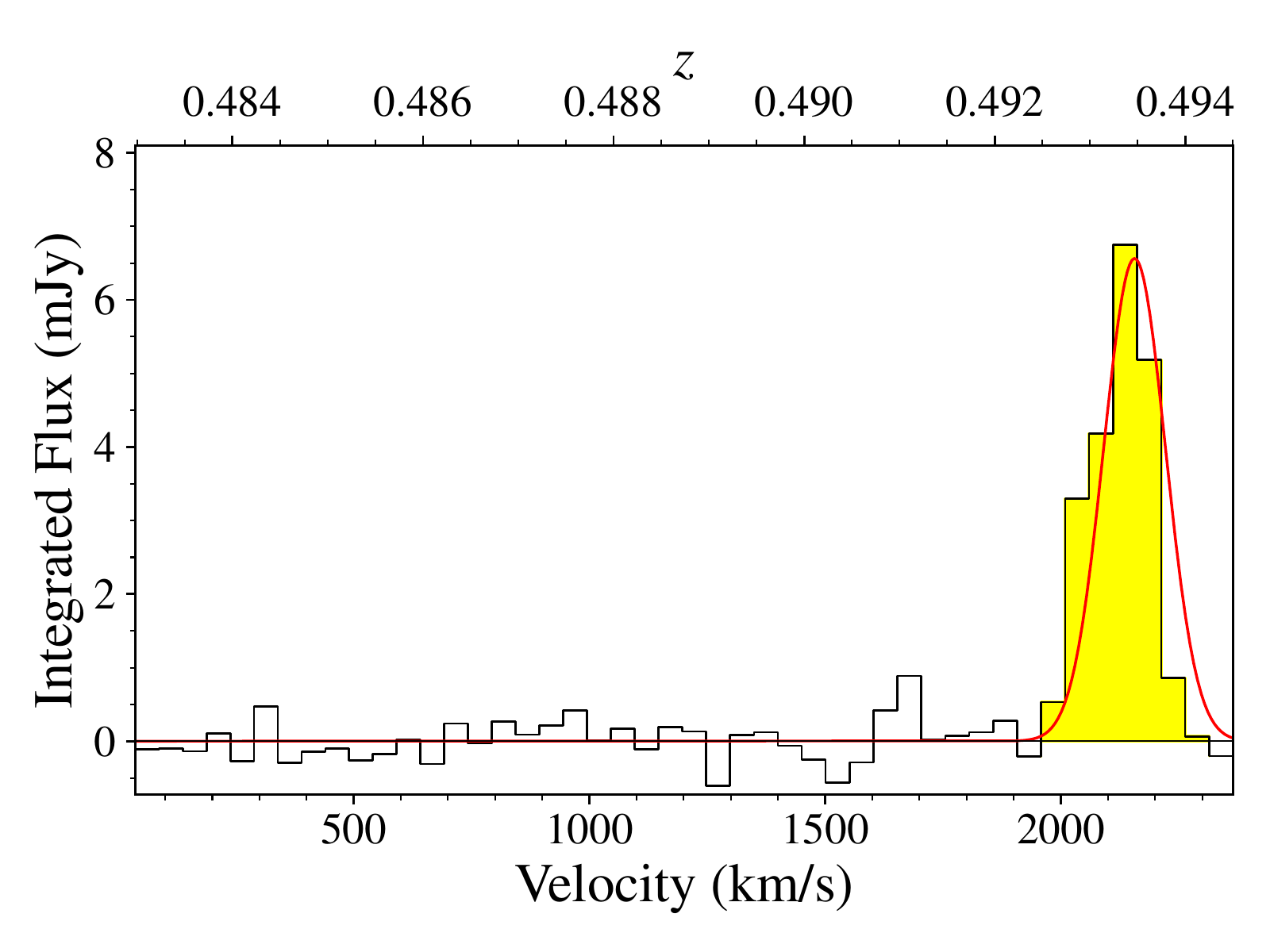}

	\includegraphics[scale=0.3]{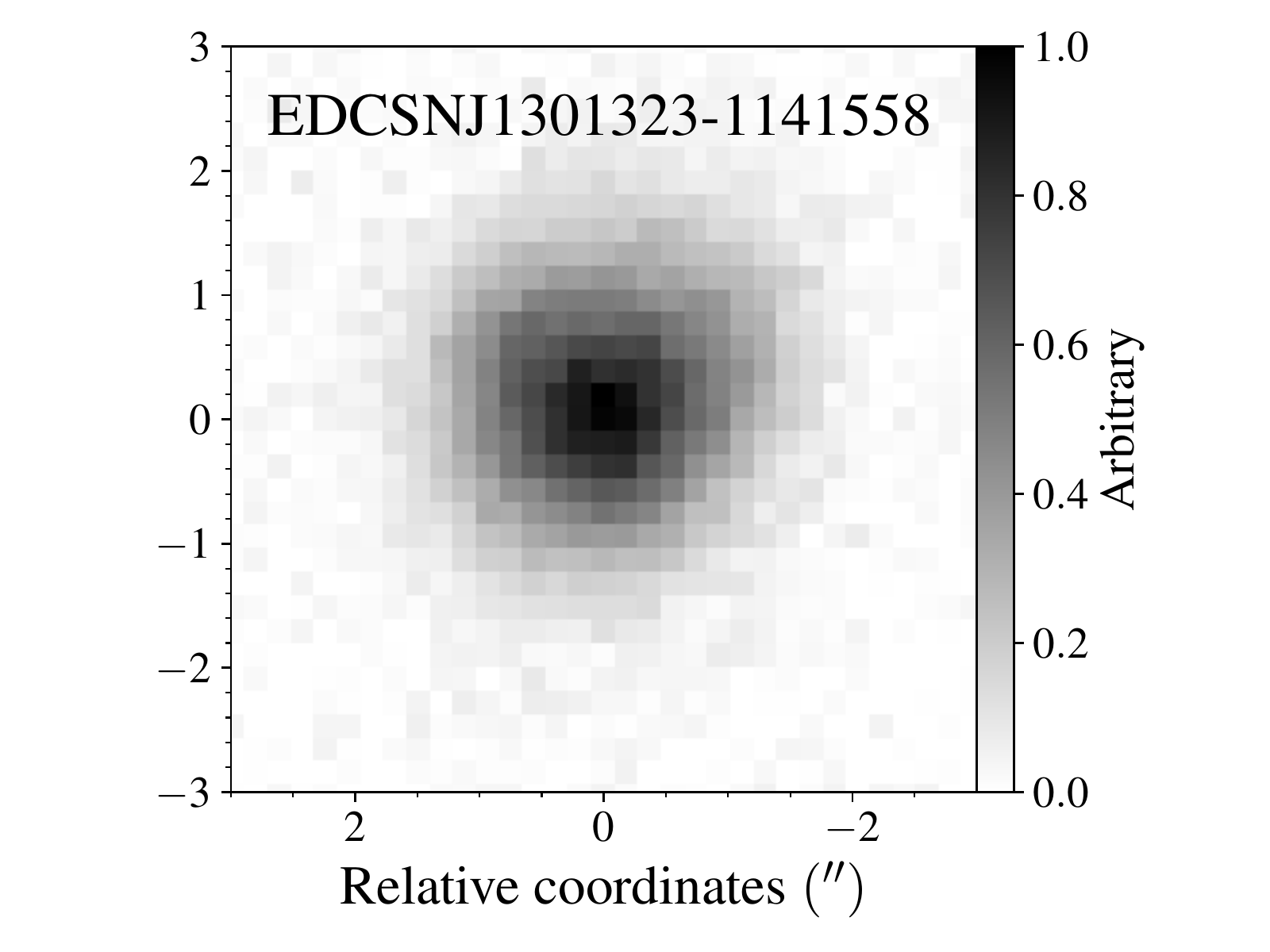}
	\includegraphics[scale=0.3]{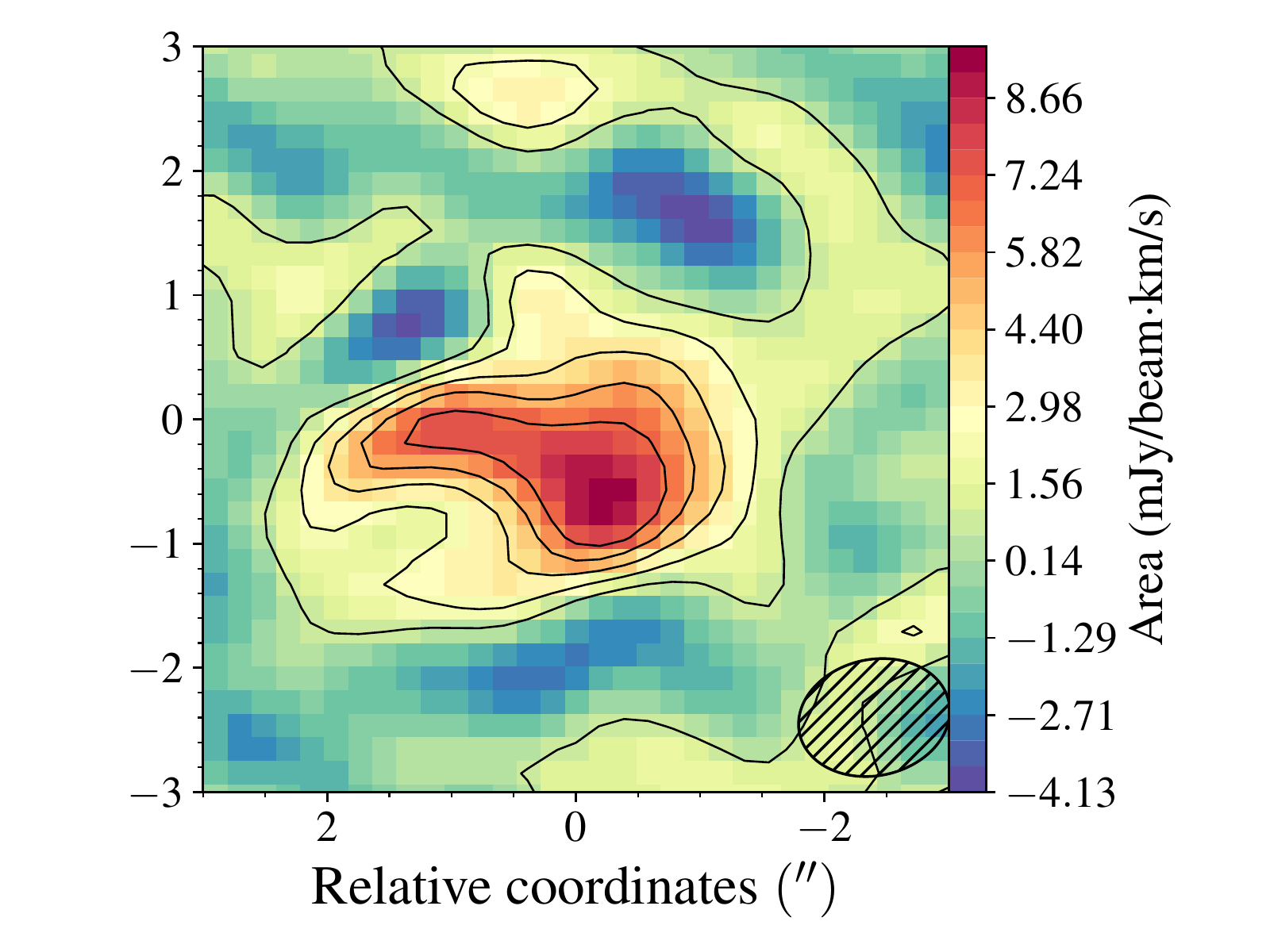}
	\includegraphics[scale=0.3]{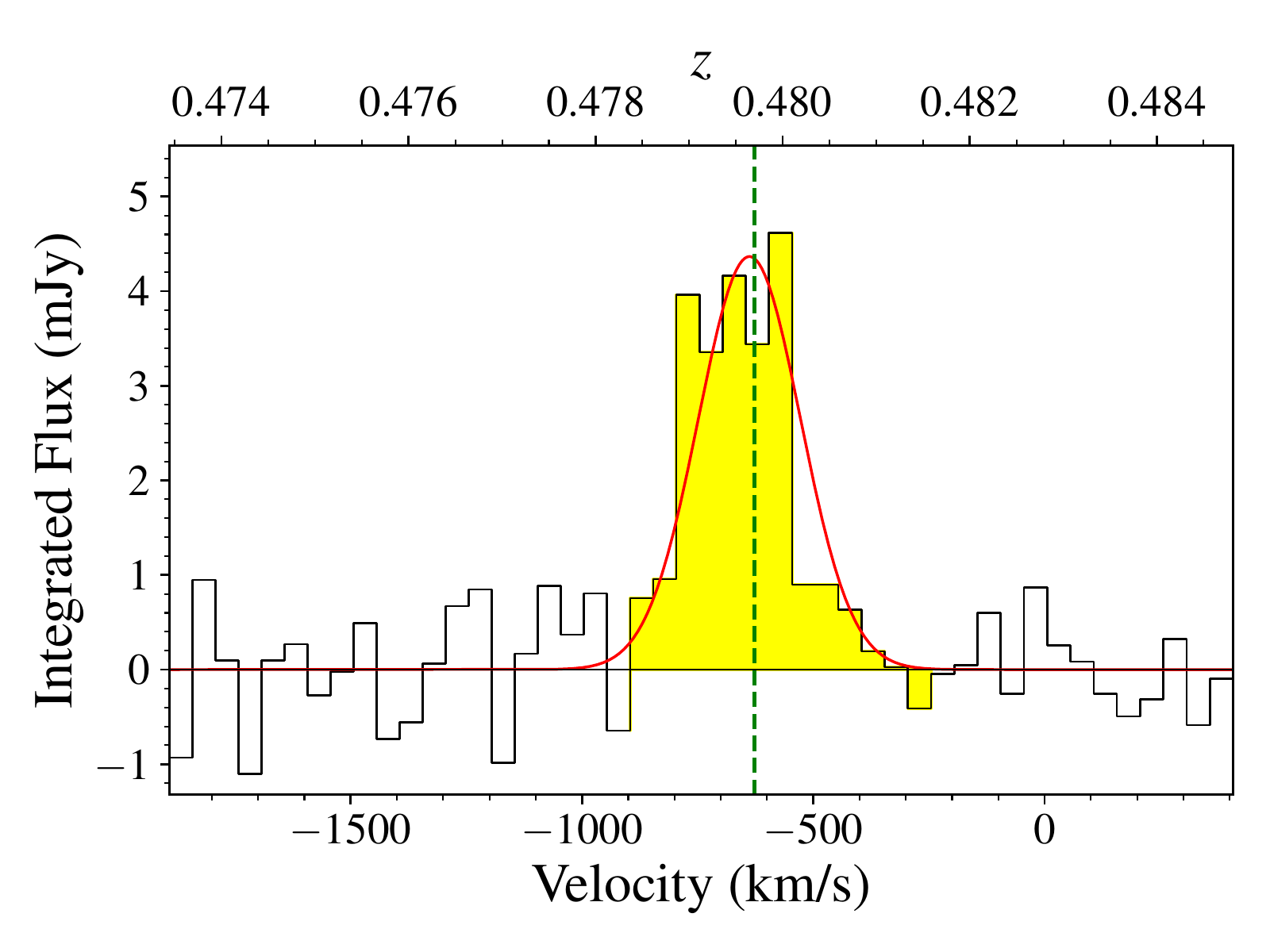}

	\includegraphics[scale=0.3]{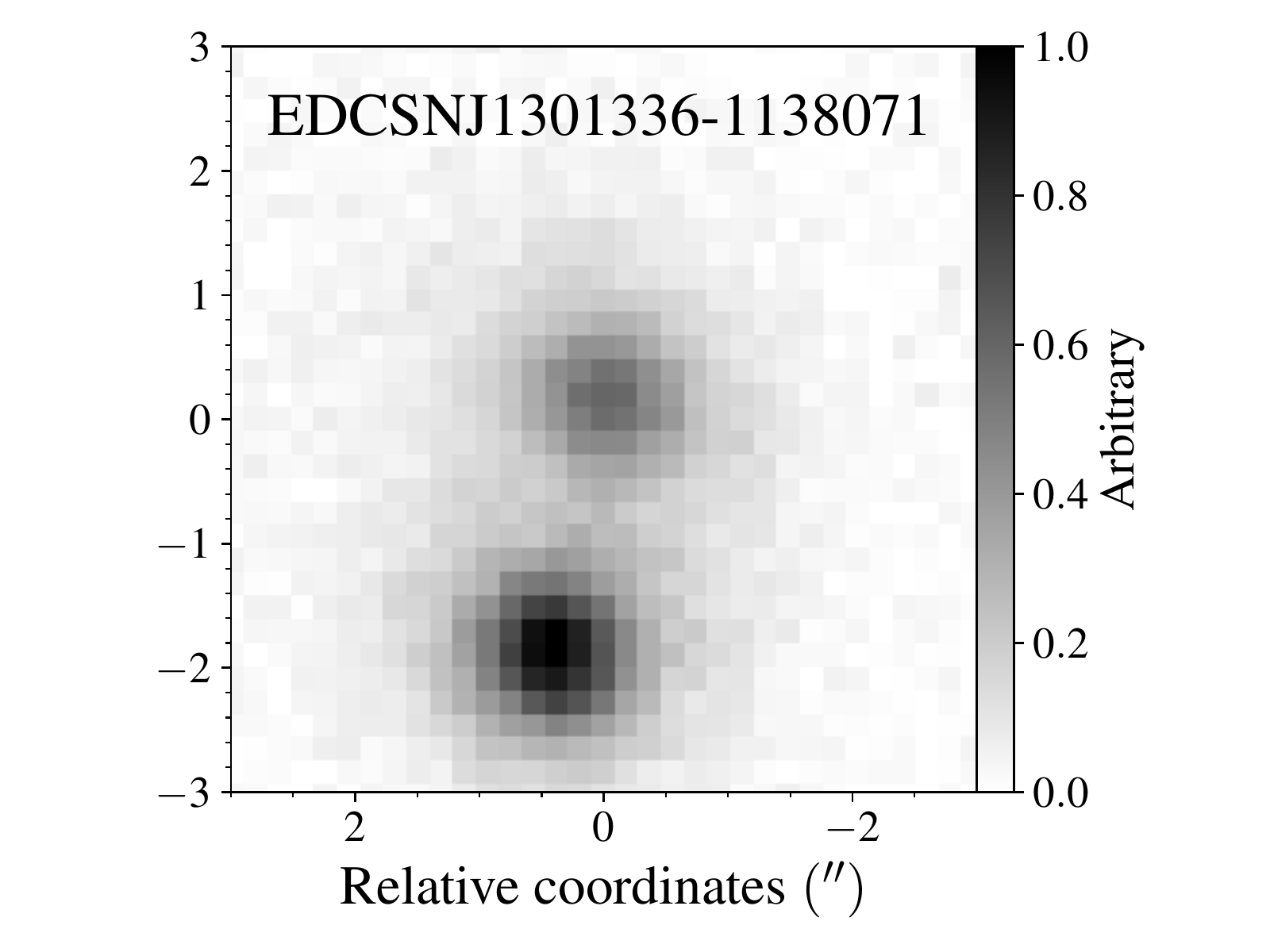}
	\includegraphics[scale=0.3]{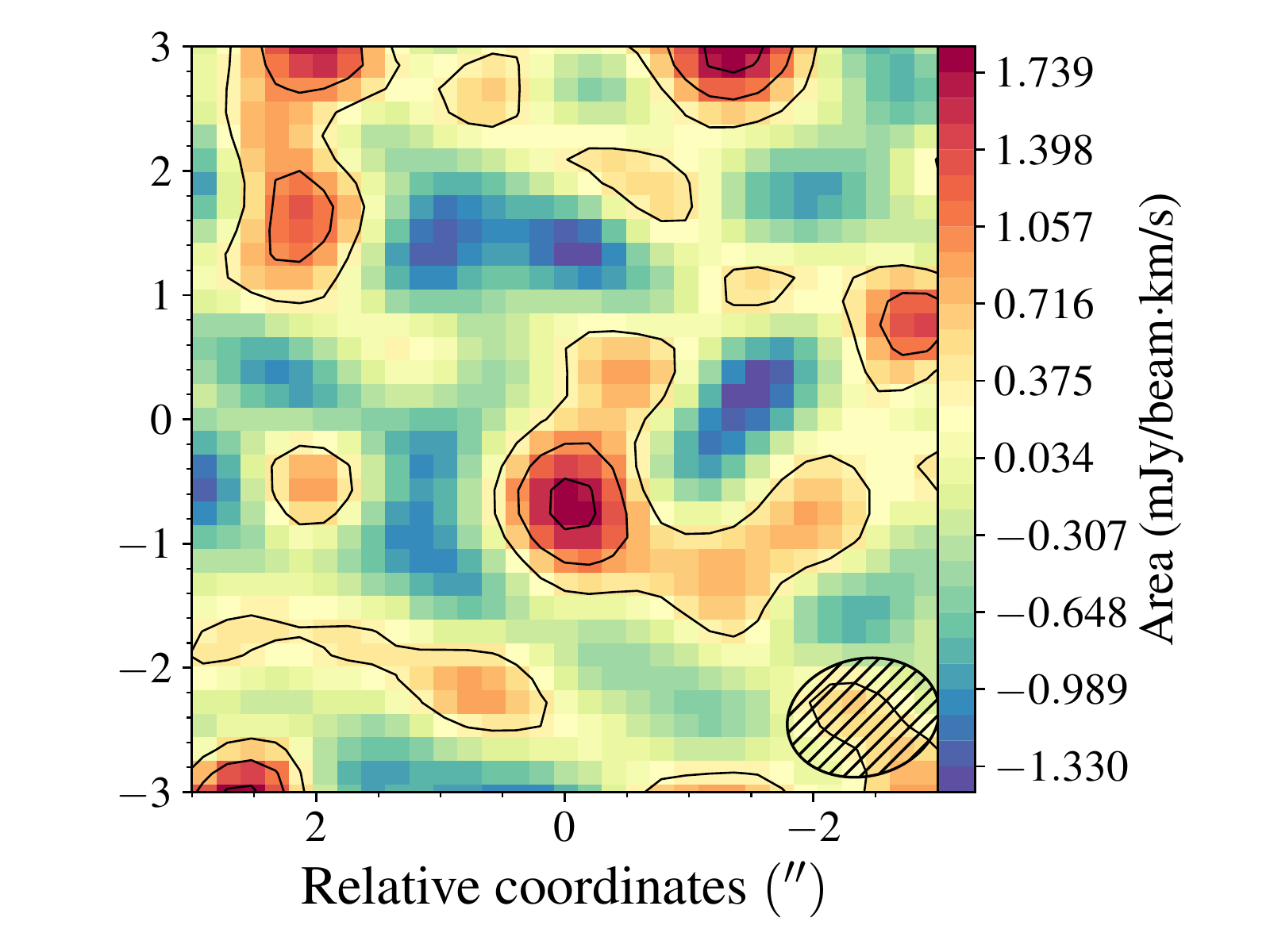}
	\includegraphics[scale=0.3]{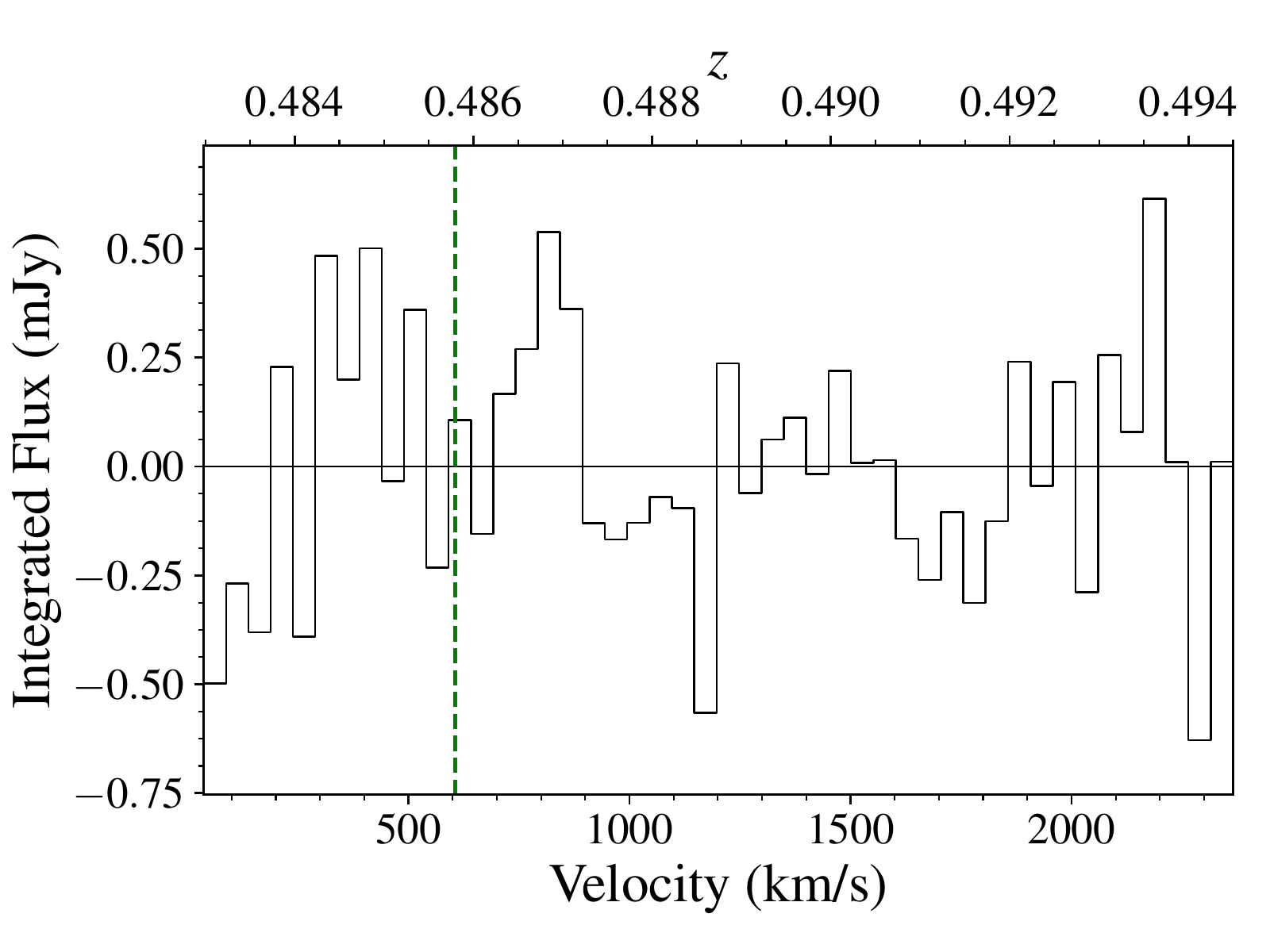}
	
    \caption{Continued.}
\end{figure*}

\begin{figure*}[htbp]\ContinuedFloat
\centering
	\includegraphics[scale=0.3]{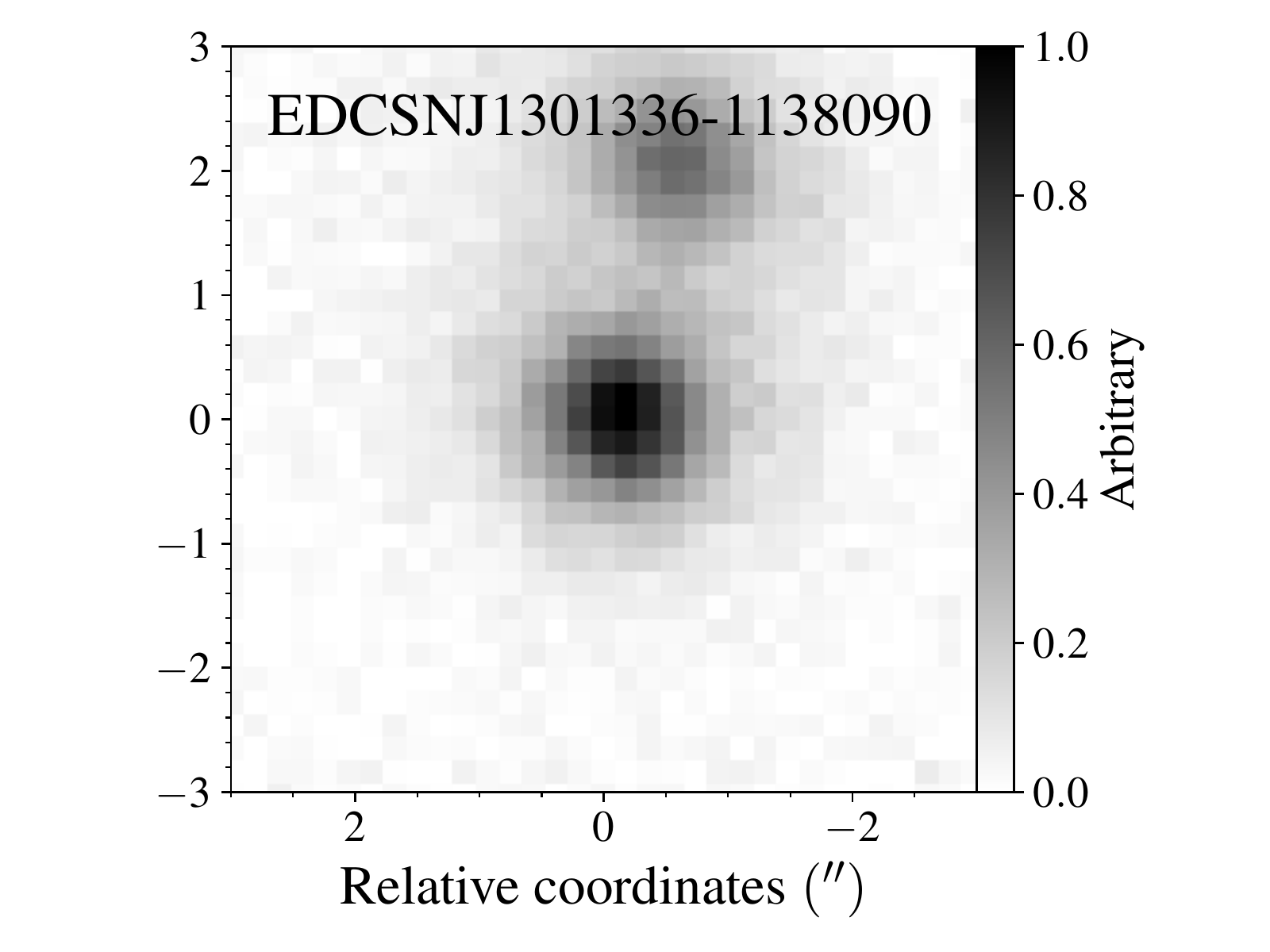}
	\includegraphics[scale=0.3]{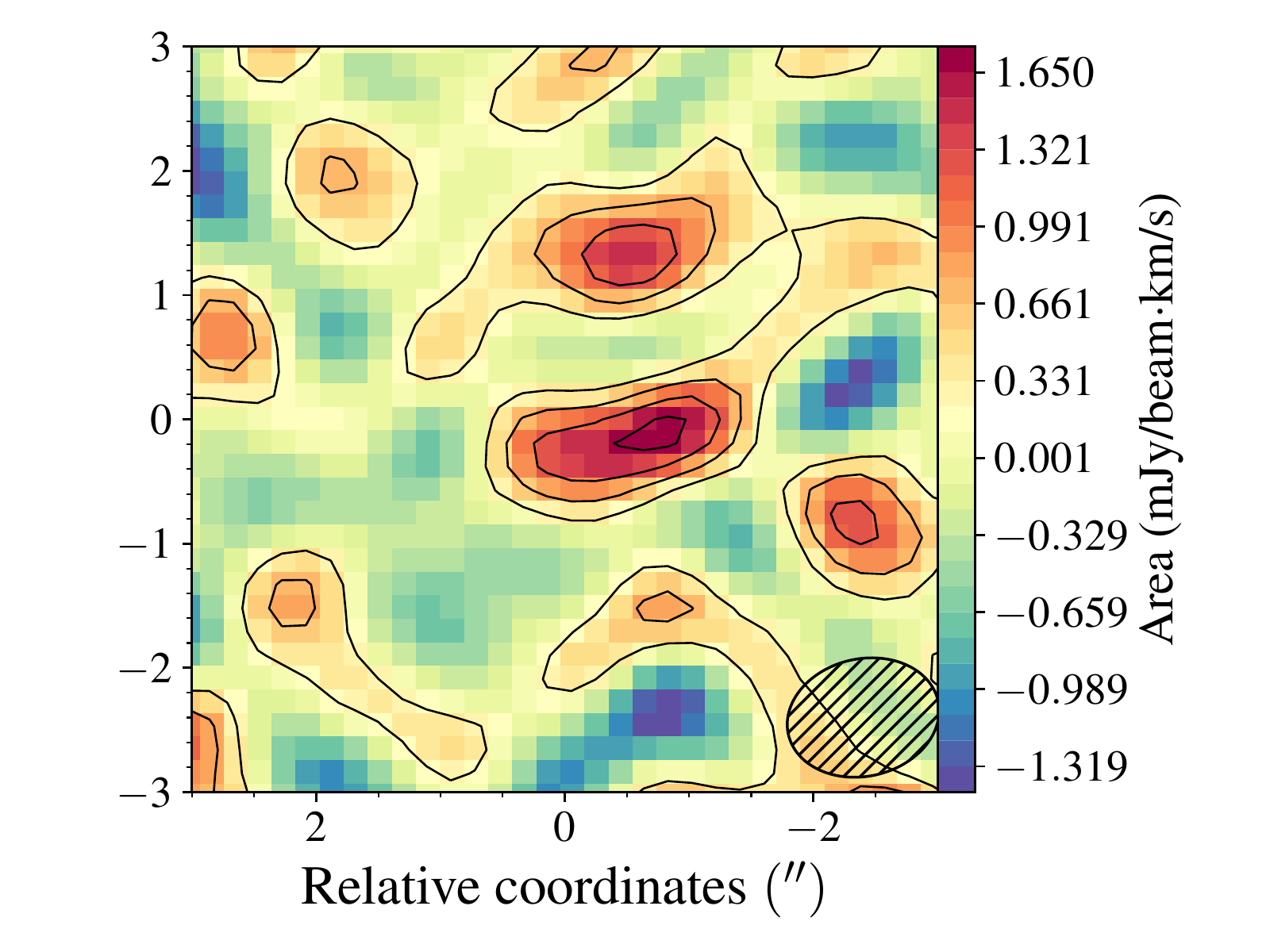}
	\includegraphics[scale=0.3]{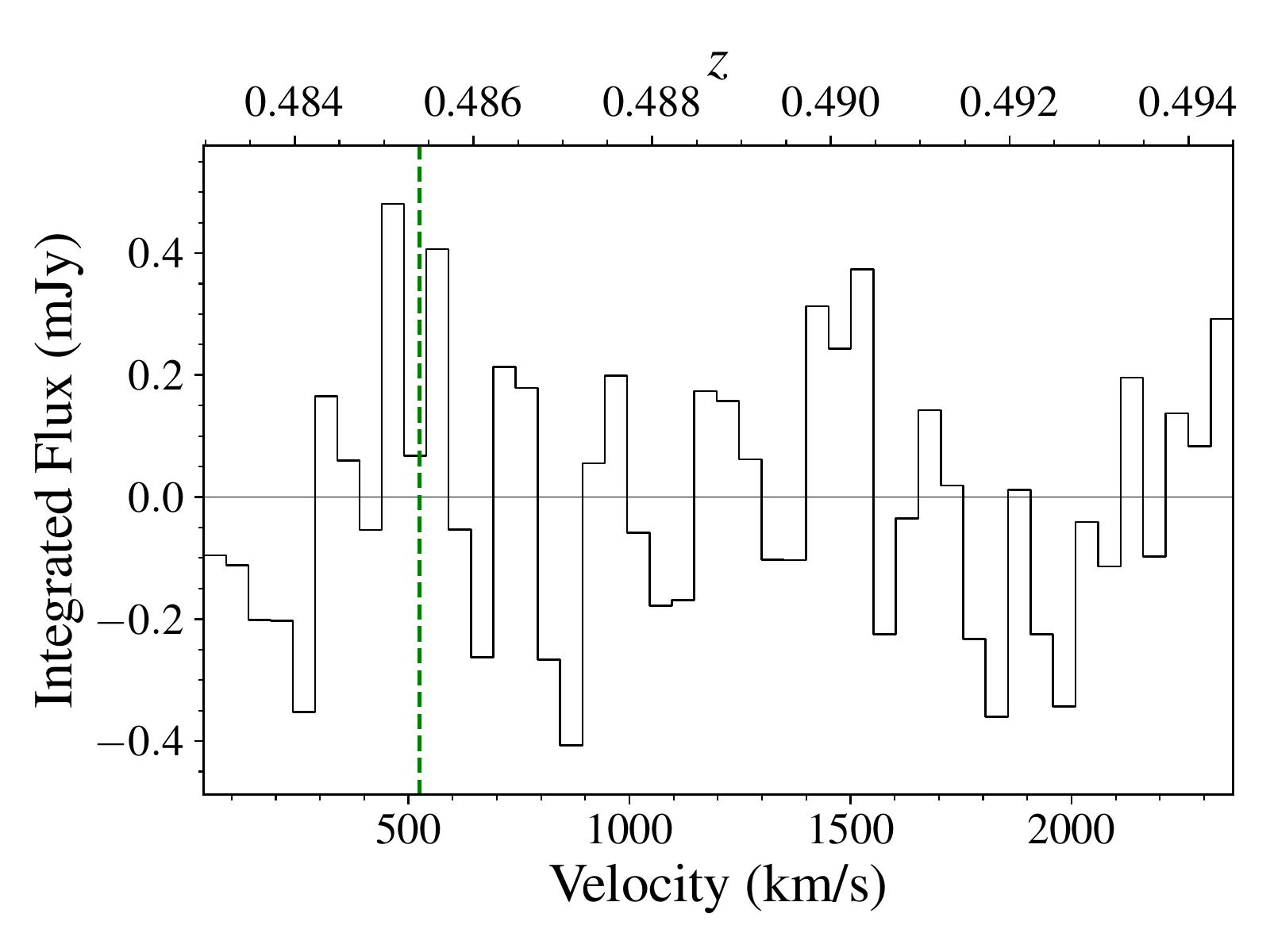}

	\includegraphics[scale=0.3]{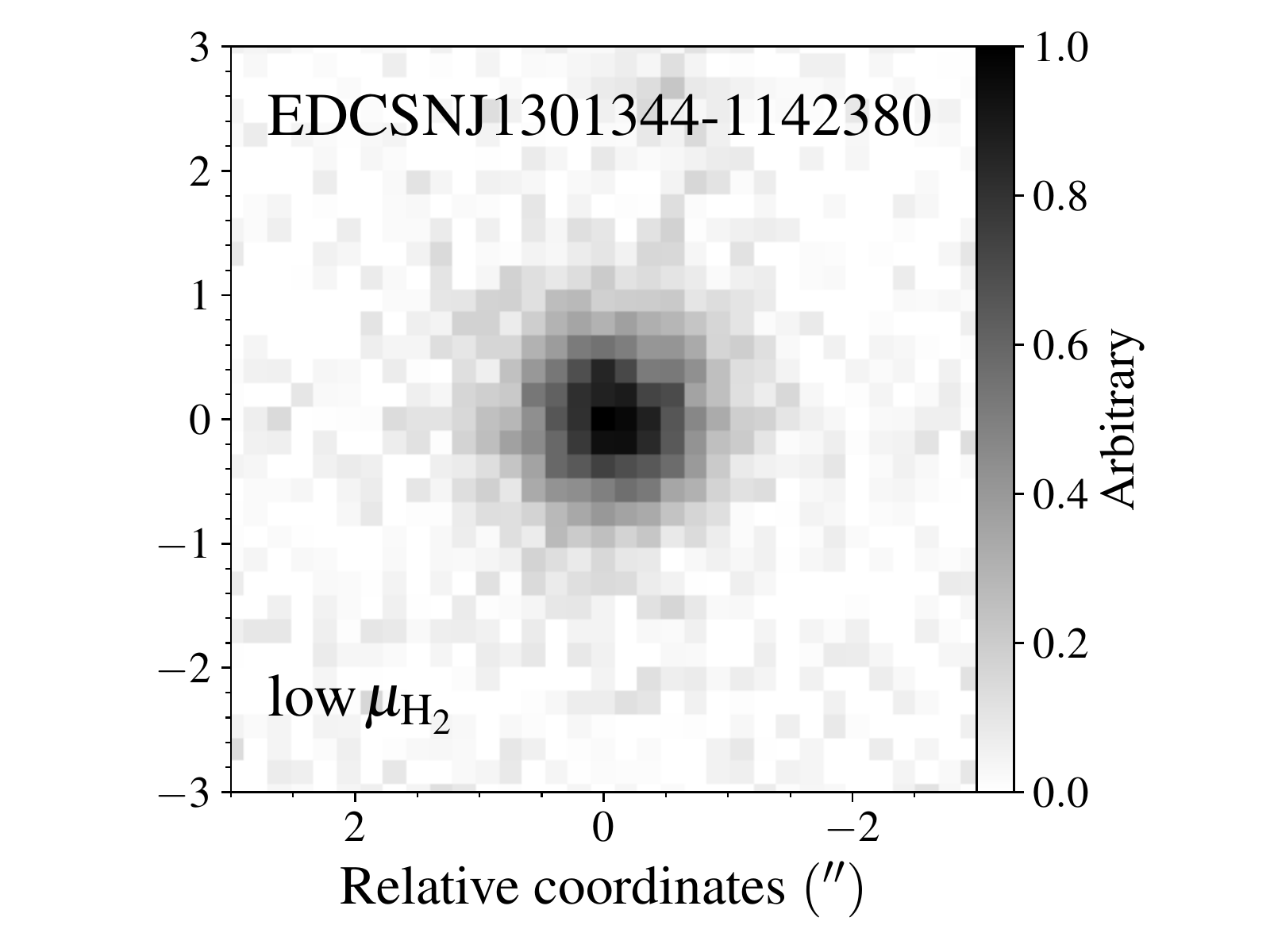}
	\includegraphics[scale=0.3]{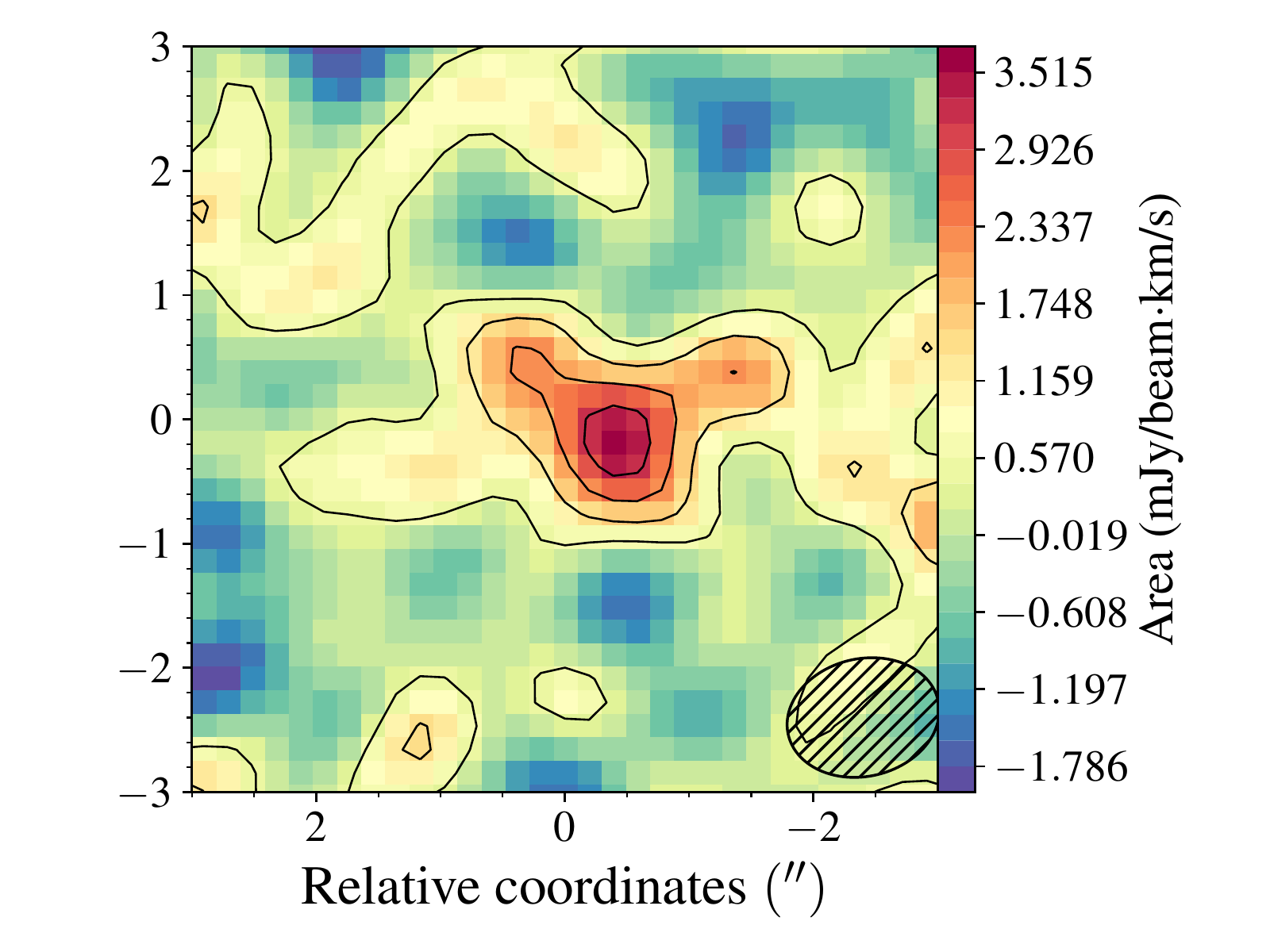}
	\includegraphics[scale=0.3]{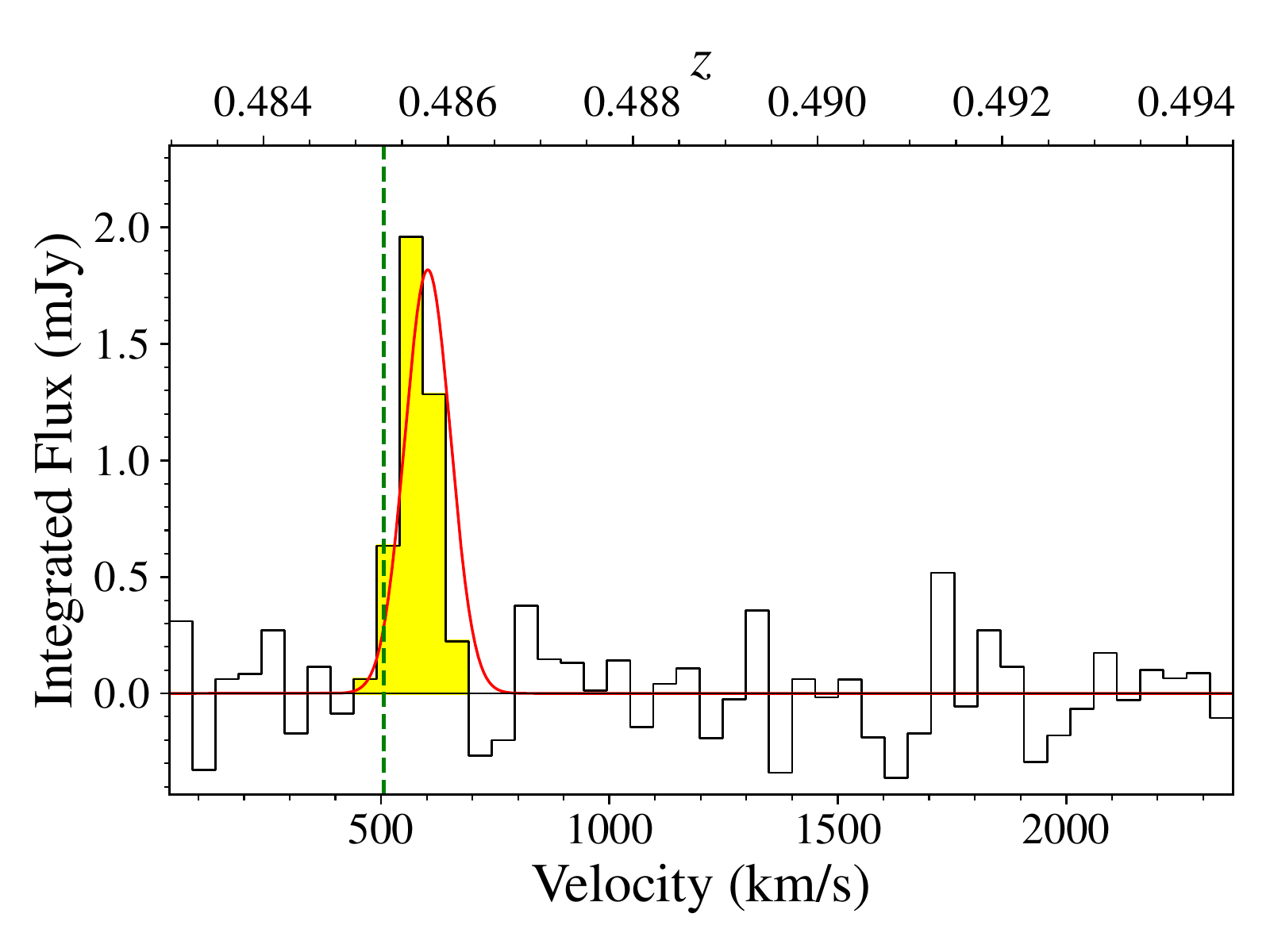}

	\includegraphics[scale=0.3]{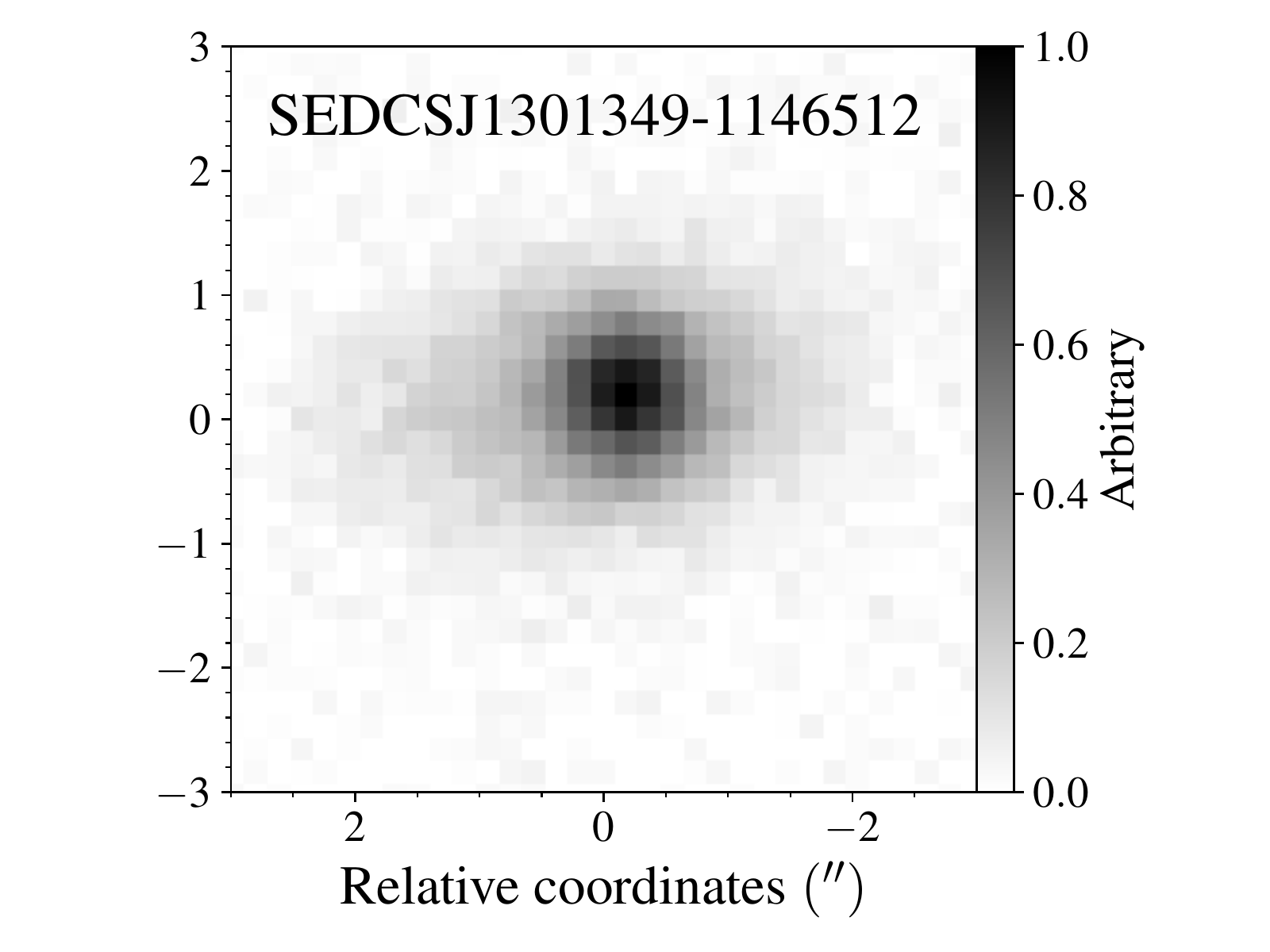}
	\includegraphics[scale=0.3]{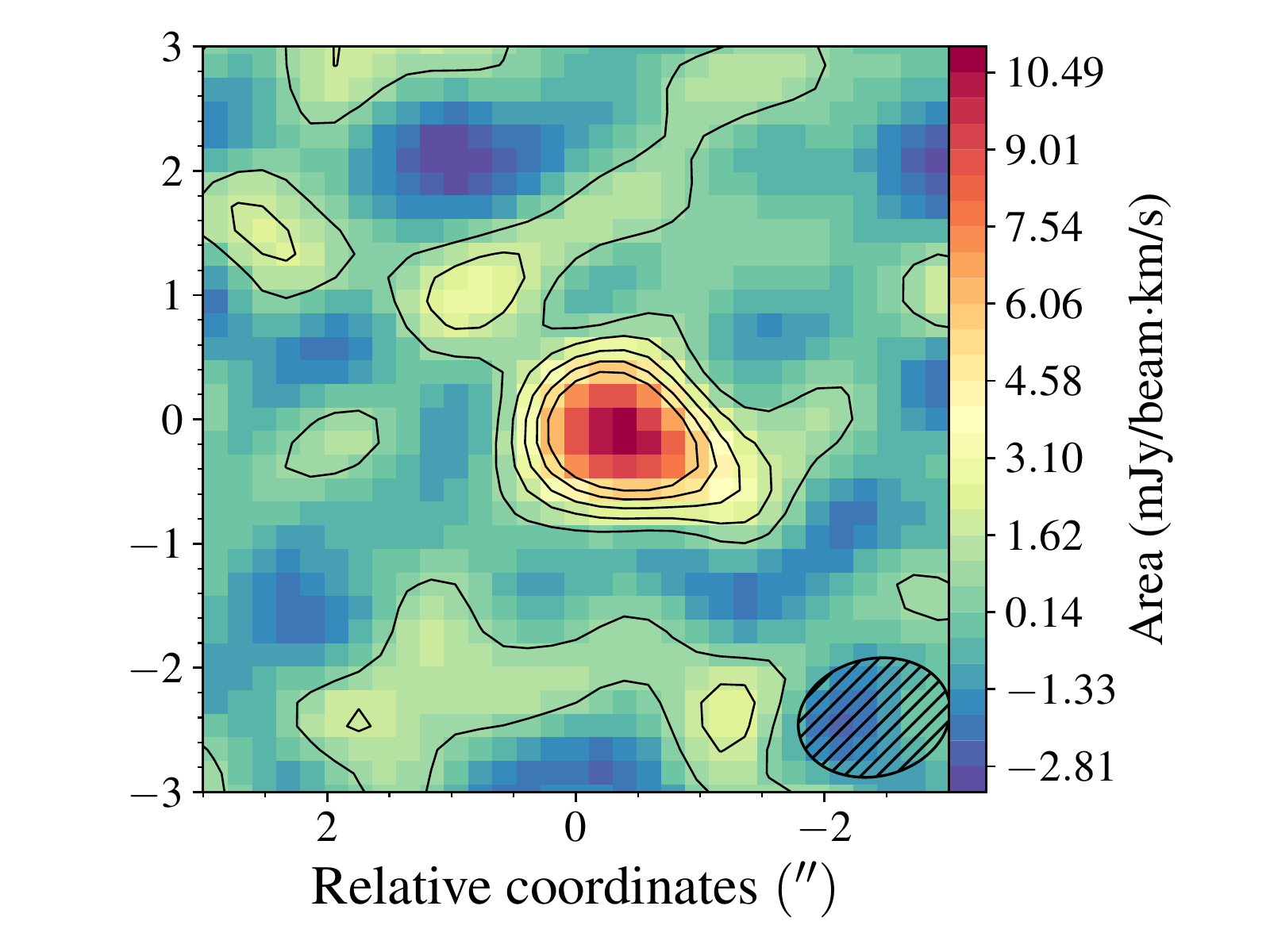}
	\includegraphics[scale=0.3]{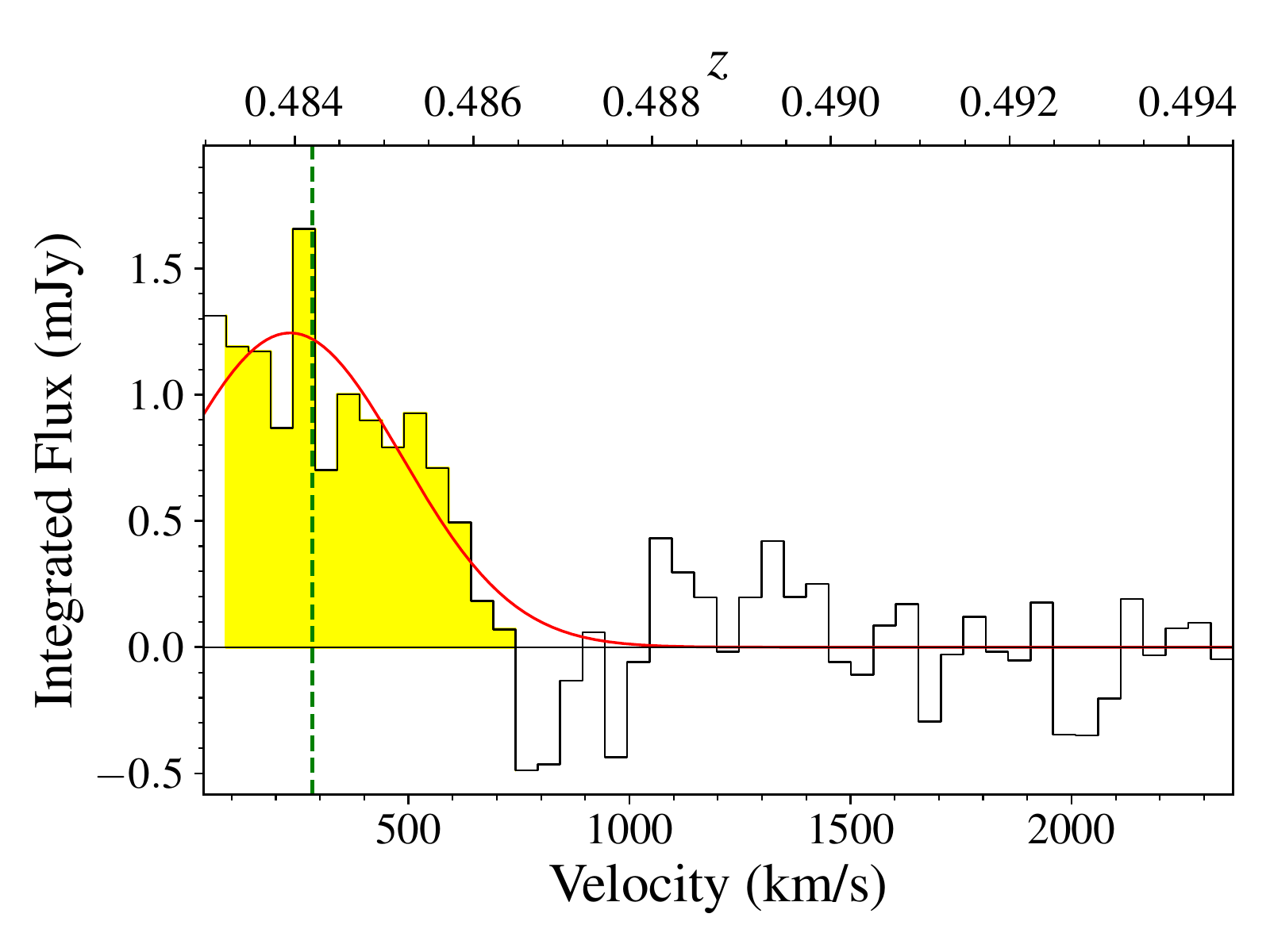}

	\includegraphics[scale=0.3]{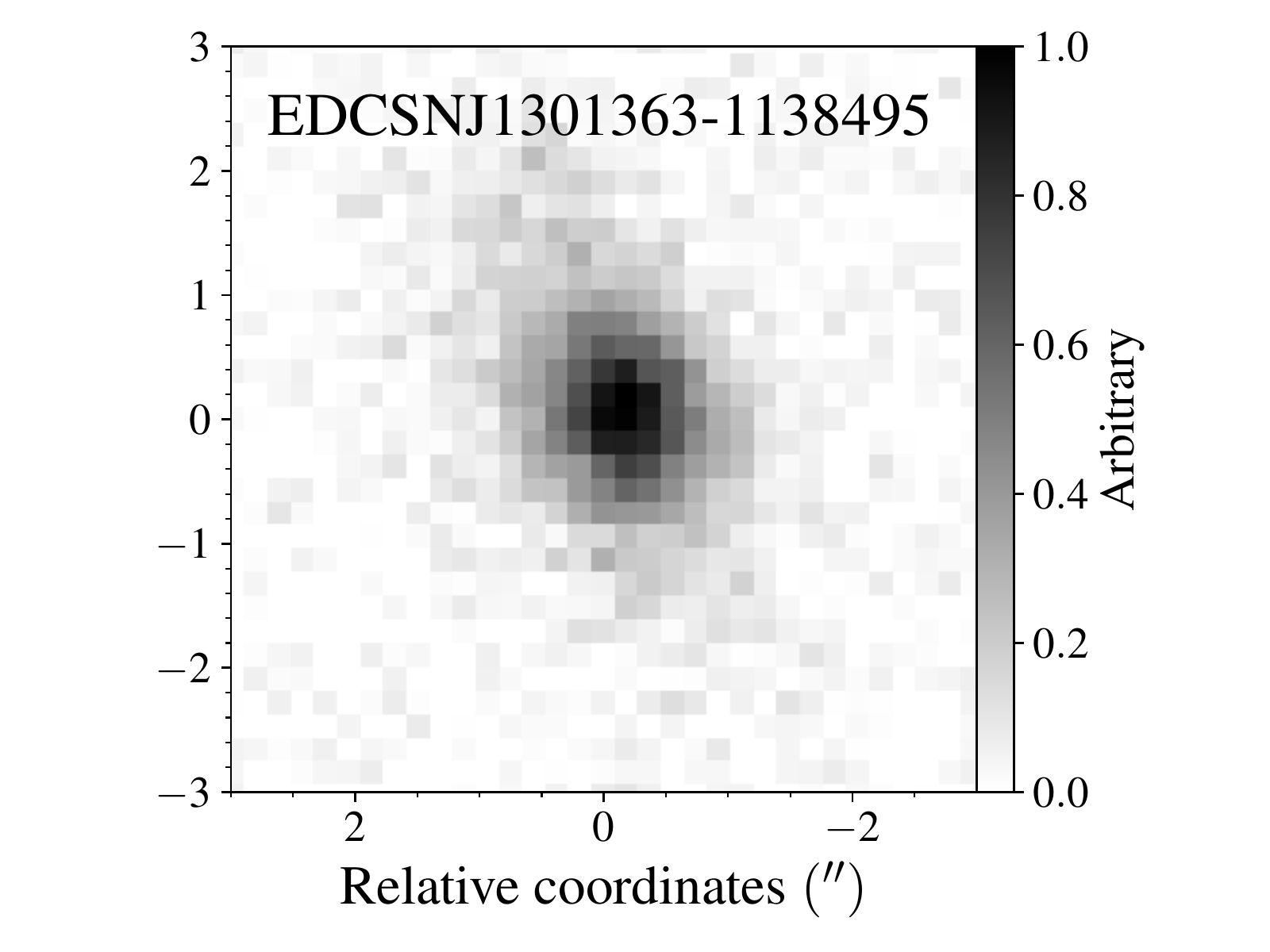}
	\includegraphics[scale=0.3]{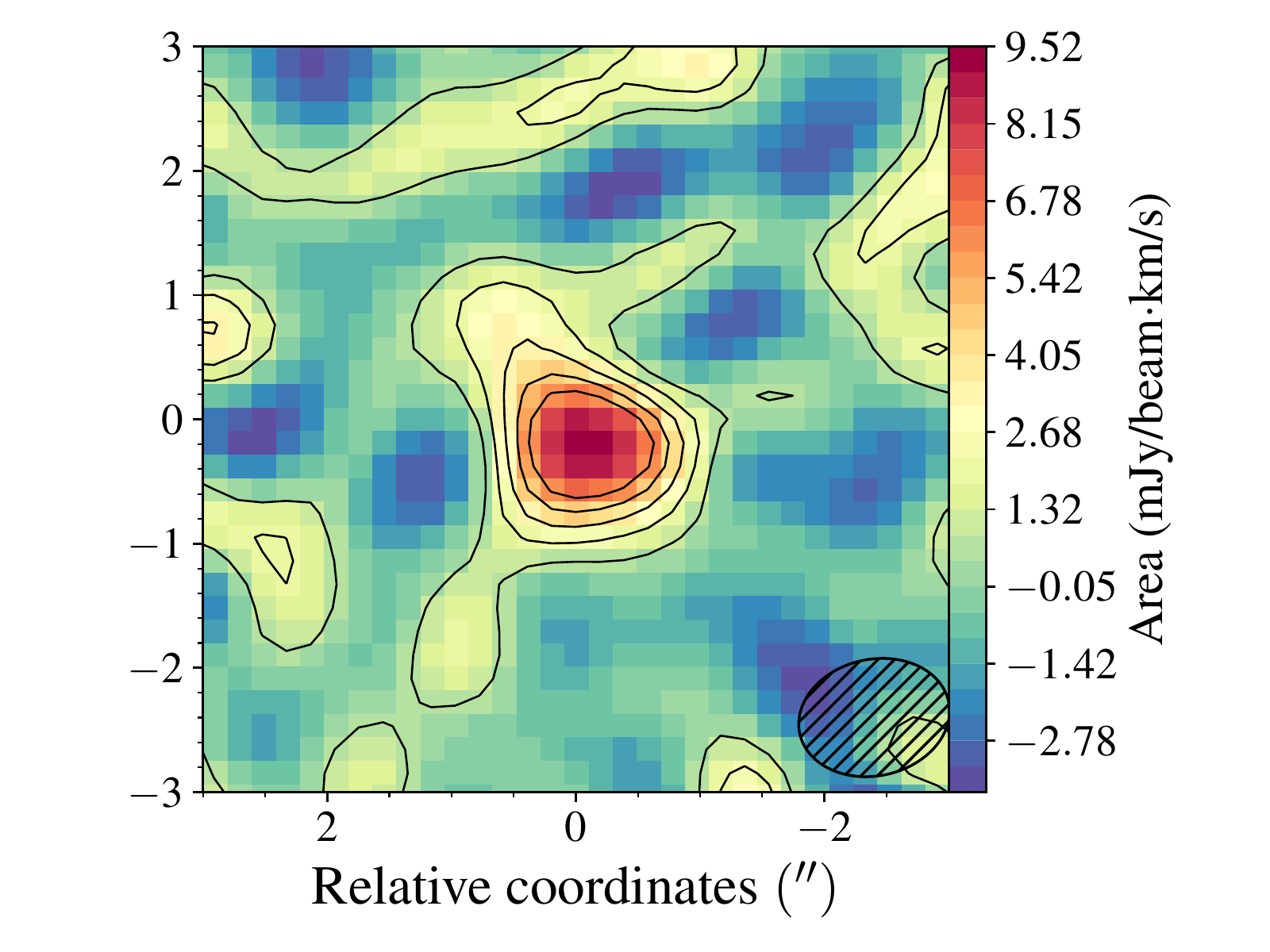}
	\includegraphics[scale=0.3]{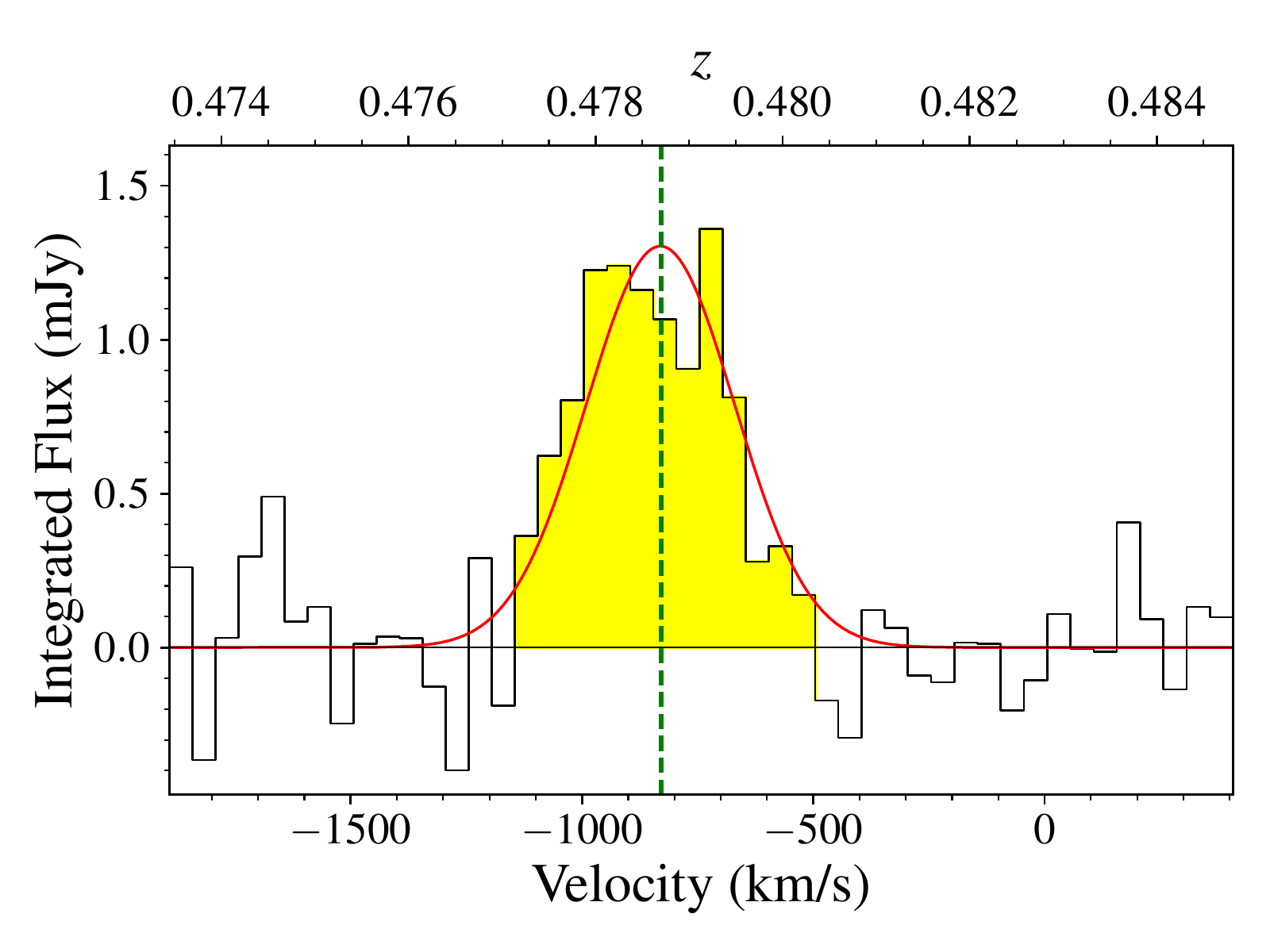}

	\includegraphics[scale=0.3]{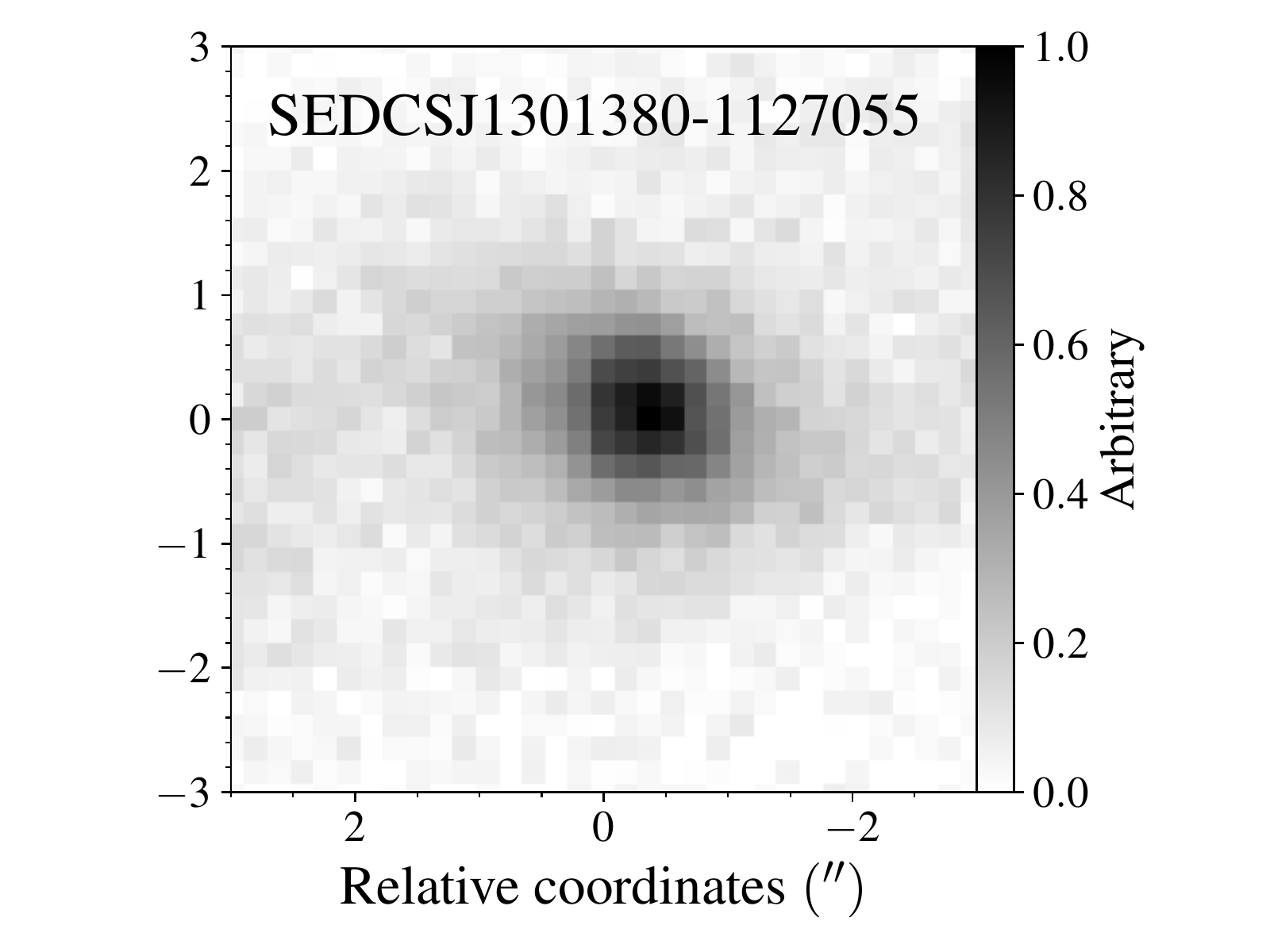}
	\includegraphics[scale=0.3]{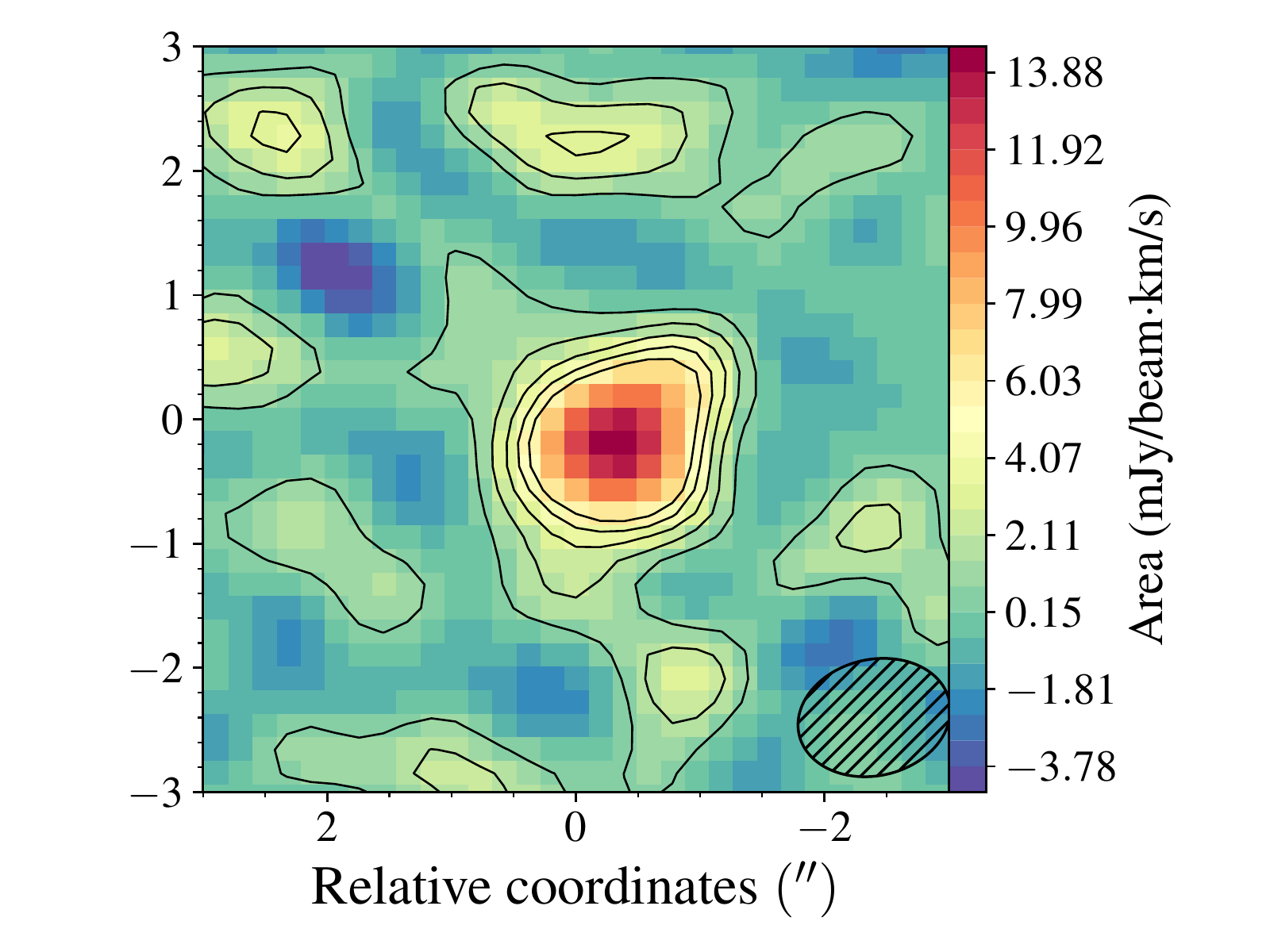}
	\includegraphics[scale=0.3]{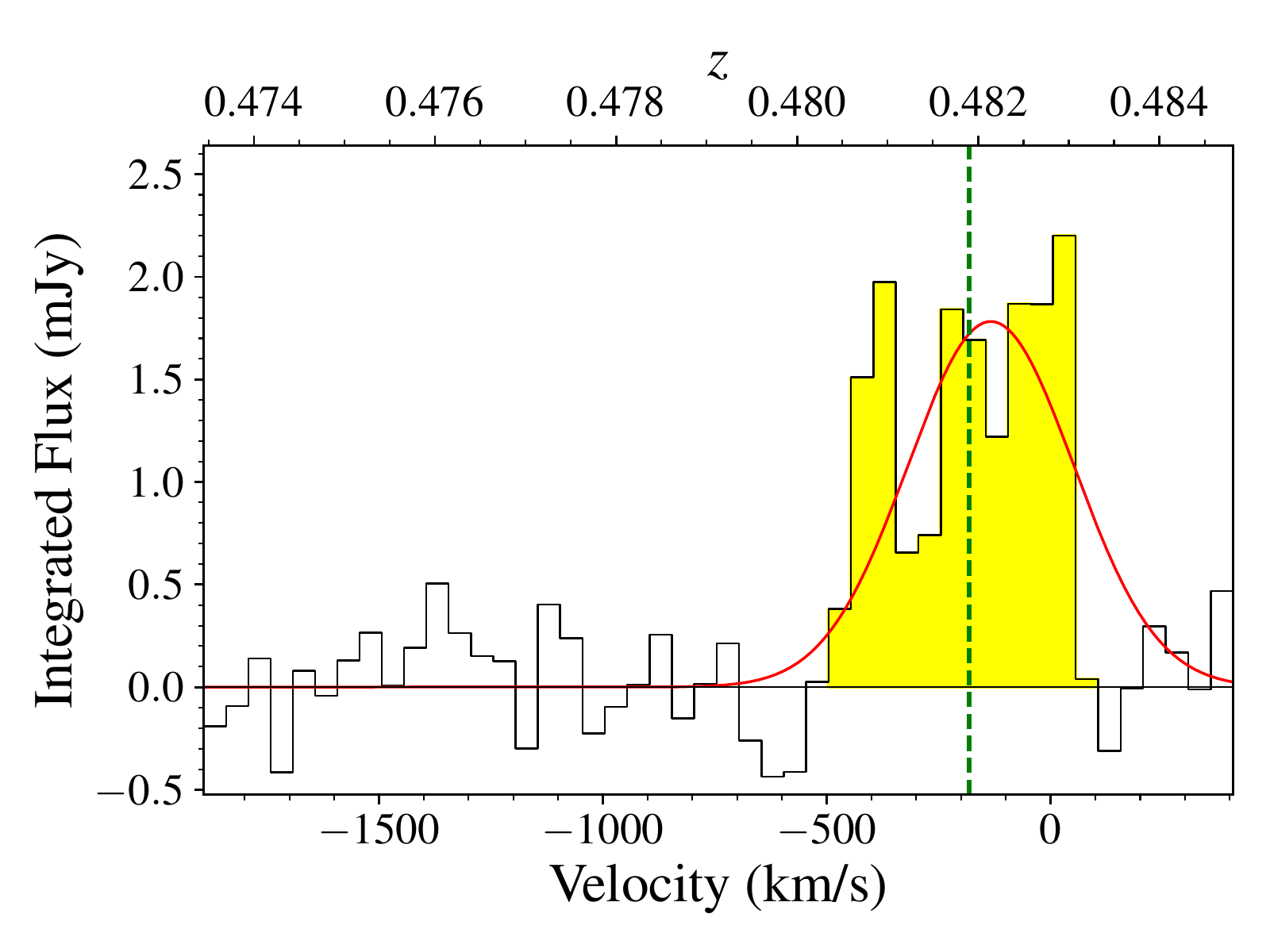}

	\includegraphics[scale=0.3]{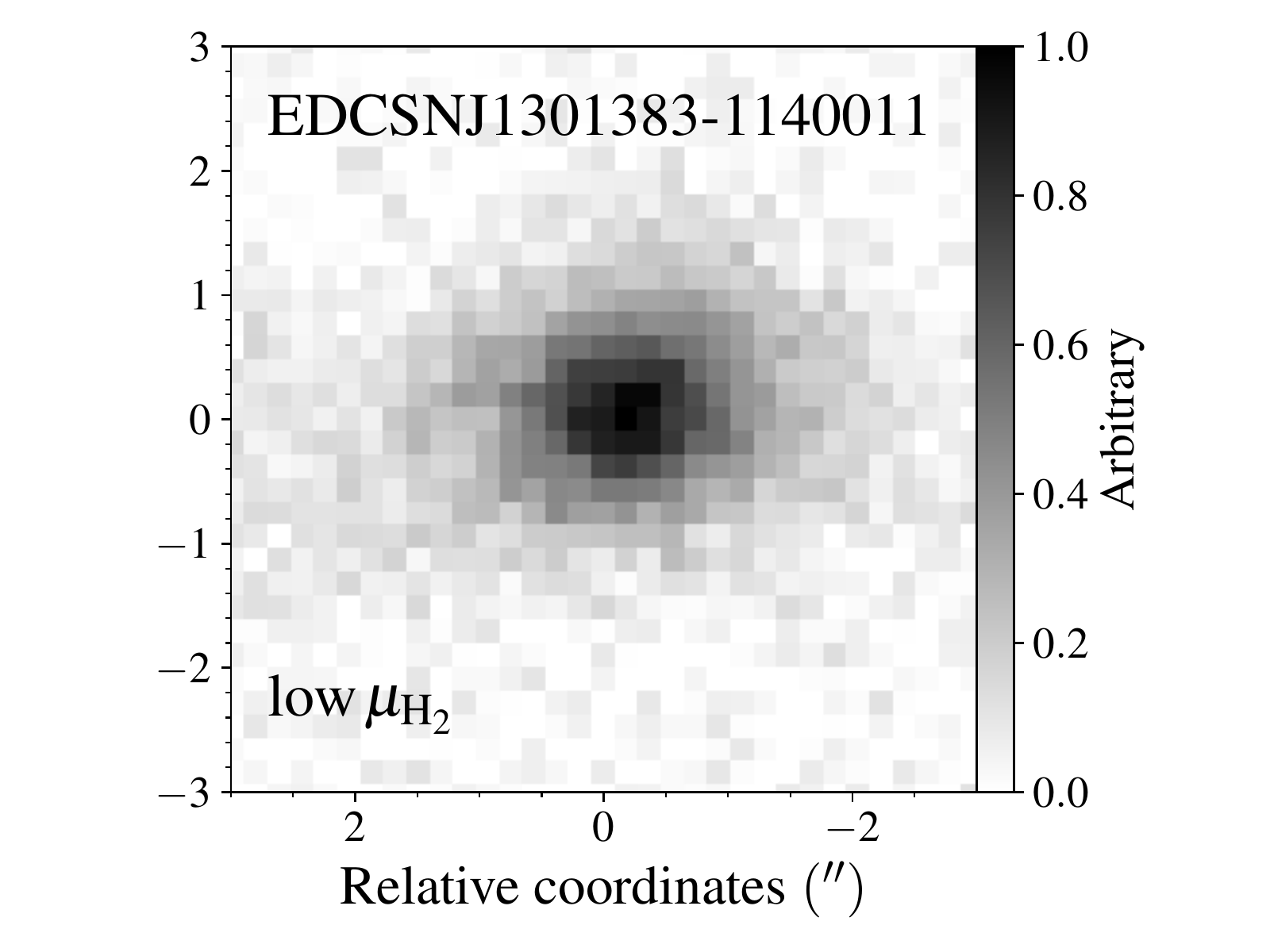}
	\includegraphics[scale=0.3]{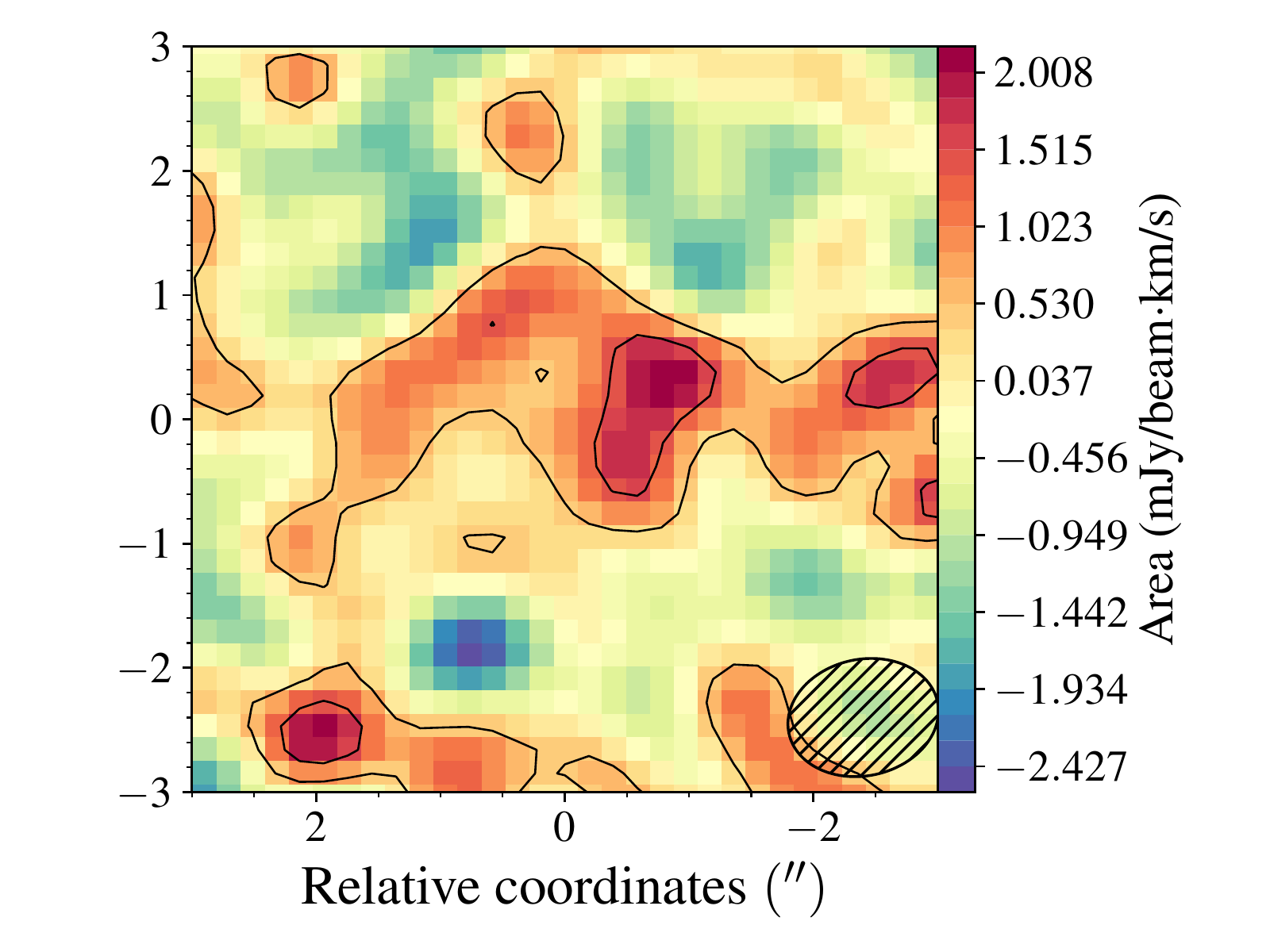}
	\includegraphics[scale=0.3]{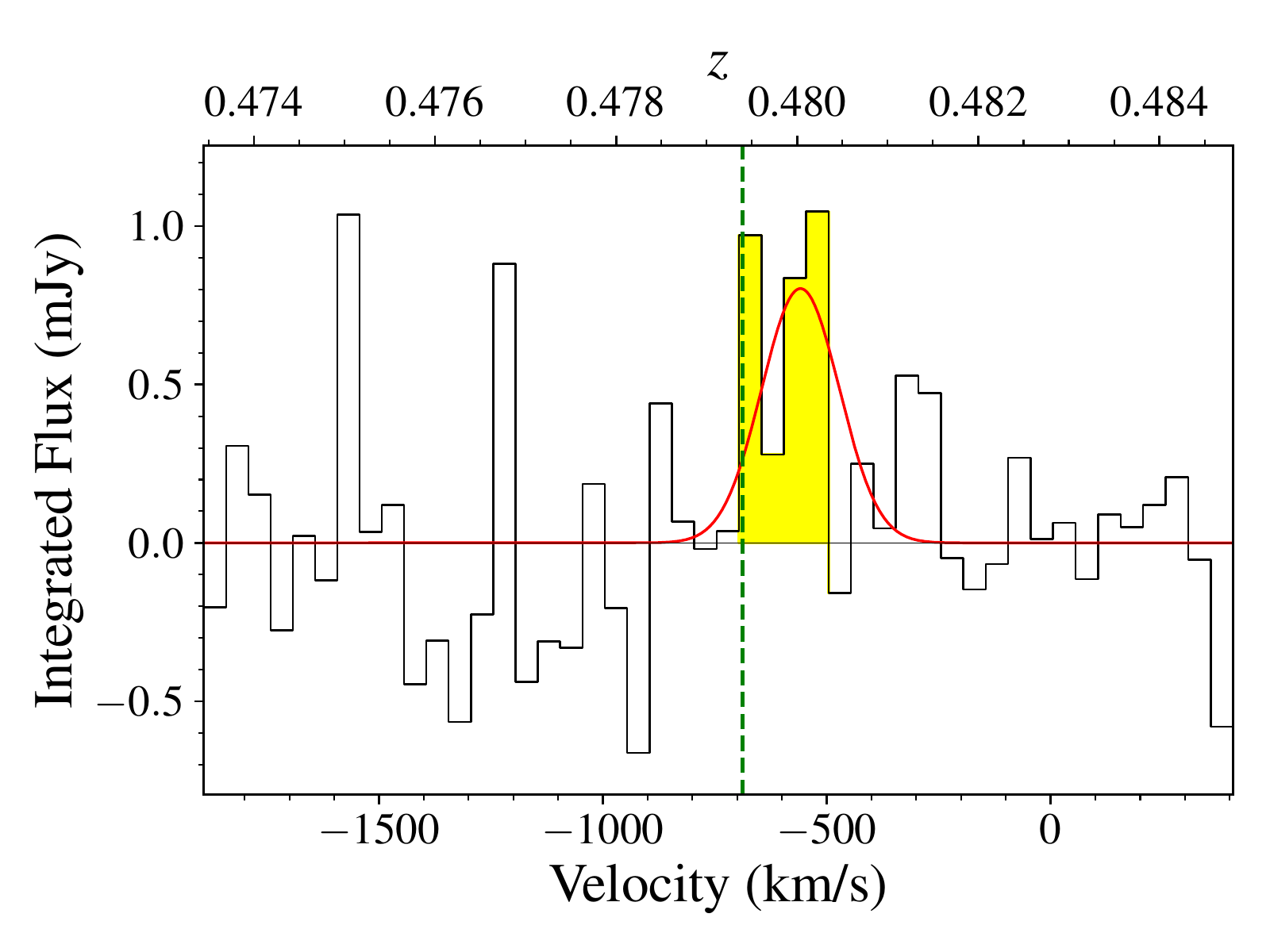}

    \caption{Continued.}
\end{figure*}

\begin{figure*}[htbp]\ContinuedFloat
\centering

	\includegraphics[scale=0.3]{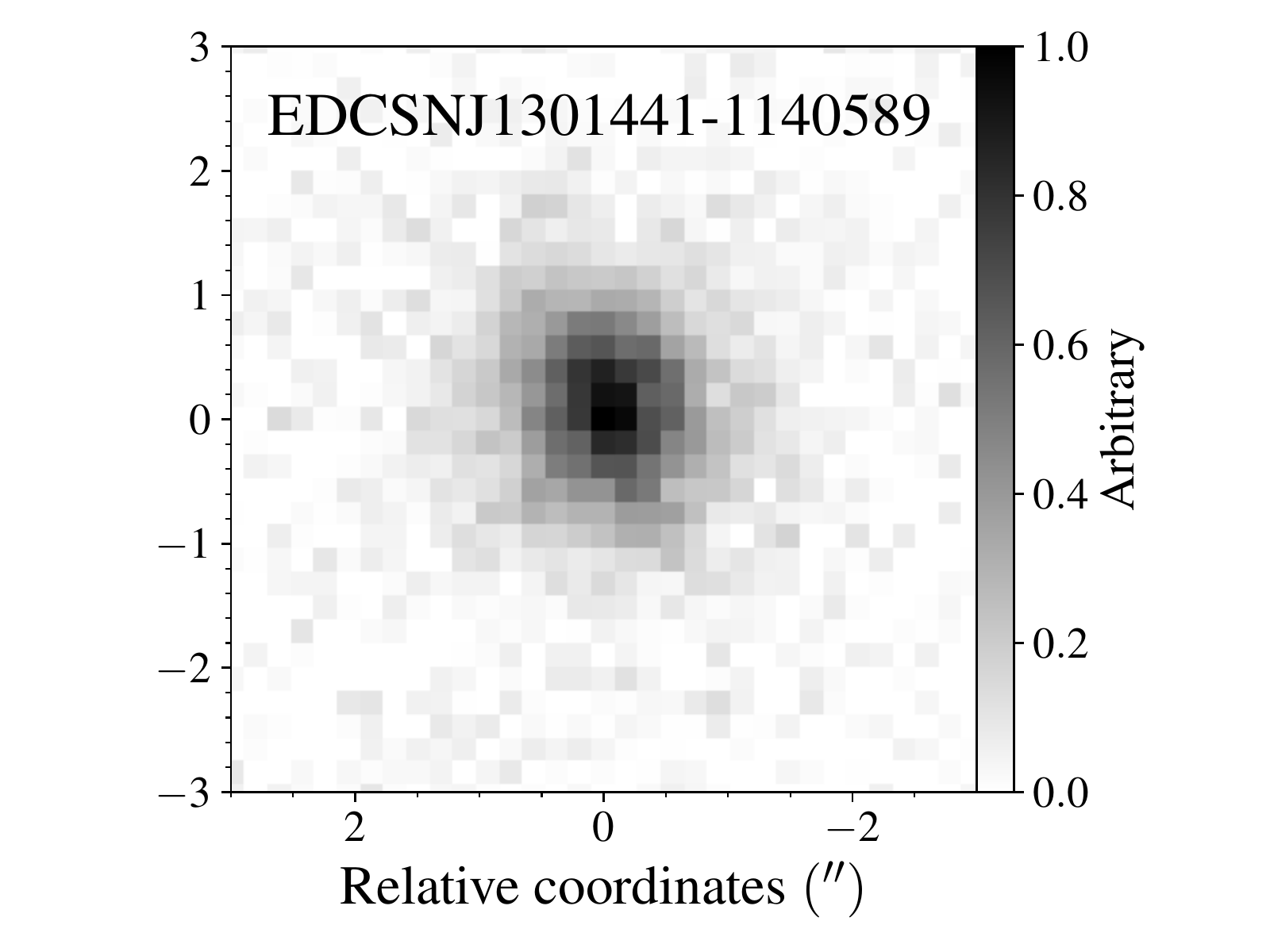}
	\includegraphics[scale=0.3]{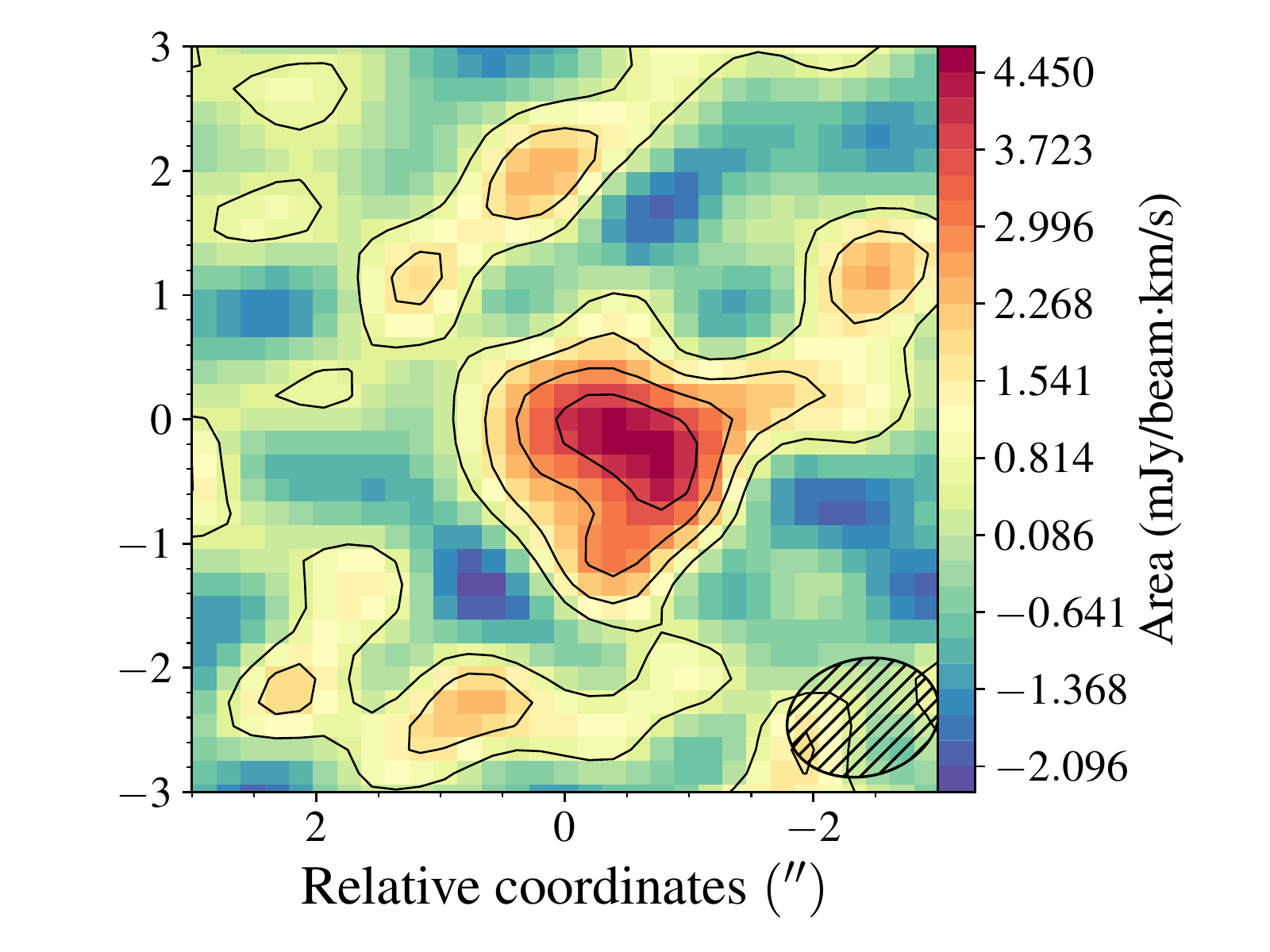}
	\includegraphics[scale=0.3]{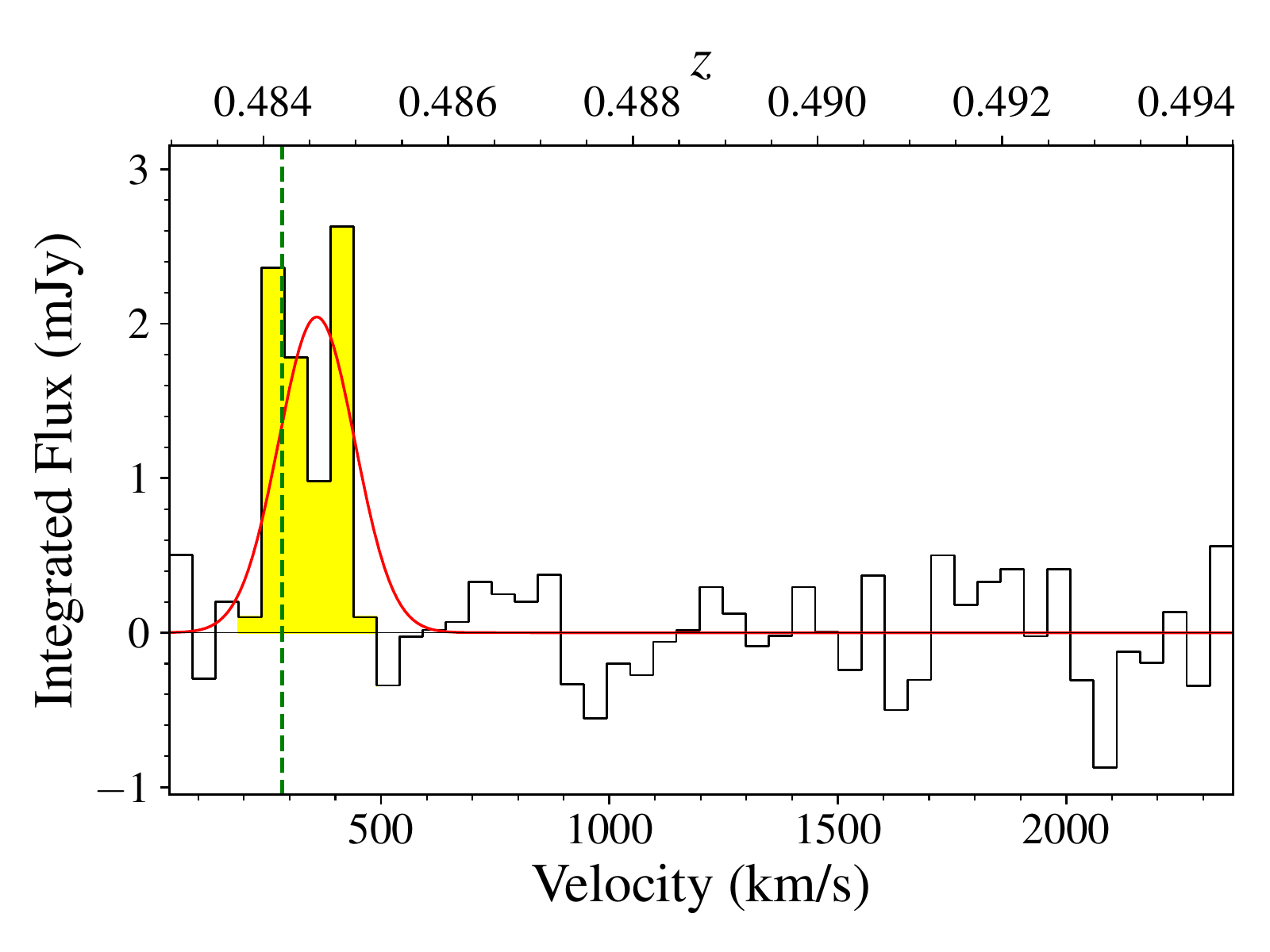}

	\includegraphics[scale=0.3]{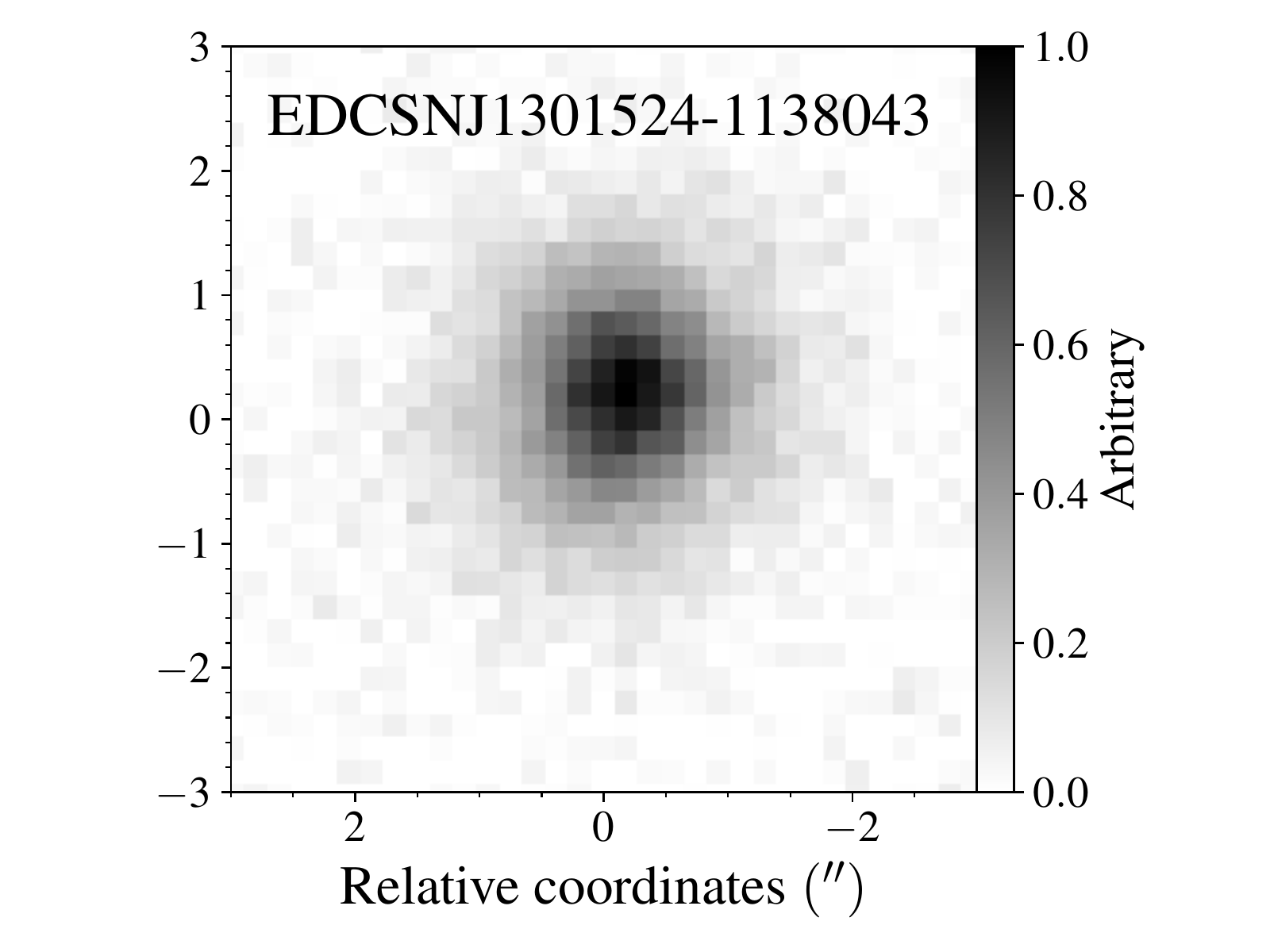}
	\includegraphics[scale=0.3]{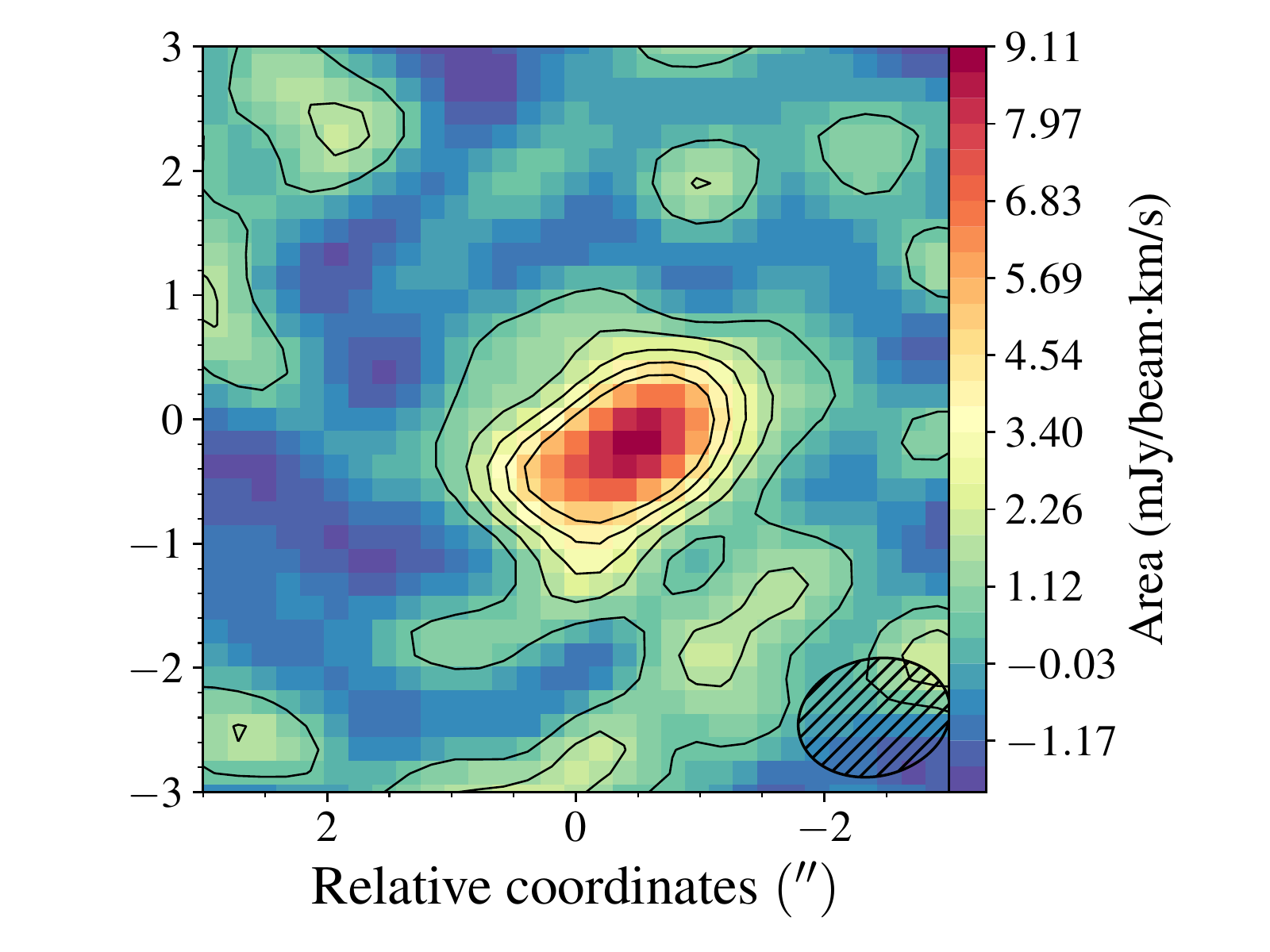}
	\includegraphics[scale=0.3]{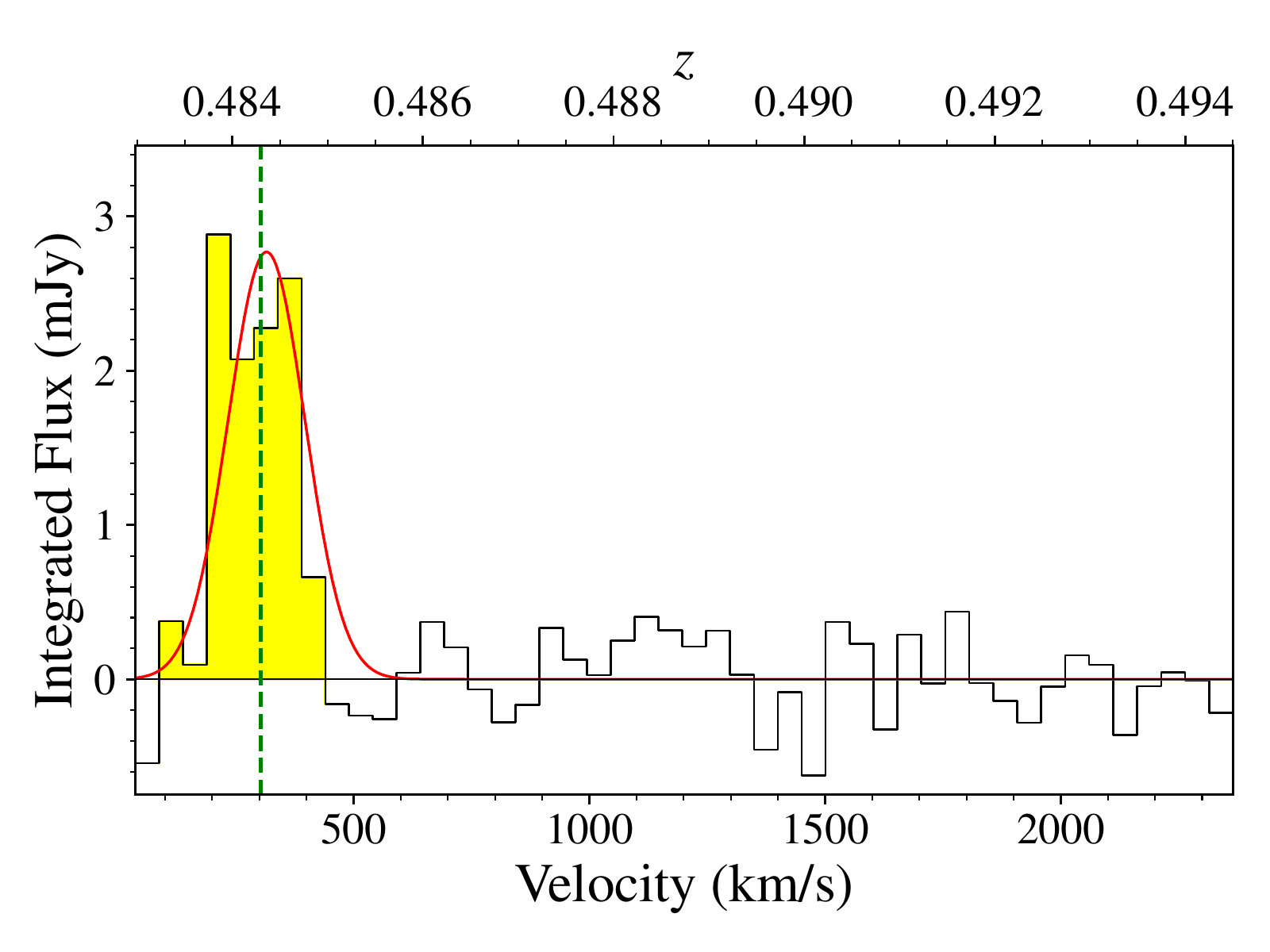}

	\includegraphics[scale=0.3]{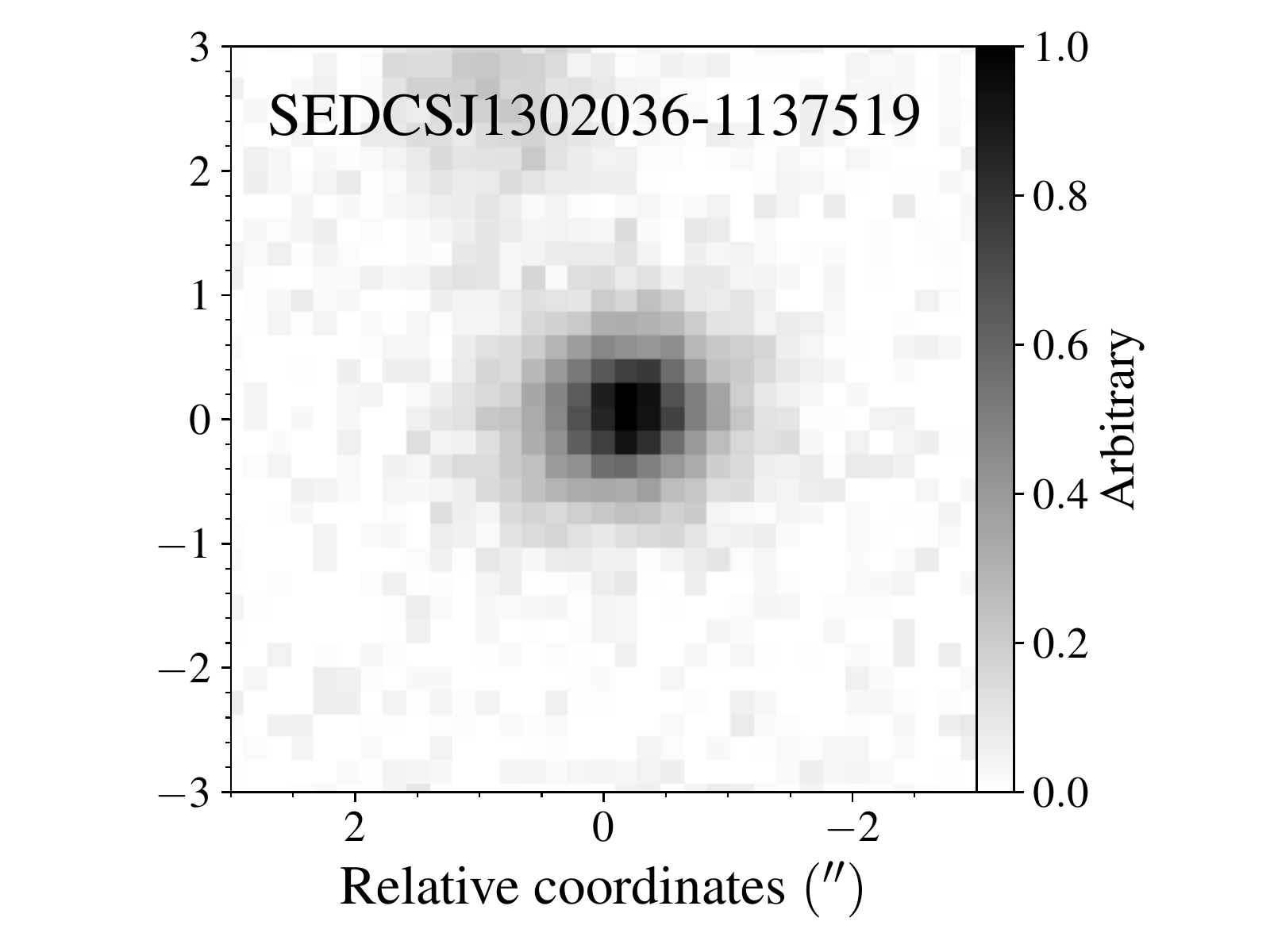}
	\includegraphics[scale=0.3]{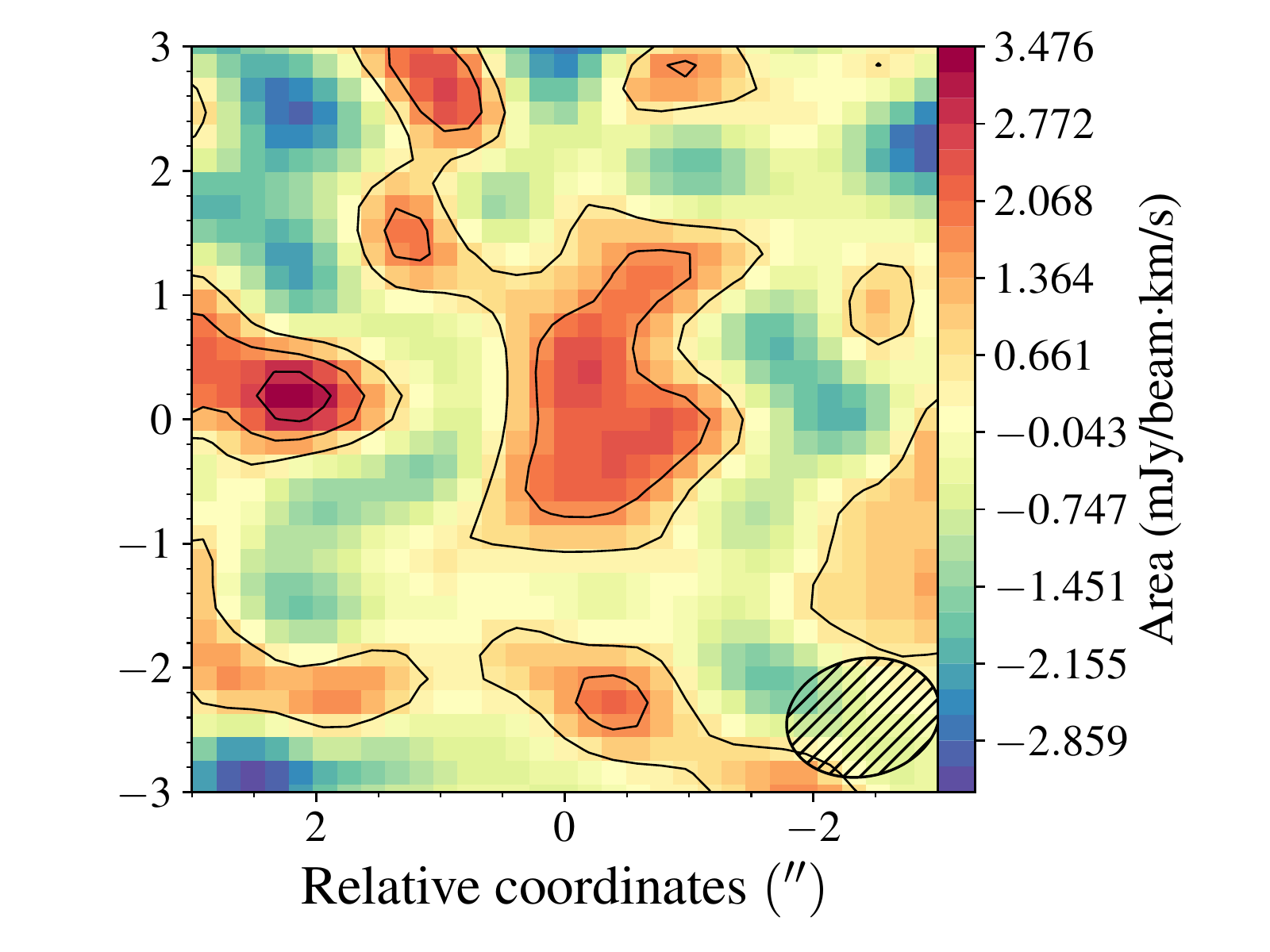}
	\includegraphics[scale=0.3]{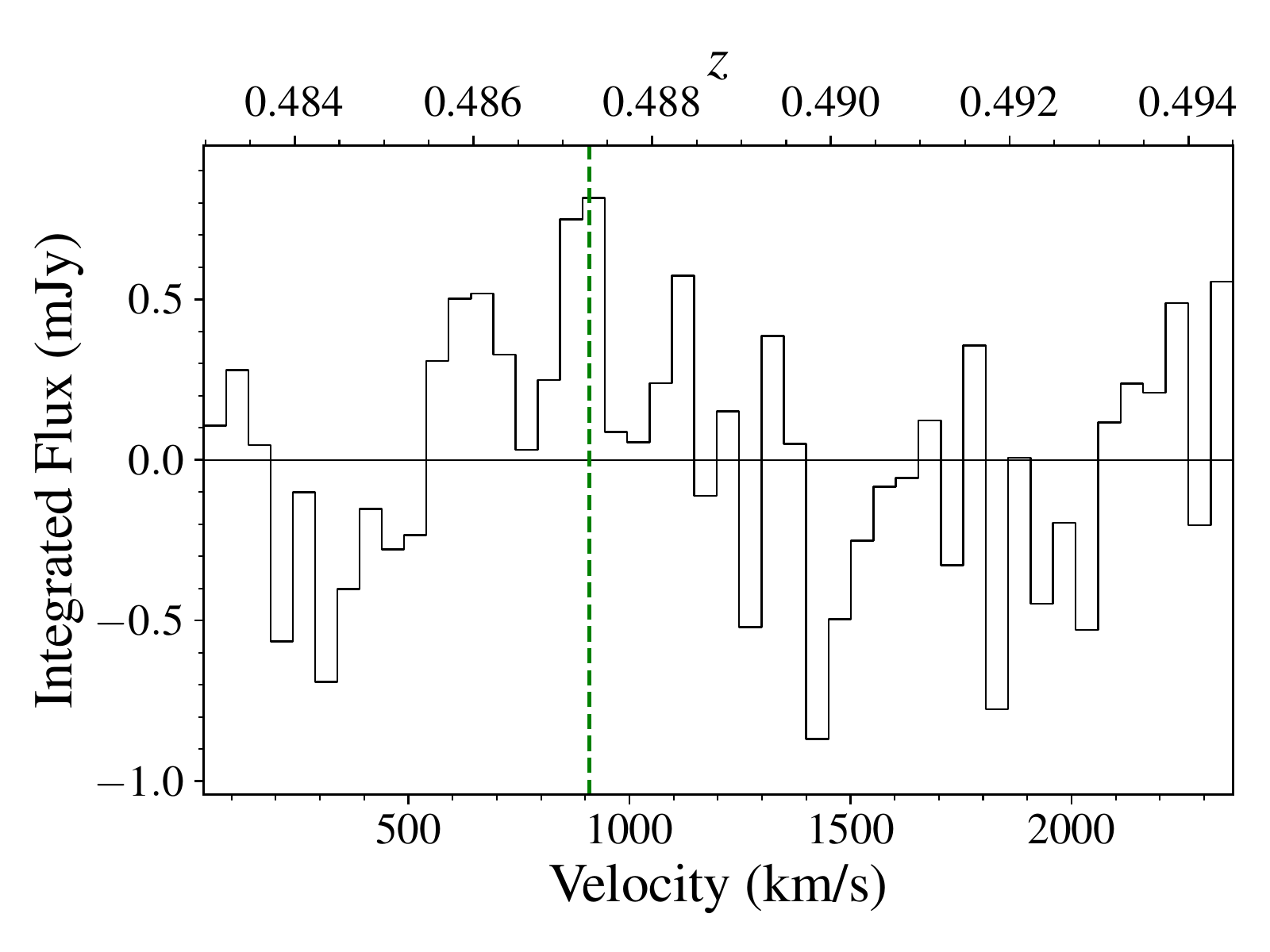}

	\includegraphics[scale=0.3]{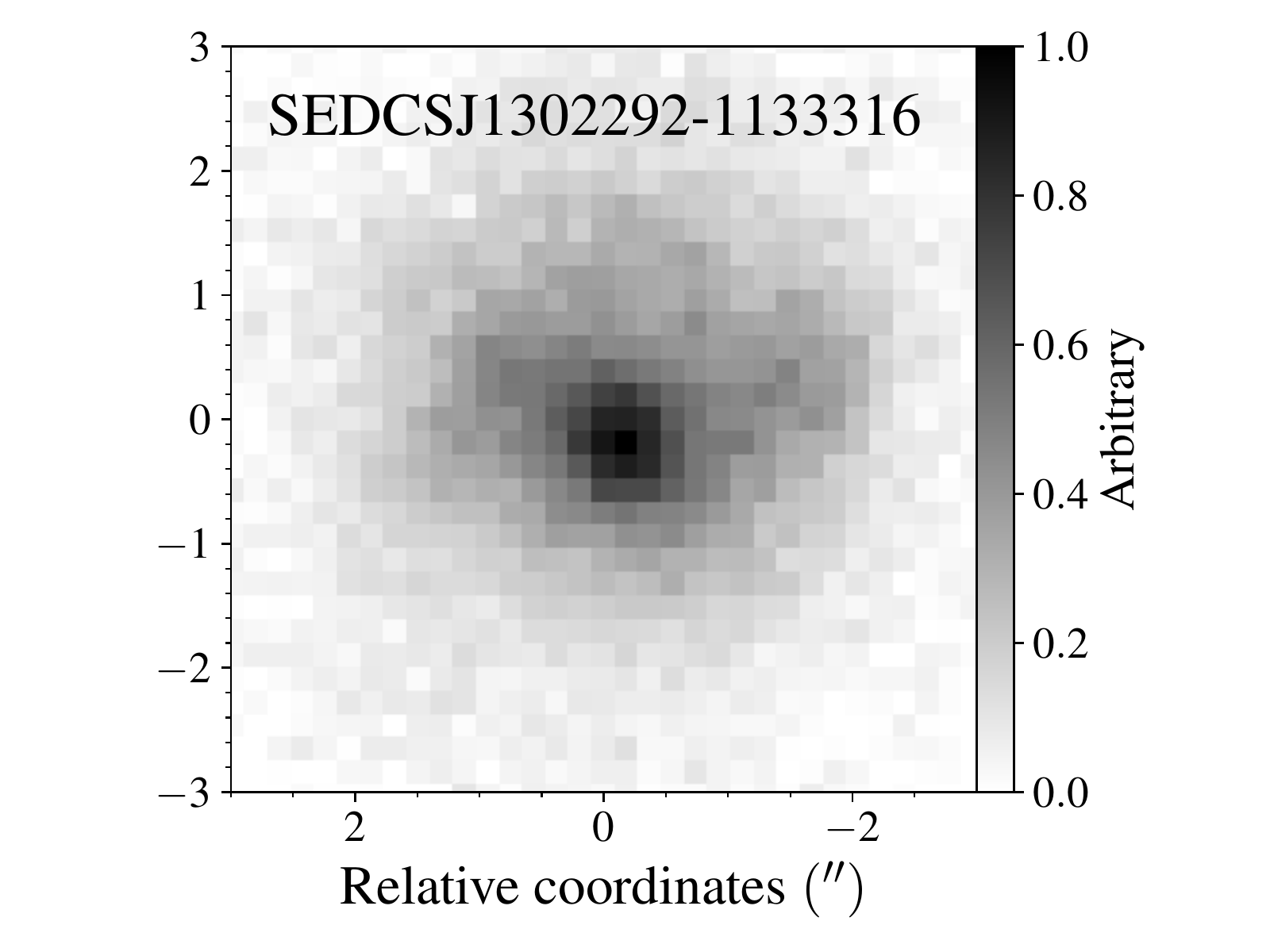}
	\includegraphics[scale=0.3]{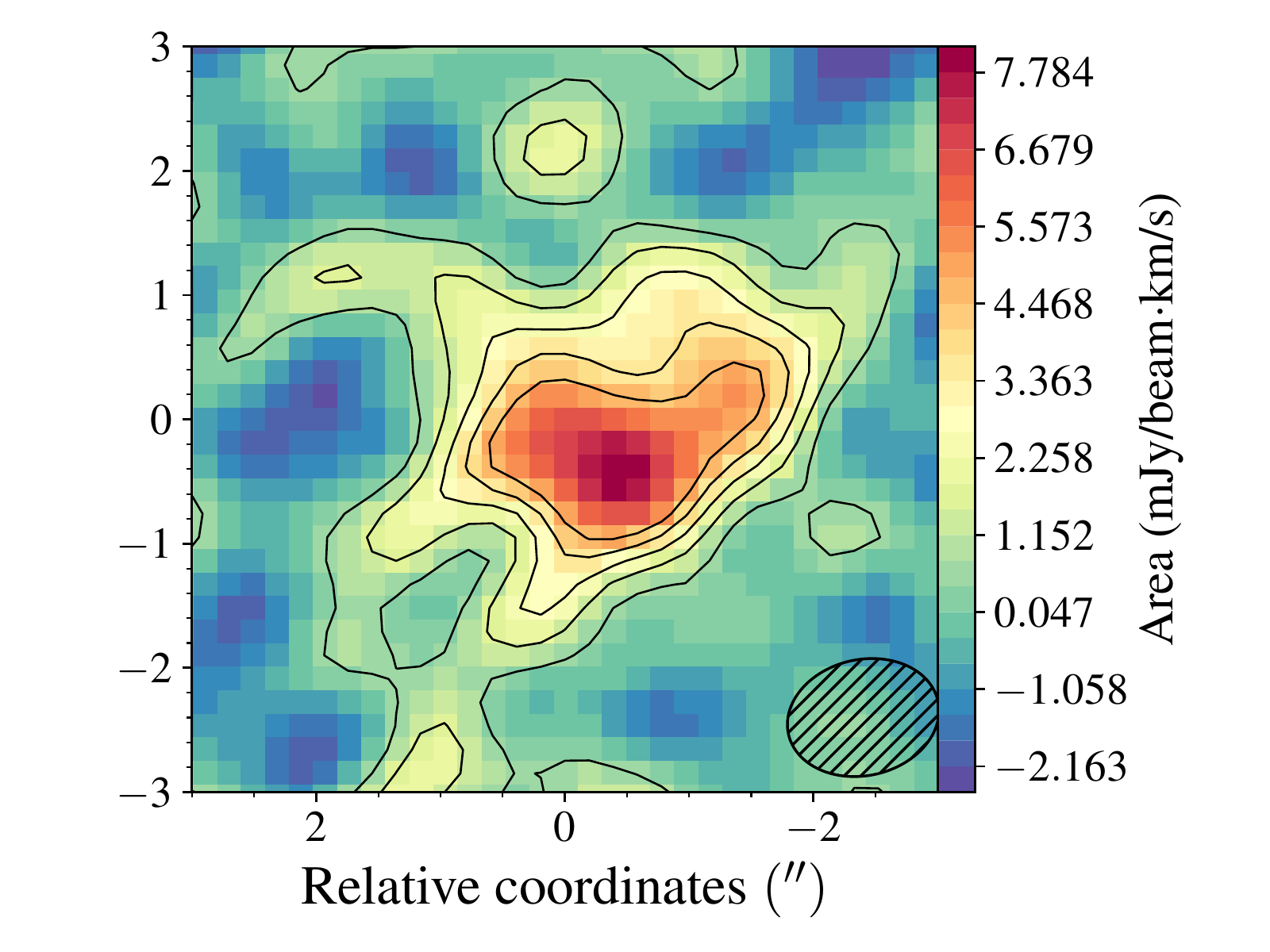}
	\includegraphics[scale=0.3]{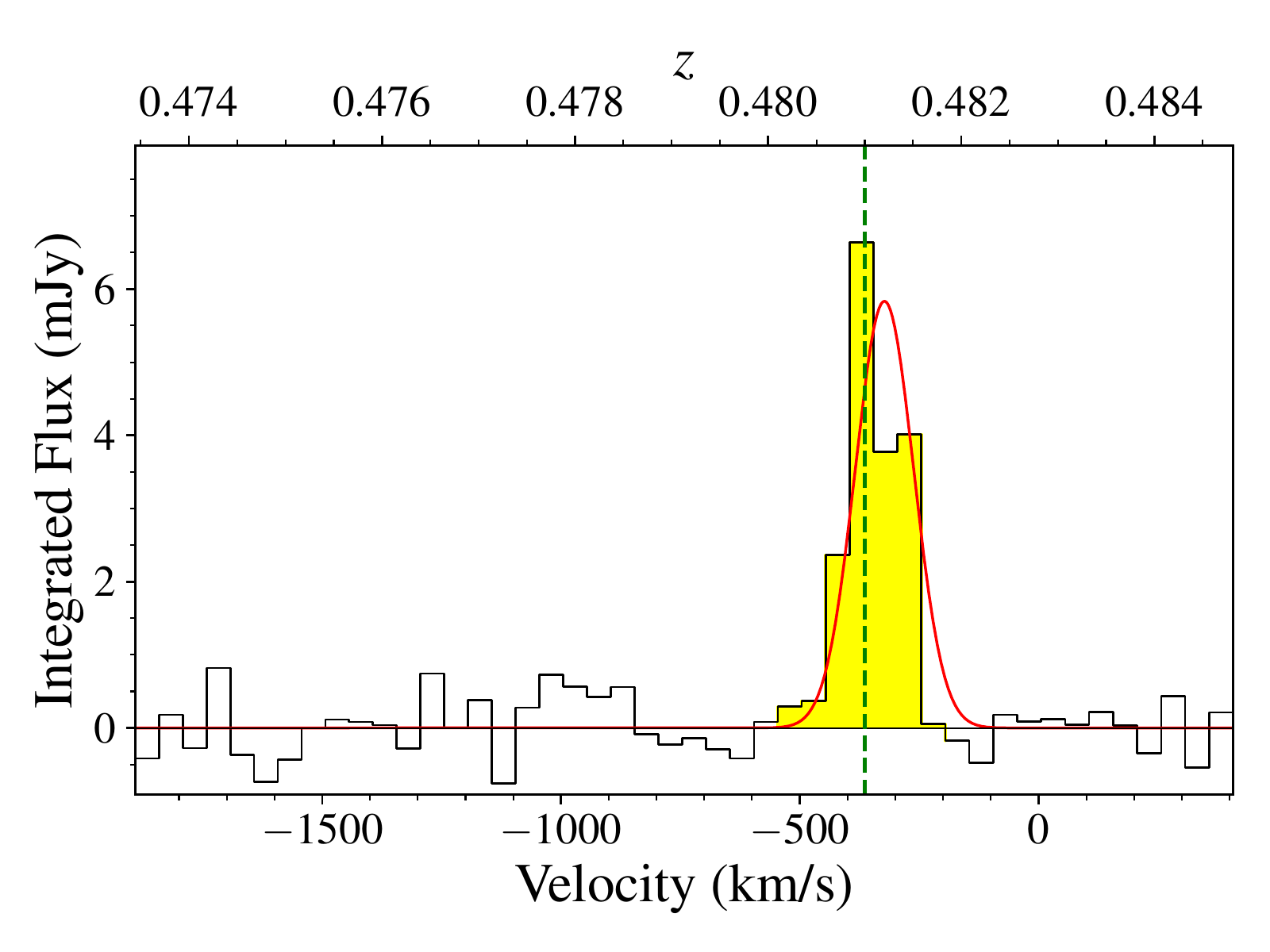}

	\includegraphics[scale=0.3]{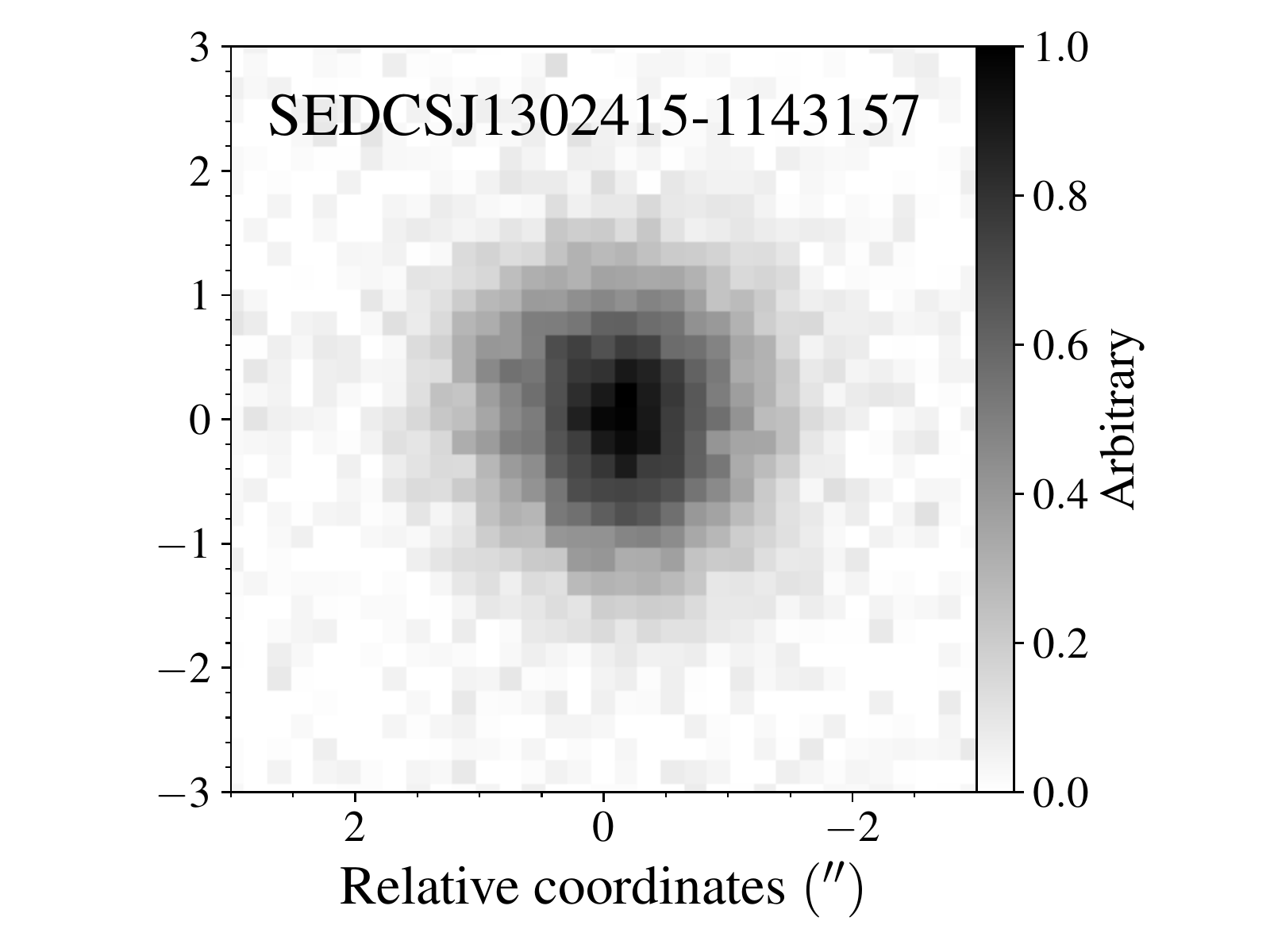}
	\includegraphics[scale=0.3]{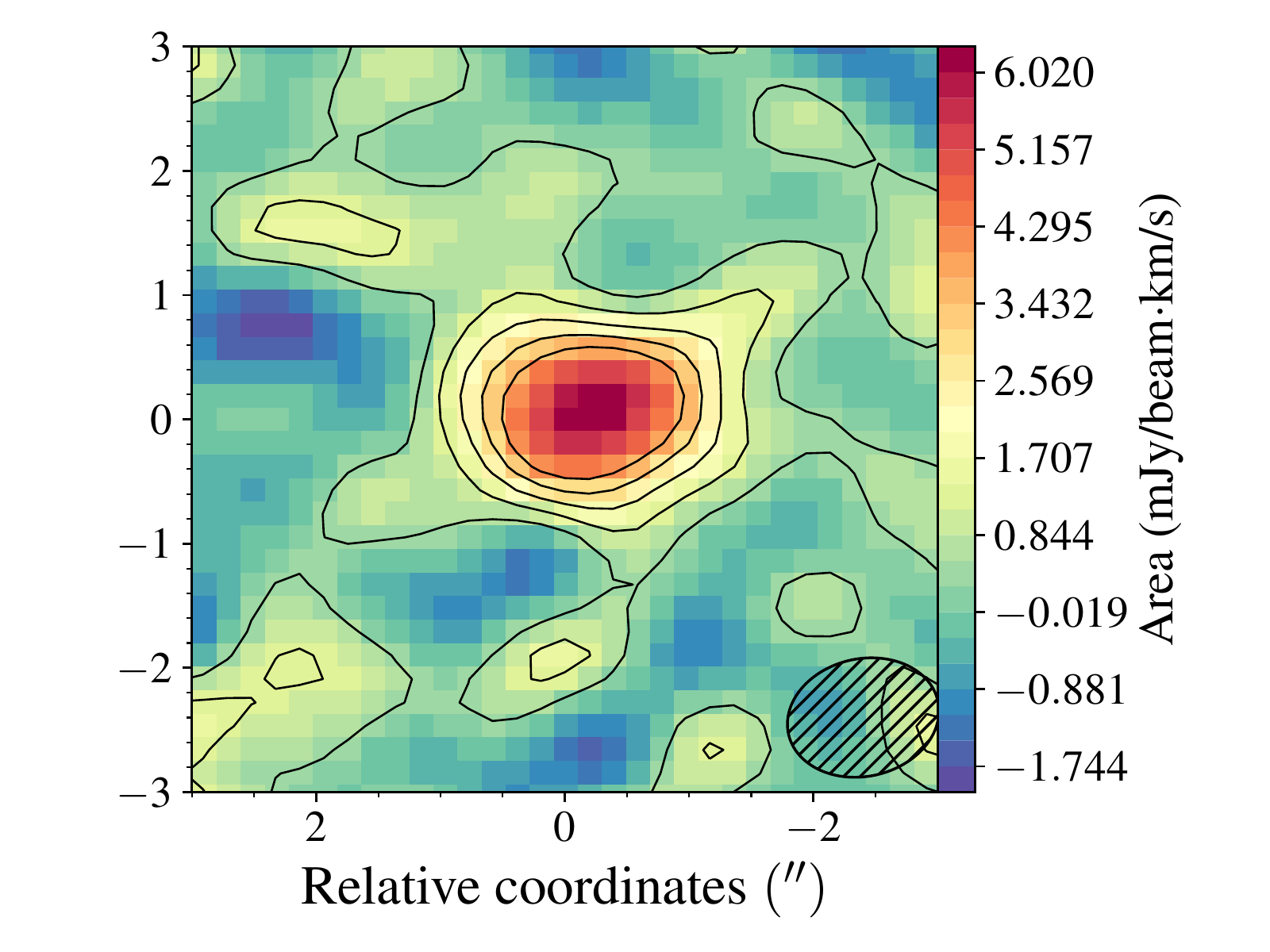}
	\includegraphics[scale=0.3]{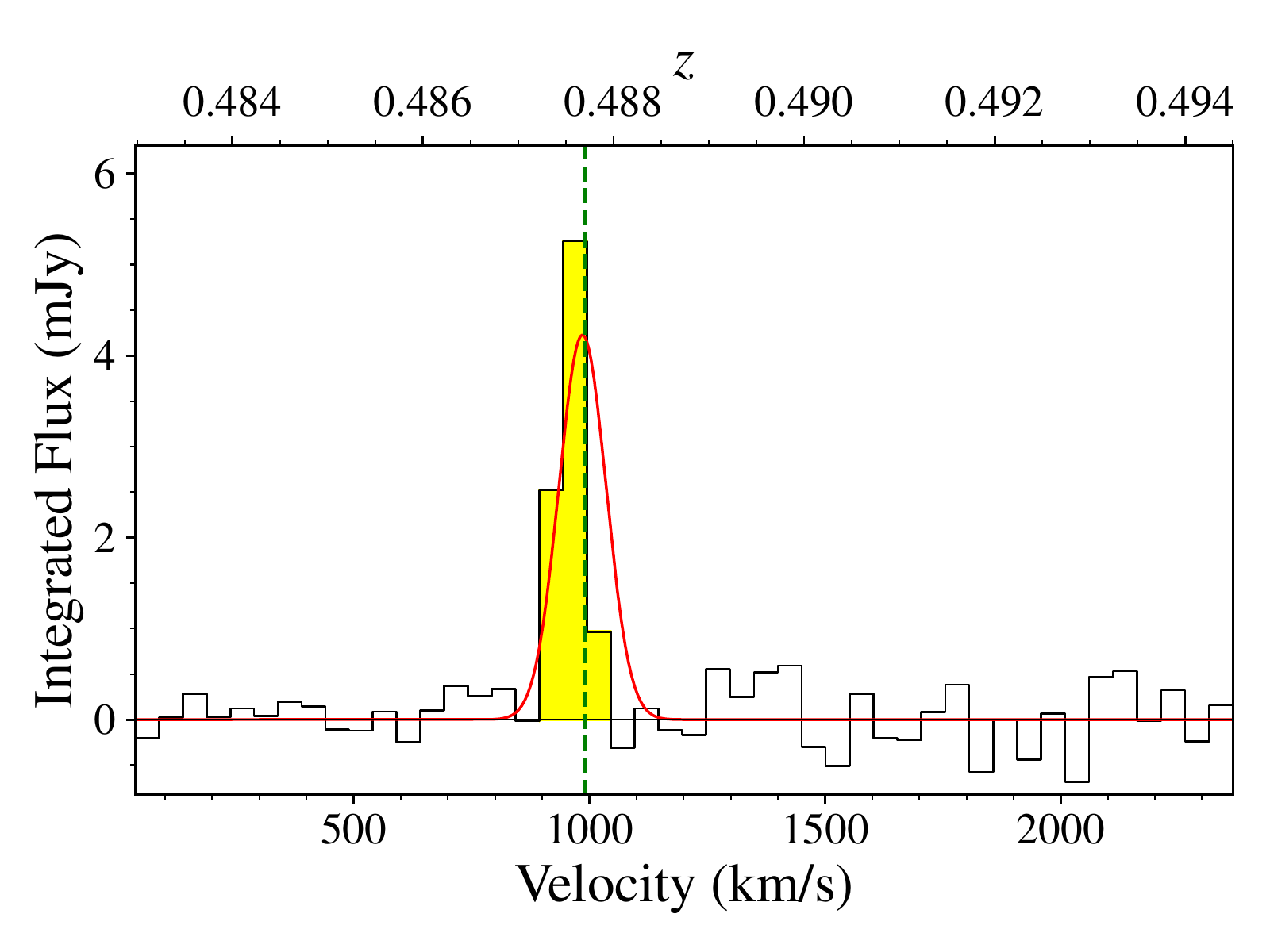}

    \caption{Continued.}
\end{figure*}

\end{appendix}

\end{document}